\documentclass[preprint,12pt]{elsarticle}

\usepackage{graphicx}

\usepackage{amssymb}
\usepackage{amsmath}
\usepackage{amsthm}

\newcommand{\Tr}{\mathrm{Tr}}
\newcommand{\STr}{\mathrm{STr}}

\newcommand{\be}{\begin{eqnarray}}
\newcommand{\ee}{\end{eqnarray}}

\newcommand{\el}{\nonumber \hfill \\}
\newcommand{\nn}{\nonumber }
\newcommand{\fslash}{\hspace{-1.5ex} \slash }

\journal{Physics Reports}

\begin{document}

\begin{frontmatter}



\title{Modeling Finite-Volume Effects and Chiral Symmetry Breaking in Two-Flavor QCD Thermodynamics}
%
%
\author{Bertram Klein}
\ead{bklein@ph.tum.de}
\address{Physik Department, Technische Universit\"at M\"unchen,\\ James-Franck-Stra{\ss}e 1, 85748~Garching, Germany}
\begin{abstract}
Finite-volume effects in Quantum Chromodynamics (QCD) have been a subject of much theoretical interest for more than two decades. They are in particular important for the analysis and interpretation of QCD simulations on a finite, discrete space-time lattice.
Most of these effects are closely related to the  phenomenon of spontaneous breaking of the chiral flavor symmetry and the emergence of pions as light Goldstone bosons. These long-range fluctuations are strongly affected by putting the system into a finite box, and an analysis with different methods can be organized according to the interplay between pion mass and box size. The finite volume also affects critical behavior at the chiral phase transition in QCD.
In the present review, I will be mainly concerned with modeling such finite volume effects as they affect the thermodynamics of the chiral phase transition for two quark flavors.

I review recent work on the analysis of finite-volume effects which makes use of the quark-meson model for dynamical chiral symmetry breaking.  To account for the effects of critical long-range fluctuations close to the phase transition, most of the calculations have been performed using non-per\-tur\-ba\-tive Renormalization Group (RG) methods. I give an overview over the application of these methods to a finite volume.
The method, the model and the results are put into the context of related work
in random matrix theory for very small volumes, chiral perturbation theory for larger volumes, and related methods and approaches. They are applied towards the analysis of finite-volume effects in lattice QCD simulations and their interpretation, mainly in the context of the chiral phase transition for two quark flavors. 
\end{abstract}


\end{frontmatter}


\tableofcontents

\newpage 

\section{Introduction}
\label{sec:introduction}
The study of finite-volume effects in Quantum Chromodynamics (QCD) has already a long history. Many different methods have been used to analyze very different types of phenomena. From an experimental point of view, such investigations appear to be of little value: it is hard to conceive of systems that are small enough to lead to observable finite-volume effects, since the length scales involved are so small compared to the typical extent of the system. 

The necessity of performing simulations of QCD on finite, discrete Euclidean space-time lattices~\cite{Wilson:1974sk} is then what mainly motivates the study of finite-volume effects. Since the gauge coupling in QCD is large on low momentum scales, it requires non-perturbative methods, and Monte-Carlo simulations of the QCD path integral~\cite{Creutz:1979dw,Wilson:1979wp} remain an important benchmark for any non-perturbative method and for model calculations. Originally, even here the investigation of finite-volume effects was more of academic interest and served mainly as a check on simulation method and algorithm. 

However, with the steady improvements of lattice simulations of full QCD in numerical calculations, the question of finite-volume effects has become more and more relevant. Historically, such simulations are hampered by discretization effects from the finite lattice spacing, by large quark masses from an imperfect implementation of the chiral symmetry of QCD, and from the limited extent of the simulation volumes.
The calculational advances have made it possible to bring the masses of the lightest degrees of freedom in these simulations, the pions, down to or even below their actual physical values~\cite{Fodor:2004nz,Aoki:2005vt,Ejiri:2009ac,Durr:2010vn,Durr:2010aw,Colangelo:2010et,Bhattacharya:2014ara}. As a result, long-range effects have become more and more important: If the wavelength of the pions approaches the extent of the simulation volume, finite-volume effects are going to become more and more significant.
This is especially true close to the so-called chiral phase transition, where the critical behavior is dominated by the critical fluctuations of light degrees of freedom. 

The ground state of QCD is characterized by two phenomena that fundamentally shape nuclear physics: The confinement of color charges, and the spontaneous breaking of the chiral flavor symmetry. Both are 
closely connected with symmetries of the QCD Lagrangian.
The ${\rm U}(N_f) \times {\rm U}(N_f)$ flavor symmetry of the massless QCD Lagrangian is broken explicitly by the axial anomaly, and then further broken spontaneously by the formation of a chiral condensate in the QCD vacuum. As a consequence of the spontaneous breaking of a continuous symmetry, light Goldstone bosons appear which can be identified as the pions (if we consider only the lightest two quark flavors up and down) and kaons (if we also consider the strange quark flavor). 
We will restrict ourselves in the following to the case of two light quark flavors and a small symmetry-breaking mass. 
The confinement of quarks in turn is closely connected to the spontaneous breaking of the $\mathrm{Z}(3)$ center symmetry of the $\mathrm{SU}(3)$ gauge group of QCD~\cite{Svetitsky:1982gs}: In the pure $\mathrm{SU}(3)$ gauge theory, the center symmetry is spontaneously broken in a phase with deconfined quarks, and restored in the normal vacuum in which no free colored quark states appear. 

Since only the spontaneous breaking of the continuous chiral flavor symmetry involves light degrees of freedom which propagate over long distances, it is this symmetry breaking which is closely connected with finite volume effects in QCD~\cite{Gasser:1986vb}. In fact, the interplay of finite volume and chiral symmetry can tell us much about the mechanisms of the spontaneous symmetry breaking and indeed provides a window on effects on different momentum scales.
Finite-volume effects need to be accounted for and corrected, but exact finite-volume results and finite-size scaling can also be a valuable tool for the analysis of chiral symmetry breaking in QCD.

This review is structured as follows: 
Since two-flavor QCD thermodynamics provide most of the discussed applications, I will briefly review some aspects of the thermodynamics of two-flavor QCD and the QCD phase diagram at finite temperature and baryon chemical potential in the next section.

In the subsequent sections, we will proceed from effects that are observed in systems of very small size to effects in systems of much larger sizes. For very small volumes, where pion modes are effectively static, Random Matrix Theory (RMT) provides an exact description of the QCD Dirac operator spectrum and can provide information about chiral symmetry breaking and its mechanisms~\cite{Shuryak:1992pi,Verbaarschot:1994qf}. Finite-volume RMT investigations have initially provided an important motivation for the application of non-perturbative Renormalization Group methods to finite-volume QCD model systems, which will be considered in the bulk of this review.

In somewhat larger volume sizes, chiral perturbation theory is the method of choice, since it is the effective field theory of QCD at low energy~\cite{Gasser:1983yg,Gasser:1984gg,Gasser:1987zq}. It captures the finite-volume effects in this region, where the pions are the relevant dynamical degrees of freedom. Chiral perturbation theory, however, is predicated on the spontaneous breaking of chiral symmetry and thus not applicable at the transition to a phase with restored symmetry~\cite{Gasser:1986vb,Gasser:1987ah}. 

In order to describe the transition between phases with broken and restored chiral symmetry, models such as the Nambu--Jona-Lasinio model~\cite{Nambu:1961tp} and its derivatives still prove extremely useful. They can be employed to provide insight into the relevant mechanisms. 
Since critical fluctuations become essential close to the critical temperature~\cite{Wegner:1972ih,Wilson:1973jj}, these need to be taken into account. They are also essential for capturing finite-size scaling effects~\cite{Fisher:1972zza}. For this reason, non-perturbative Renormalization Group (RG) methods~\cite{Wetterich:1992yh,Liao:1994fp} have been applied to these models, which also allow a consistent description across a wide range of length scales and temperatures~\cite{Berges:2000ew}. In this context we also review the application of RG methods to finite-volume systems developed in \cite{Braun:2004yk,Braun:2005gy,Braun:2005fj,Braun:2007td,Braun:2008sg}. 
Incidentally, here we also need to raise the question of the choice of spatial boundary conditions for the different fields, and we find a significant effect on the low-energy structure of the models, and unexpected effects for the most common choice of quark boundary conditions in lattice QCD simulations.  

Finally, the results of these investigations of the relevant finite-size scaling behavior and of phenomenological finite-volume effects are applied towards the interpretation of lattice QCD simulation results. They give guidance on the pion mass and volume size regime in which a finite-size scaling analysis is applicable and on the deviations from infinite-volume behavior to be expected in the finite-volume simulation results for a given pion mass. Lastly, they can help to explain discrepancies between lattice simulation results for different volume sizes.  


\section{QCD thermodynamics for $N_{\mathrm{f}}=2$ flavors} 

In order to provide some context for the phenomena discussed in the following, I will briefly review some features of the QCD phase diagram for $N_{\mathrm{f}}=2$ flavors. This is neither intended to be nor can it be a full overview over the state of knowledge about QCD thermodynamics, which is beyond the scope of this review. For more complete and comprehensive reviews of QCD phases see e.g.~\cite{Fukushima:2010bq,Huang:2010nn,Satz:2009hr,Ding:2015ona}.

In the low-temperature regime, quarks and gluons are confined, i.e. free color charges cannot be observed, and instead we observe baryonic and mesonic degrees of freedom. From the existence of light mesons with much heavier parity partners in this regime, it can also be concluded that the chiral flavor symmetry in the quark sector of QCD is spontaneously broken in the ground state of the low-temperature regime. 
From general arguments it then follows that there must be at least one phase transition and possibly two phase transitions between the low and high temperature phases of QCD. 
These deconfinement and chiral phase transitions are not necessarily at the same temperature, although there are indications from lattice simulations~\cite{Aoki:2005vt} that both transitions might coincide.

Our expectations for the QCD phase diagram as a function of temperature and baryon density, or alternatively baryon chemical potential, are shaped by the symmetries of QCD. These are in particular the chiral flavor symmetry, and the $\mathrm{Z}(3)$ center symmetry of the $\mathrm{SU}(3)$ gauge group. Since both these symmetries are broken explicitly in the presence of quarks with finite mass, from a theoretical point of view it is useful to consider the masses of the three lightest u, d, and s quarks, $m_{\mathrm{u}}$,  $m_{\mathrm{d}}$, and $m_{\mathrm{s}}$, as free parameters, and to think about the limiting cases of vanishing or  infinite quark masses.  Physically, $m_{\mathrm{u}} \approx m_{\mathrm{d}} \ll m_{\mathrm{s}}$, and the strange quark mass $m_{\mathrm{s}}$ is of the same order of magnitude as the temperatures for chiral symmetry restoration and deconfinement, so that these two (or three) lightest quarks dominantly shape the behavior of QCD at these phase transitions, with very little contribution from heavier quarks. 

\begin{figure}
\begin{center}\includegraphics[scale=0.55,clip=true]{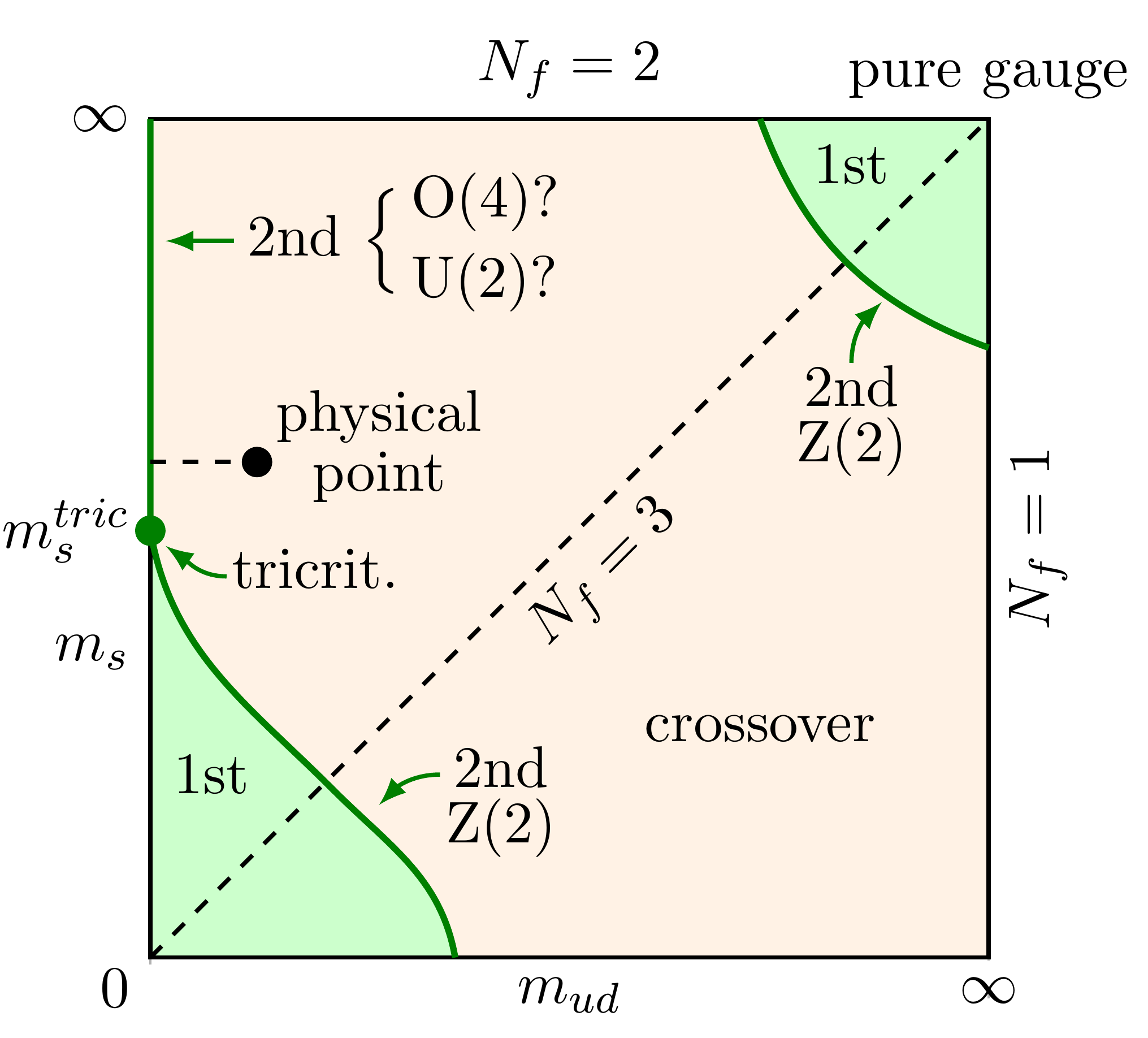}\end{center}
\caption{The \emph{Columbia plot}, here in a representation from Brandt \emph{et al.}~\cite{Brandt:2016daq}, indicates the order of the QCD phase transition as a function of the masses of the three lightest quarks, $m_u$, $m_d$, and $m_s$, where the masses of the lightest quarks are taken to be degenerate for simplicity, $m_u=m_d$. In the standard rendition, there is a tricritical value $m_s^{\mathit{tric}}$ of the strange quark mass, above which the transition is of second order and, depending on the strength of the anomaly,  of the $\mathrm{O}(4)$ or the $\mathrm{U}(2)$ universality class, possibly then also a first order transition. \label{fig:Columbia}}
\end{figure}

The relevance of the QCD symmetries for the phase diagram is nicely illustrated in the so-called \emph{Columbia plot}, which depicts the order of the QCD phase transitions and their dependence on the values of the u, d, and s quark masses. Fig.~\ref{fig:Columbia} shows a current rendition of the Columbia plot from~\cite{Brandt:2016daq}.  The masses of the lightest quarks are taken to be degenerate for simplicity, $m_u=m_d$, and the axes of the diagram correspond to the values of the light ($m_u=m_d$) and strange ($m_s$) quark masses, respectively.
For three light flavors (lower left corner, $m_u=m_d\to 0$, $m_s\to 0$), the chiral phase transition is of first order. For three infinitely heavy flavors (upper right corner, $m_u=m_d\to \infty$, $m_s\to \infty$), the pure gauge theory has a first order phase transition associated with the spontaneous breaking of the $\mathrm{Z}(3)$ center symmetry of the $\mathrm{SU}(3)$ gauge group. In between, the phase transition is a crossover, since both the center symmetry of the gauge group and the chiral flavor symmetry are explicitly broken by finite quark masses, and there is no symmetry restoration associated with the transition. These regions are separated by lines on which a second order critical phase transition takes place, which in each case is expected to be of the $\mathrm{Z}(2)$ Ising universality class, since no additional symmetry is associated with these transitions. 
Current indications are that at physical values of the quark masses, the transition is a crossover.
For $N_f=2$ massless quark flavors, on the axis $m_u=m_d=0$, and for sufficiently large values of $m_s$, the phase transition is expected to be of second order and in the $\mathrm{O(}4)$ universality class (assuming a sufficiently strong axial anomaly). In this case, the second order $\mathrm{Z}(2)$ critical line and the second order $\mathrm{O}(4)$ critical line must meet in a tricritical point at some value $m_s^{tric}$ of the strange quark mass. 
Depending on the physical value $m_s^{phys}$ of the strange quark mass in comparison to the tricritical value $m_s^{tric}$, one expects to see different remnants of critical behavior at the physical point. In the case that $m_s^{phys}>m_s^{tric}$, one expects that the crossover behavior at the physical point can be related to $\mathrm{O}(4)$ critical behavior, and that in the limit of $m_u=m_d \to 0$, one recovers a second-order phase transition in the $3$-$d$ $\mathrm{O}(4)$ universality class.

There is an important argument about the order of restoration of the axial symmetry and the restoration of the chiral symmetry at large temperatures and the consequences of one occurring before the other with increasing temperature~\cite{Pisarski:1983ms,Butti:2003nu,Pelissetto:2013hqa}. 
While a detailed discussion of this topic and related finite-volume effects is beyond the scope of this review, 
the question is, however, not completely unrelated to the investigations reviewed here: The possibility of an $\mathrm{O}(4)$-type chiral phase transition depends on a scenario with sufficiently strong breaking of the axial $\mathrm{U}_{A}(1)$ symmetry at the transition temperature.
For a recent discussion of the interplay between the strength of the $\mathrm{U}_{A}(1)$ breaking and the nature of the chiral phase transition and its universality class see e.g. \cite{Pelissetto:2013hqa,Cossu:2014aua,Ding:2015ona,Brandt:2016daq}.
Indications that the axial $\mathrm{U}_A(1)$ symmetry is not restored at the physical point of the quark masses at the chiral crossover come from lattice simulation, e.g.~\cite{Bhattacharya:2014ara}. Other lattice investigations find evidence that disfavors an $\mathrm{O}(4)$ transition~\cite{Brandt:2016daq}. In a simulation with a fermion action minimizing chirality violations~\cite{Cossu:2014aua,Cossu:2015lnb}, indications of a suppression of the $\mathrm{U}_{A}(1)$ symmetry breaking compatible with an effective restoration of the axial symmetry at large temperatures have been observed.

While there are strong indications from lattice simulations for a crossover in the chiral phase transition for physical quark masses~\cite{Aoki:2005vt,Borsanyi:2010cj,Bazavov:2011nk,Bhattacharya:2014ara}, 
the order of the phase transition in the continuum limit for two massless flavors is still considered to be an open issue~\cite{Philipsen:2016hkv} with important implications for the overall phase structure of QCD.
Current attempts to reach the chiral limit for $N_f=2$ from QCD with an imaginary chemical potential \cite{Philipsen:2015eya,Philipsen:2016hkv} try to shed additional light on this issue, as do attempts with $2$ light and  $N_f$ heavy quark flavors, simulated at finite chemical potential, using a reweighting method~\cite{Ejiri:2016oit}.
Lattice simulation results in the context of a scaling analysis will be discussed below in Section~\ref{sec:scalinglattice}.

Starting from the Columbia plot, we can consider what happens to the phase transition if we introduce the baryon chemical potential $\mu_B$ as an additional parameter. This extends the plot in a third dimension. 
Taking the quark masses at their physical values, we can read off from the plot what happens to the QCD phase transition if we increase the baryon chemical potential. For this it is mainly relevant what happens to the boundary of the first-order transition region in the lower left corner  of the plot, i.e. the line on which the transition becomes a second-order $\mathrm{Z}(2)$ transition, and the surface that it maps out with increasing baryon chemical potential. 
If this surface bends towards the origin ($m_u=m_d=m_s$), the phase transition at finite chemical potential would always remain a crossover. If it bends towards larger quark masses, the transition at the physical quark masses would eventually become first order~\cite{deForcrand:2003hx,deForcrand:2008vr}.
The conventional expectation is that the first-order transition region in the lower left corner of the Columbia plot increases in size with increasing baryon chemical potential, and at some value of $\mu_B$ its boundary reaches the physical point. At this point, the phase transition would become second order in the $3$-$d$ $\mathrm{Z}(2)$ universality class, and above this value it would be of first order.
\begin{figure}
\begin{center}\includegraphics[scale=0.55,clip=true]{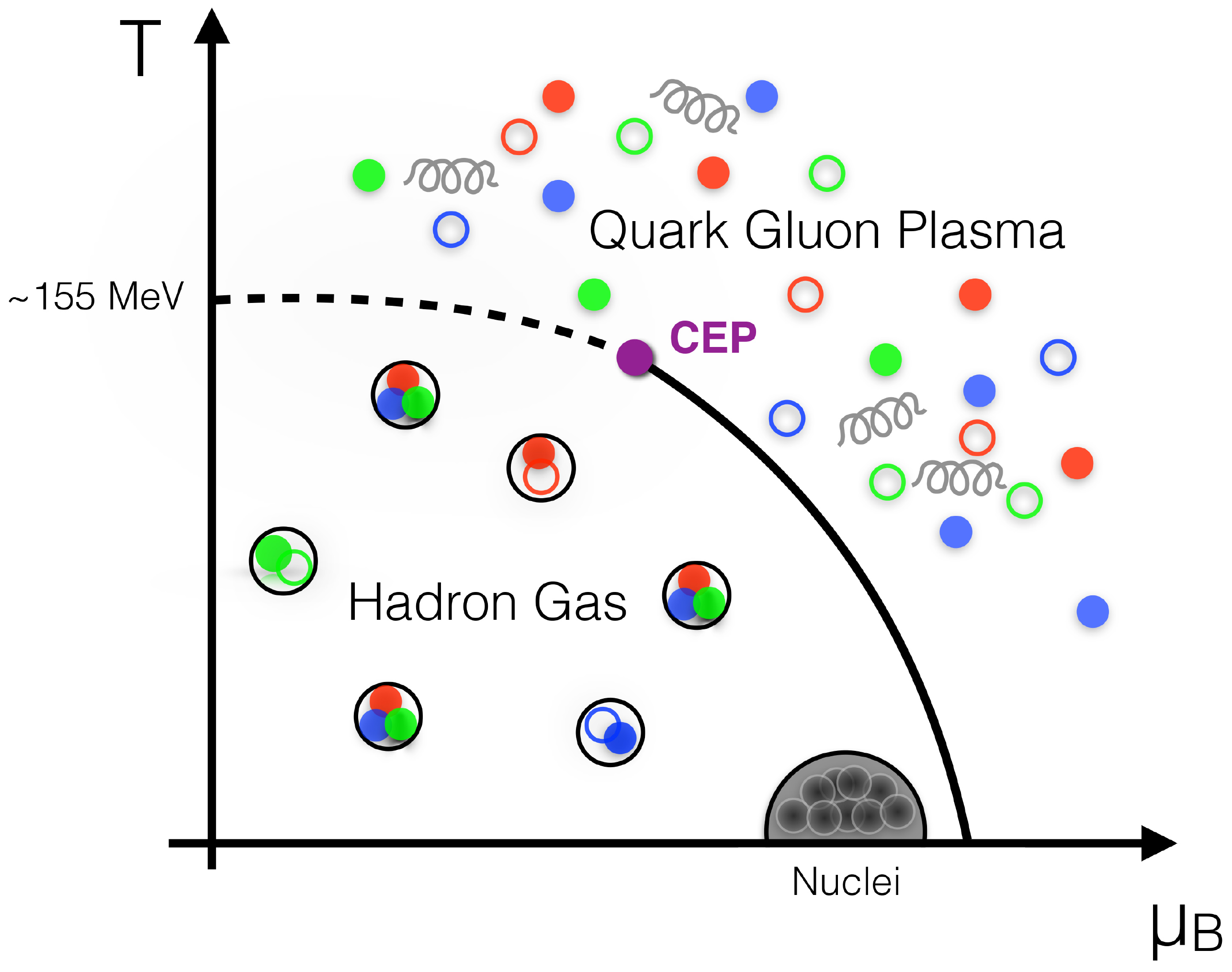}\end{center}
\caption{Schematic phase diagram of QCD in the baryon chemical potential ($\mu_B$)  temperature ($T$) plane from~\cite{Ding:2015ona}. At vanishing baryon chemical potential $\mu_B$ and for physical quark masses, lattice simulations indicate that chiral symmetry restoration and deconfinement take place in a continuous crossover, not a proper phase transition. The behavior at this transition is expected to reflect the critical behavior at the phase transition in the chiral limit of vanishing quark masses. At large baryon chemical potential, the transition is expected to be of first order, with the line of first-order transitions terminating in a second-order critical end point (CEP).\label{fig:pdmuTschematic}}
\end{figure}
The resulting phase diagram of QCD in the baryon chemical potential ($\mu_B$) temperature ($T$) plane for this standard scenario and physical quark masses is shown in Fig.~\ref{fig:pdmuTschematic}, taken from the review~\cite{Ding:2015ona}.  
The search for the first order transition line and its critical end point (CEP) is the subject of numerous theoretical studies and of numerous experiments and a question that has commanded considerable interest in recent years.

At large baryon densities and low temperatures, many new exotic phases of dense baryonic matter have been proposed and investigated, see e.g.~\cite{Fukushima:2010bq} for a review. The stability of such phases against fluctuations has also been researched by means of Renormalization Group methods.

Lattice simulations are a very important method to explore the QCD partition function and its properties from the theoretical side. 
Quark and gluon fields are put on a finite, discrete space-time lattice and the partition function, correlation functions and other quantities are calculated by Monte-Carlo sampling with the QCD action.
Since it is non-local and connects many sites of such a lattice, the determinant in the partition function, which arises from the fermionic (quark) path, is numerically costly to calculate. For this reason, in many early calculations it has been neglected, which means that effects due to quark loops are not present in these "quenched" calculations, and there is e.g. no dynamical pion exchange. 
In the presence of a finite baryon chemical potential, the quark Dirac operator looses its hermiticity properties, and the eigenvalues become complex. Since this leads to a complex phase in the partition function, it makes Monte-Carlo sampling difficult (this is the so-called "sign" or phase-problem, which arises at finite density or chemical potential for the quark fields). This has in the past provided an additional motivation to perform approximate, "quenched" calculations~\cite{Stephanov:1996ki}.
Nowadays, lattice simulation methods have become much more capable and sophisticated and can be performed at physical values of the pion mass, and they have also been extended to include a finite baryon chemical potential by various methods.
However, mainly due to the challenges associated with discretizing fermion fields on a lattice, the extrapolation from finite lattice spacing to the continuum remains difficult and discrepancies between different calculations remain. Despite these challenges, lattice simulations are an important source of our present knowledge about the features of the QCD phase diagram.
Some lattice results from extending the simulations to finite chemical potential will be discussed below in Sections~\ref{sec:curvature} and \ref{sec:qsusc}. 
For a review of recent lattice results for QCD thermodynamics, see e.g.~\cite{Ding:2015ona}. 
With regard to finite-volume effects, the presence of light, dynamical pions in current lattice simulations implies an increased importance of finite-volume effects due to long-range fluctuations.


\section{Random Matrix Theory}
In extremely small volumes, it is possible to describe QCD in terms of a static theory. This description relies purely on the symmetries of QCD and their breaking pattern in the QCD ground state. One method used in such a description is Random Matrix Theory (RMT), which has been successfully used to predict the microscopic eigenvalue spectrum of the QCD Dirac operator. A comprehensive review of RMT and chiral symmetry breaking in QCD is given in~\cite{Verbaarschot:2000dy}.
We briefly review some of the central ideas, since they have provided an important motivation for the first attempts to model chiral symmetry breaking in a finite volume with Renormalization Group methods~\cite{Spitzenberg:2002tq}.

Random matrix theory was originally developed in the context of correlations in nuclear energy levels of large nuclei and goes back to work by Wigner~\cite{Wigner:1951,Wigner:1958js} and Dyson~\cite{Dyson:1953zz}. For a review of these applications see~\cite{Weidenmuller:2008vb}.
The basis of RMT is a hypothesis of \emph{spectral universality}, according to which  spectral correlations in certain quantum-chaotic systems depend only on the symmetries of the system \cite{Bohigas:1983er}. 

In QCD, the eigenvalue spectrum of the Dirac operator is of physical interest, since it is closely connected to the spontaneous breaking of this symmetry. The order parameter in the phase with broken chiral symmetry, the chiral condensate $\langle \bar{\psi} \psi\rangle$, is closely related to the spectral density at the origin by the Banks-Casher-relation~\cite{Banks:1979yr}
\be
\Sigma = |\langle \bar{\psi} \psi \rangle| = \lim_{\lambda \to 0} \lim_{m_c \to 0} \lim_{V \to \infty} \frac{\pi}{V} \rho(\lambda).
\ee
A non-zero spectral density at the origin of the spectrum therefore implies spontaneous breaking of chiral symmetry. The spectral density itself is defined as
\be
\rho(\lambda) &=& \left\langle \sum_k \delta(\lambda - \lambda_k) \right\rangle, 
\ee
where the brackets indicate an averaging with regard to the QCD action.
The Banks-Casher relation implies that the eigenvalues close to the origin of the spectrum are spaced as $\sim 1/V$ in the phase with broken chiral symmetry and $\Sigma \neq 0$. 

The importance of the low-lying eigenvalues was further emphasized by the work of Leutwyler and Smilga, who calculated sum rules for inverse powers of the Dirac operator eigenvalues~\cite{Leutwyler:1992yt}. These sum rules are dominated by the low-lying eigenvalues, and are sensitive to the topological charge of the QCD vacuum. Originally intended to explore the relation between gauge field topology and confinement, these sum rules proved a valuable tool to investigate the relation between the Dirac operator spectrum and chiral symmetry breaking.

These Leutwyler-Smilga sum rules, which rely on the contribution of the low-momentum modes to the partition function and the chiral symmetry breaking pattern, provided the motivation for random matrix models with the structure appropriate for the symmetries of QCD and the same symmetry breaking pattern~\cite{Shuryak:1992pi}.
The spectral sum rules are exactly reproduced by those chiral Random matrix models \cite{Shuryak:1992pi}, and this was an incentive to apply the models also to spectral correlations~\cite{Verbaarschot:1993pm}.

Because the QCD Dirac operator anti-commutes with $\gamma_5$, i.e. $\{\slash\!\!\!\! D, \gamma_5\}=0$, the non-zero eigenvalues of the Dirac operator appear in pairs, $\pm \lambda_k$. As a consequence, close to the origin of the spectrum $\lambda=0$, the universal eigenvalue correlations between the eigenvalue pairs determine the spectral density as such. 

This can be seen at the level of the \emph{microscopic} spectral density. Since the average distance between two eigenvalues $\lambda_k$ of the Dirac operator in a finite volume scales as $\sim 1/V$ in the vacuum with broken chiral symmetry, $\Sigma \neq 0$, it is natural to look at the spectral density in terms of a rescaled eigenvalue variable $z = \lambda V \Sigma$. The microscopic spectral density is then defined by the infinite-volume limit
\be
\rho_S(z) = \lim_{V \to \infty} \frac{1}{V\Sigma} \rho \left(\frac{z}{V\Sigma}\right).
\ee
The microscopic spectral density is therefore in a sense the spectrum close to the origin put under a ``microscope" and magnified by a volume factor. The finite-volume analysis thus allows to resolve individual eigenvalues.

This quantity is of particular interest for a comparison to lattice simulations of the discretized QCD action. Random Matrix Theory provides an analytical prediction for the microscopic spectral density in QCD~\cite{Verbaarschot:1993pm}, dependent on the the number of colors, the number of quark flavors and the resulting symmetry breaking pattern, and on the topological structure of the vacuum~\cite{Verbaarschot:1993pm,Verbaarschot:1994qf,Verbaarschot:1994te}.
The QCD vacuum allows for instanton solutions which carry a finite topological charge \cite{Belavin:1975fg,tHooft:1976up}. These solutions are associated with exact zero-mode solutions for the QCD Dirac equation, and therefore the eigenvalue spectrum of the Dirac operator is sensitive to the topological charge of the vacuum state~\cite{Shuryak:1981ff,Shuryak:1989cx,Verbaarschot:1993ze,Verbaarschot:1994te}.     
For QCD with $N_\mathrm{c}=3$ colors, the microscopic spectral density is given by~\cite{Verbaarschot:1993pm,Verbaarschot:1994te}
\be
\rho_s(z)&=& \frac{z}{2} \big[ J^2_{N_\mathrm{f}+|\nu|} (z) - J_{N_\mathrm{f}+|\nu|+1}(z) J_{N_\mathrm{f}+|\nu|-1}(z)   \big]
\ee
where the $J_\mu(z)$ are Bessel functions, $N_\mathrm{f}$ are the number of massless flavors, and $\nu$ is the topological charge of the vacuum.
The result for the spectrum must therefore be classified according to the topological charge for a comparison to QCD lattice simulations.     
   
\begin{figure}
\includegraphics[scale=1.5]{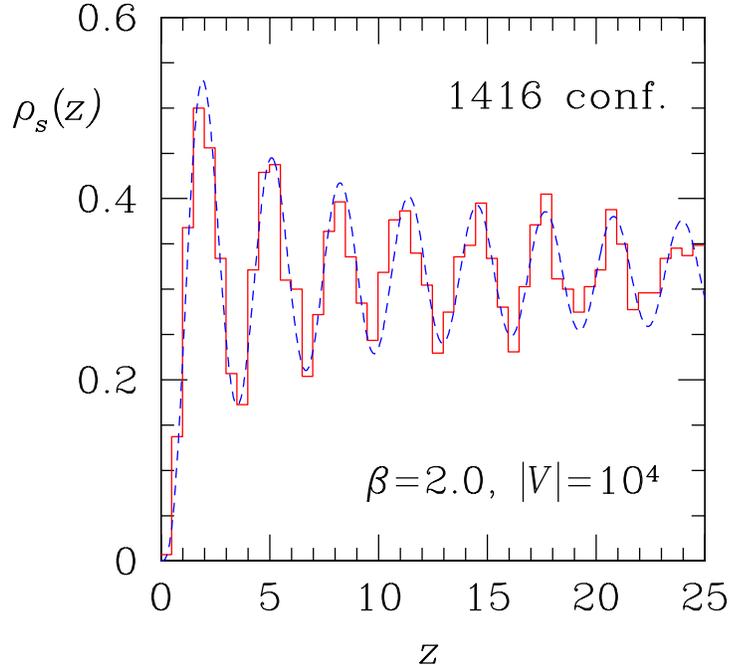}
\caption{Microscopic spectrum of the QCD Dirac operator for an $\mathrm{SU}(2)$ gauge theory with one quark flavor in the staggered formulation~\cite{BerbenniBitsch:1997tx}. The density of eigenvalues from the simulation of the theory in quenched approximation are compared to the prediction for the microscopic spectral density from RMT. Because of the symmetries in this case, the appropriate comparison is to chiral Symplectic Ensemble in RMT. Figure reprinted with permission from Berbenni-Bitsch \emph{et al.}, Phys. Rev. Lett. 80, 1146 (1998). Copyright (1998) by the American Physical Society.}
\label{fig:micspec}
\end{figure}

Fig.~\ref{fig:micspec} from \cite{BerbenniBitsch:1997tx} shows such a comparison for the microscopic spectral density of a quenched $\mathrm{SU}(2)$ gauge theory with $N_\mathrm{c}=2$ colors and one quark flavor in the staggered representation, and in the sector with zero topological charge. 
The RMT prediction, given by the dashed line, agrees perfectly with the histogram of the Dirac operator eigenvalues obtained from the lattice simulation.
In this case, for the gauge theory with $N_\mathrm{c}=2$ colors, an additional anti-unitary symmetry is present in the Dirac operator, leading to a stronger anti-correlation between the eigenvalues and a more distinct structure in the spectral density than in QCD with $N_\mathrm{c}=3$ colors. The appropriate comparison for $N_\mathrm{c}=2$ is therefore to the chiral Symplectic Ensemble (chSE) in RMT, in contrast to the chiral Unitary Ensemble (chUE), which governs QCD with $N_\mathrm{c}=3$ and fermions in the fundamental representation~\cite{Verbaarschot:1994qf}.
Because only the volume size and the value of the chiral condensate enter into the eigenvalue spacing, the spectrum can in turn be used to estimate the chiral condensate $\Sigma$ from lattice simulation results for the spectrum~\cite{BerbenniBitsch:1997fj}.

Random matrix theory is valid in a \emph{mesoscopic} volume range which is limited by two requirements: On the one hand, it needs to be small enough such that the pion wavelength is larger than the size of the box, $1/m_\pi \gg L$, so that a \emph{static} description of the partition function is valid. On the other hand, the volume needs to be large enough such that hadronic effects are suppressed, $\sim \exp (- \Lambda L)$, where $\Lambda$ is the typical hadron mass scale. This  presupposes confinement of light elementary degrees of freedom  \cite{Klein:2000pj}. Taken together, these two conditions mean that RMT provides a valid description in the volume range
\be
\frac{1}{m_\pi} \gg L \gg \frac{1}{\Lambda}.
\ee

In an analogy to condensed matter physics, it is possible to define a \emph{Thouless} energy scale $E_\mathrm{c}$~\cite{Thouless:197493}, below which the eigenvalue spectrum of the QCD Dirac operator in a finite volume can be described by a static chiral Lagrangian or RMT \cite{Verbaarschot:1995yi,Osborn:1998nf,Osborn:1998nm,Janik:1998ki}. 
In condensed matter physics, this scale is given by the inverse diffusion time of an electron through a sample of length $L$. 
The energy (eigenvalue) up to which the spectrum of the Dirac operator is still given by a static description is
\be
E_c \sim \frac{F^2}{\Sigma L^2}
\ee
where $F$ is the pion decay constant in the chiral limit of massless quarks, $\Sigma$ is the value of the chiral condensate in vacuum, and  $L$ is the linear extent of the volume. This relation follows from the above condition, requiring that the pion wavelength exceeds the volume size, combined with the leading-order relation between the pion mass and the current quark mass in QCD \cite{GellMann:1968rz}.
In this region, chiral Random Matrix Theory and the static chiral partition function provide equivalent descriptions of the Dirac operator spectrum~\cite{Osborn:1998qb,Damgaard:1998xy}.

The effective static partition function valid in this domain is given by an integral over the Goldstone manifold $G/H$ associated with the spontaneous breaking of chiral symmetry from $G \to H$, where $H$ is the residual symmetry group~\cite{Gasser:1987ah,Leutwyler:1992yt}:
\be
Z^{\mathrm{eff}}_{N_{\mathrm{f}}}( {\mathcal M}, \theta) = \int_{U \in SU(N_f)} \mathrm{d} U \exp \Big [V \Sigma\,  \mathrm{Re}\,  \Tr {\mathcal M} U \mathrm{e}^{\mathrm{i}\theta/N_f} \Big].
\ee
For actual QCD with $N_{\mathrm c} = 3 $ colors and quarks in the fundamental representation, the Goldstone manifold is $G/H = SU_L(N_f) \times SU_R(N_{\mathrm f})/SU(N_\mathrm{f})$. ${\mathcal M}$ is the current quark mass matrix in flavor space. The partition function depends also on the vacuum angle  $\theta$ which appears in the Euclidean QCD Lagrangian as the coefficient of the topological term $\tilde F F$, where $F$ and $\tilde F$ are the gauge field tensor and its dual.
By Fourier decomposition, the QCD partition function can be written as a sum of contributions with definite topological charge
\be
Z^{\mathrm{eff}}_{N_{\mathrm{f}}}( {\mathcal M}, \theta) &=& \sum_{\nu=-\infty}^\infty \mathrm{e}^{\mathrm{i}\nu \theta} Z^{\mathrm{eff}}_{N_{\mathrm{f}}}( {\mathcal M}, \nu).
\ee
The finite-volume partition function in a sector with a given topological charge $\nu$ of the gauge field can be identified as
\be
Z^{\mathrm{eff}}_{N_\mathrm{f}} (\mathcal{M}, \nu) =  \int_{U \in SU(N_f)} \mathrm{d} U \det{}^\nu U  \exp \Big [V \Sigma\,  \mathrm{Re} \, \Tr {\mathcal M} U \Big ]
\ee
The effects of the topology on the chiral condensate and the microscopic Dirac operator spectrum in a finite volume have been considered in~\cite{Damgaard:1999ij}, where explicit expressions for these quantities are derived.

Results for the behavior of the chiral condensate from chiral RMT were first obtained in~\cite{Verbaarschot:1995yi}, where it was shown that lattice simulation results for the finite-volume condensate as a function of the quark mass $m$ fall onto a universal scaling curve, when plotted as a function of the scaling variable $mV\Sigma$.
The behavior of the chiral condensate as a function of quark mass and volume has more recently again been tested in lattice QCD simulations, and agreement with the predictions of the effective chiral description has been observed, as well as agreement of the eigenvalues with the Leutwyler-Smilga sum rules \cite{Giusti:2007cn}.   
   
In recent years, the focus of RMT calculations has shifted towards the QCD partition function at finite baryon density, i.e. in the presence of a finite chemical potential \cite{Stephanov:1996ki,Osborn:2004rf,Osborn:2005aj,Osborn:2005ss,Akemann:2006ru}. Complex eigenvalue spectra for the resulting non-Hermitian Dirac operator calculated from the method can help to solve the complex phase problem (``sign problem") in lattice QCD simulations~\cite{Splittorff:2006fu,Splittorff:2007ck,Splittorff:2008tu,Bloch:2012ye,Bloch:2012bh}: They provide analytic results against which lattice QCD results can be directly compared. Such results are of significant interest to practitioners of QCD lattice simulations, see e.g. \cite{Splittorff:2006fu,Splittorff:2007ck,Splittorff:2008tu,Lombardo:2009jq,Lombardo:2009aw,Lombardo:2010fj}. 
Also in the case of a finite chemical potential, QCD and chiral Random Matrix Theory are equivalent in the $\varepsilon$-regime below a certain scale~\cite{Basile:2007ki}.
A review of applications or random matrix ensembles to QCD at finite chemical potentials can be found in \cite{Akemann:2007rf}.

In addition, RMT can serve as a schematic model for the chiral phase transition~\cite{Jackson:1995nf}.
The spectral density for such random matrix models at finite temperature has been investigated in \cite{Guhr:1997uf,Jackson:1997ud,Seif:1998id}.
It can also be used to explore the phase structure of QCD~\cite{Stephanov:1996ki,Halasz:1998qr}, insofar as it is determined by the symmetries of the QCD partition function. 
Such schematic models have been extended to finite isospin chemical potential~\cite{Klein:2003fy} and related QCD-like theories~\cite{Kogut:1999iv,Kogut:2000ek,Klein:2004hv}.
A review of RMT models as schematic models for the QCD phase diagram can be found e.g. in~\cite{Vanderheyden:2011iq}.
These applications do not rely on finite-volume arguments and use only the symmetry properties of the partition function.

QCD in finite volume at extremely high densities, where one expects color superconductivity, is considered in \cite{Yamamoto:2009ey}, and sum rules \`a la Leutwyler-Smilga for the complex eigenvalues are derived. These ideas are further developed with RMT methods in~\cite{Kanazawa:2009en} for two-color QCD. These results are reviewed extensively in~\cite{Kanazawa:2011tt} and put into context with other RMT results at finite chemical potential. 

 \begin{figure}
 \begin{center}
\hspace*{-10mm} \includegraphics[scale=1.00, clip=true]{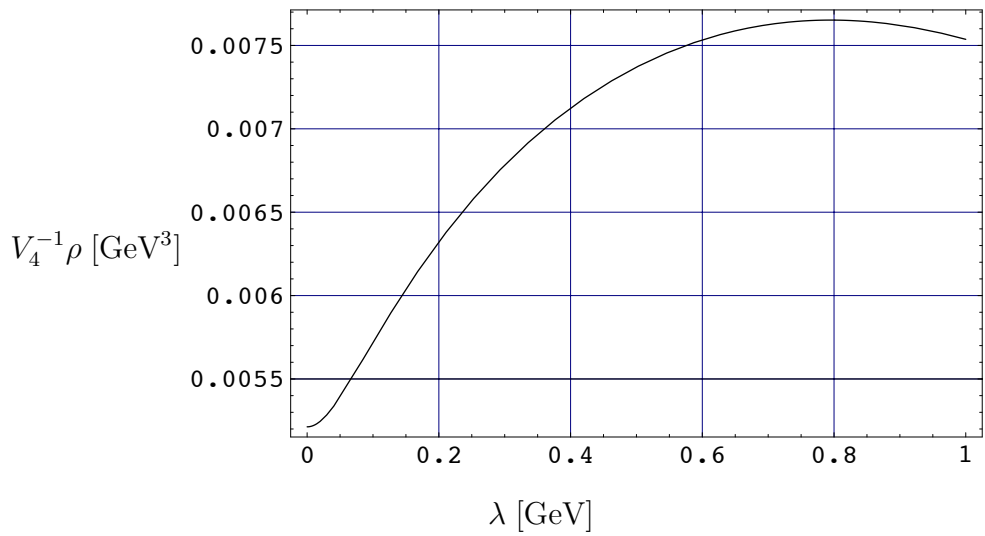}
 \end{center}
 \caption{Eigenvalue spectrum of the quark-meson model Dirac operator in an infinite volume~\cite{Spitzenberg:2002tq}. A possible linear term in the spectrum for small values of the eigenvalues $\lambda$: For $N_\mathrm{f}=2$ quark flavors is absent: This is the expected behavior, since the coefficient $\sim N_\mathrm{f}^2-4$ of the linear term in the spectral density vanishes \cite{Smilga:1993in} and $\rho(\lambda) \sim \lambda^2$ for small eigenvalues $\lambda \ll 1$ GeV. Figure reprinted with permission from Spitzenberg \emph{et al.}, Phys. Rev. D 65, 074017 (2002). Copyright (2002) by the American Physical Society.}
 \label{fig:spectrum-spitzenberg}
 \end{figure}  
The ideas on the Dirac operator spectrum at the heart of RMT provide an important motivation for the work on finite-volume effects in QCD. 
Considering the macroscopic spectrum of the Dirac operator, Smilga and Stern derived the next term in the spectral density for small eigenvalues $\lambda \ll \Lambda$ beyond the result of Banks and Casher ~\cite{Smilga:1993in}:
\be
\frac{\pi}{V}\rho(\lambda) &=& \Sigma + \frac{N_\mathrm{f}^2-4}{32 \pi N_\mathrm{f} }\frac{\Sigma^2}{F^4} |\lambda| + {\mathcal O}(\lambda^2).
\ee
The result has been re-derived in \cite{Toublan:1999hi} from a chiral Lagrangian beyond the validity region of chRMT for eigenvalues $E_c \ll \lambda \ll \Lambda$. 
This spectral behavior is indeed shared by different theories and models for QCD that have the same chiral symmetry properties \cite{Verbaarschot:1993ze,Shuryak:1989cx}. For this reason it should be possible to observe it in other models with the correct symmetries which might lend themselves more easily to an extension to larger scales and to investigations of dynamical behavior at the chiral symmetry restoration transition.  

In \cite{Spitzenberg:2002tq}, the authors calculated a \emph{global} spectrum of the Dirac operator in a chiral quark-meson model, see Fig.~\ref{fig:spectrum-spitzenberg}. Since it implements the chiral symmetries of QCD, the authors expected to find universal spectral features in such a calculation. An important result of the investigation in \cite{Spitzenberg:2002tq} was the observation that the global spectral density for a theory with two light quark flavors behaved as $\rho(\lambda) \sim \lambda^2$, in agreement with the predictions in \cite{Smilga:1993in} that the term linear in $\lambda$ should be absent for {$N_{\mathrm f}=2$}. 
However, a calculation in infinite volume is not enough to resolve the microscopic spectrum: In order to resolve individual eigenvalues close to the origin of the spectrum, finite-volume results are required. Only with such finite-volume results can discrete individual eigenvalues be resolved and a microscopic spectral density can be defined for the model. 

This observation has provided the initial inspiration for the work investigating the behavior of chiral-symmetry breaking models in a finite volume. Spontaneous symmetry breaking of a continuous symmetry in a finite volume always requires to implement an explicit breaking of this symmetry, since the interplay of the Goldstone boson length scale $1/m_\pi$ given by the explicit symmetry breaking and the system size $L$ is essential for the observed phenomena, as we have already seen from RMT. While a calculation of the spectrum requires an analytical continuation of the results to imaginary quark mass parameters \cite{Spitzenberg:2002tq} and is difficult to implement in a numerical calculation, many finite-volume effects are readily accessible from such a calculation. Starting from this initial motivation,  a wide range of interesting results for the interplay of chiral symmetry breaking and a finite volume size have been obtained, and indeed efforts in this direction have proven very fruitful. 
%


\section{Chiral Perturbation Theory}

QCD at low energy scales is fundamentally shaped by both the phenomenon of confinement of color charges and by the spontaneous breaking of chiral symmetry.  The light Goldstone bosons which arise from the spontaneous breakdown of the chiral flavor symmetry appear as the relevant low-energy degrees of freedom. Since the chiral symmetry is also broken explicitly by the small current quark masses, the pions as (pseudo-) Goldstone bosons are not exactly massless, but only very light compared to the masses of other hadronic degrees of freedom. Because of the mass gap between the Goldstone modes and the other hadrons, an effective description in terms of only the light degrees of freedom is possible.
An effective field theory description can provide important insights and is an important touchstone for other approaches. For this reason we will give a brief overview over the relevant low-energy theory and some of its finite-volume results, which provides important context for any model-based calculations. 

The effective field theory describing the low-momentum regime is \emph{Chiral Perturbation Theory} (ChPT). It is valid below a hadronic scale $\Lambda_\chi \approx 1$ GeV, which is usually defined as
\be
\Lambda_\chi \approx 4 \pi f_\pi
\label{eq:ChPT-scale}
\ee
where $f_\pi \approx 93$~MeV is the pion decay constant. 
The theory is expanded systematically in powers of the small parameters
\be
\frac{p}{\Lambda_\chi} \quad \quad \mathrm{and}\;\; \frac{m_\pi}{\Lambda_\chi}, 
\label{eq:ChPTexpansion}
\ee
where $p=|\vec{p}\,| $ and $m_\pi$ are the momenta and mass of the (pseudo-)Goldstone bosons. Consequently, the theory is only valid for sufficiently small momenta and small symmetry breaking.
The effective Lagrangian~\cite{Weinberg:1978kz,Gasser:1983yg,Gasser:1984gg,Leutwyler:1993iq}
 can then be written as
\be
{\mathcal L}_{\mathrm{eff}} &=& {\mathcal L}_{\pi}^{(2)} + {\mathcal L}_{\pi}^{(4)} +  {\mathcal L}_{\pi}^{(6)} + \ldots
\ee
where the index indicates the powers of the expansion parameters.

For QCD with two light quark flavors, and neglecting a breaking of the isospin symmetry between up- and down-quarks, the leading-order effective Lagrangian is given by \cite{Weinberg:1978kz,Gasser:1983yg,Gasser:1984gg,Leutwyler:1993iq}
\be
\mathcal {L}_{\mathrm{eff}} &=& \frac{F^2}{4} \Tr\, \partial_\mu U^\dagger \partial_\mu U + \frac{m \Sigma}{2} \Tr (U^\dagger + U) + \ldots
\ee
The fields $U(x)= \exp(\mathrm{i}  \tau^a \pi^a(x)/F)$
 parametrize the Goldstone manifold, where $\tau^a$ are the generators of $SU(2)$ and the fields $\pi^a(x)$ represent the three Goldstone bosons. The parameter $F$ is the leading-order pion decay constant and describes the scattering of the weakly interacting Goldstone bosons. The explicit breaking of chiral symmetry is controlled by the parameter $m$, which corresponds to a current quark mass in the QCD Lagrangian. As before, $\Sigma$ is the chiral condensate in the chiral limit. From the leading-order mass term, one identifies the pion mass $m_\pi$ and reads off the leading-order Gell-Mann--Oakes--Renner relation \cite{GellMann:1968rz} $m_\pi^2 F^2= m \Sigma$.

It is straightforward to extend ChPT to finite temperature to describe a hot gas of pions \cite{Gasser:1987ah} below the chiral phase transition temperature. 
In this case, describing the theory in Euclidean space-time, the boundary conditions in the Euclidean time direction are dictated by the commutation relations of the bosonic pion fields.
In the same way, ChPT can be extended to a finite volume \cite{Gasser:1987zq}. It describes QCD in a finite box of Euclidean volume $L^3 \times 1/T$ at finite temperature $T$, provided the temperature is sufficiently low and the volume is sufficiently large such that chiral symmetry is not restored.  

For a finite volume, the condition that the discretized momenta $\vec{p}=\frac{2 \pi}{L} \vec{n}$ must be small compared to the scale $\Lambda_\chi$
\be
\frac{p}{4 \pi f_\pi} \ll 1
\ee  
can be translated into a condition for volume sizes in which ChPT is applicable \cite{Colangelo:2005gd}:
\be
L \gg \frac{1}{2 f_\pi} \sim 1 \, \mathrm{fm}.
\ee
However, the dimensionless product $m_\pi L$ is unconstrained by this condition. Similar to the considerations on the applicability of RMT, one needs to distinguish two different power-counting schemes to organize the chiral expansion, depending on the relative size of the pion mass $m_\pi$ and the volume size $L$: For $m_\pi L \gg 1$ one speaks of the $p$-expansion, in which the pions are dynamical particles. For $m_\pi L \ll 1$, one organizes the power-counting according to the $\epsilon$-expansion, where the treatment of zero-modes is of paramount importance, and it has to be integrated out exactly \cite{Gasser:1987ah}.
The partition function for the zero-mode part is then given by
\be
Z_{\mathrm{ChPT}}^{(0)}(m_\pi) &=& \mathrm{const.} \int_{SU(N_f)} \mathrm{d} U \,\exp \left[ \frac{1}{2} f_\pi^2 m_\pi^2 V  \mathrm{Re} \Tr U \right], 
\ee
where the integration with the invariant group measure is over the Goldstone manifold.
The zero mode is the only fluctuation mode which is not suppressed, and in fact its contributions in the pion propagator
\be
G \sim \frac{1}{m_\pi^2 V}  + \ldots
\ee 
diverge in the chiral limit $m_\pi \to 0$.

The most natural choice of periodic spatial boundary conditions for the pion (Goldstone) fields thus implies the existence of an exact zero-momentum mode in finite volume with $\vec{p}=0$.
The treatment of these Goldstone zero-modes in finite volume is extremely important \cite{Gasser:1987zq,Gasser:1987ah}. Depending on the relation between the mass of the pion fields and the linear extent of the volume, fluctuations of the zero-momentum mode can dominate the partition function and need to be treated exactly in this case.

In principle, the effective Lagrangian of the finite-volume system need not be the same as in infinite volume: the low-energy effective Lagrangian can depend on the volume size and the boundary conditions for the fields in the underlying high-energy theory. The effective Lagrangian is, however, independent of temperature: All coupling constants at $T>0$ remain at the same values as at $T=0$, although they may depend on the volume size.
If one now chooses for all fields the same boundary conditions in the spatial directions as in the Euclidean time (temperature) direction in the underlying high-energy theory, namely anti-periodic boundary conditions for the quark (fermion) fields and periodic boundary conditions for the gluon (boson) fields, this has powerful consequences for the low-energy effective theory in finite volume.
In this case, the Euclidean space and time directions are physically equivalent, and a permutation symmetry holds among them. Consequently, in this case the effective low-energy Lagrangian must have the same space-time permutation symmetry. Since the coupling constants are temperature-independent, they cannot depend on the volume size, either~\cite{Gasser:1987ah}. 
Conversely, if different boundary conditions in time and space are chosen that break this permutation symmetry, such a dependence cannot be excluded. This applies to the choice of periodic boundary conditions for the quark fields in spatial directions, which is popular in QCD lattice simulations. Empirically, such a choice appears to minimize finite-volume effects, e.g. for hadron masses~\cite{Aoki:1993gi}. In early lattice QCD simulation work, the effects of the choice of quark boundary conditions on meson propagators~\cite{Carpenter:1984dd} and hadron masses~\cite{Fukugita:1992jj,Aoki:1993gi} have been investigated, and it has been observed that this choice leads to quite different behavior. In most of the current simulations, the choice of periodic quark boundary conditions has become a \emph{de facto} standard. 
As we will argue below, this is an important \emph{caveat} which should be kept in mind when comparing results from QCD lattice simulations and ChPT.

For lattice simulations performed in a finite box, the extrapolation of the results from a finite to infinite volume is obviously very important. The results of ChPT have found numerous applications.
ChPT has been successful in predicting the volume dependence of the chiral condensate~\cite{Hansen:1990un}, many hadron properties, e.g. for the nucleon mass~\cite{AliKhan:2003cu}, see also~\cite{Koma:2004wz}, where L\"uscher's approach~\cite{Luscher:1983rk,Luscher:1985dn} is re-visited for the nucleon. 
A lattice-regularized version of ChPT has also been used to calculate the shift in the pion mass, charge radius and decay constant in a finite volume \cite{Borasoy:2004ac,Borasoy:2004zf}.

In very small volumes, the so-called $\delta$-region expansion in ChPT \cite{Leutwyler:1987ak} has been used to calculate e.g. the nucleon mass shift from ChPT \cite{Bedaque:2004dt}, and to describe the pion mass in lattice simulations \cite{Bietenholz:2010az} (in both the $\epsilon$- and $\delta$-regime).
In this region, there is an overlap with the region of validity of RMT, and the static QCD partition function is given by an exact integral over the zero mode \cite{Osborn:1998nf,Osborn:1998nm,Osborn:1998qb}. 
Effects connected with the restoration of chiral symmetry in a small volume only occur in the limit $m L^2 \to 0$ \cite{Gasser:1987ah}, i.e. when the quark mass is of the order of $1/V$. This is consistent with the observation in RMT that the microscopic spectral density goes to zero for $\lambda \to 0$, since this quantity describes the spectral density for  eigenvalues $\lambda \sim 1/V$, and the quark mass $m \sim \lambda$ effectively probes the eigenvalue spectrum of the QCD Dirac operator.

The quantities that are readily accessible in a finite-volume ChPT calculation, and which are of much interest for thermodynamics, are the pion mass, the pion decay constant and the chiral condensate.
In early results, Gasser and Leutwyler directly calculated the pion mass shift in a finite box from a one-loop calculation \cite{Gasser:1986vb}. Due to the periodicity of the finite-volume pion propagator
\be
G_L(x_0, \vec{x}) &=& \sum_{\vec{n}} G_{\infty}(x_0, \vec{x} + \vec{n} L)
\ee
there is an infinite number of contributions to the correction, which are summed in the result. (Pictorially, these contributions corresponds to pions leaving the box and re-entering from the opposite side  $n$ times due to the periodic boundary conditions.)
The result of the one-loop calculation for the shift in the pion mass and in the pion decay constant is \cite{Gasser:1986vb}
\be
m_\pi(L) &=& m_\pi \left[1+\frac{1}{2N_f} \xi g_1(\lambda) + {\mathcal O}(\xi^2)  \right], \el
f_\pi(L) &=& f_\pi \left[1-\frac{N_f}{2} \xi g_1(\lambda) + {\mathcal O}(\xi^2)  \right],
\ee
where $\lambda = m_\pi L$, $\xi = \left(\frac{m_\pi}{ 4 \pi f_\pi}\right)^2$ and 
\be
g_1(\lambda) = \sum_{r=1}^\infty \frac{4 m(r)}{\sqrt{r} \lambda} K_1(\sqrt{r} \lambda),
\label{eq:multiplicity}
\ee
with $m(r)$ being the multiplicity of a vector $\vec{n}$ of magnitude $r = \vec{n}^2$.

There is an additional, powerful method to determine the mass shift of a particle in a finite volume much larger than the characteristic particle wavelength introduced by L\"uscher~\cite{Luscher:1983rk,Luscher:1985dn}. It makes use of a relation valid to all orders in perturbation theory between the \emph{finite-volume} mass shift and the \emph{infinite-volume}  forward scattering amplitudes with the lightest particles~\cite{Luscher:1985dn}.   
The original L\"uscher formula connects the shift in the mass of a particle in a finite volume to a forward scattering amplitude. In the case of the pion, the mass shift in a box is given by
\be
m_\pi(L) - m_\pi(\infty) &=& - \frac{m(1)}{32 \pi^2} \frac{1}{m_\pi L} \int_{-\infty}^\infty \mathrm{d}y \,F(\mathrm{i} y) \mathrm{e}^{-\sqrt{m_\pi^2 + y^2} L} + {\mathcal O}(\mathrm{e}^{-\bar{M} L}) \nn\el
\ee
where $F(\nu)$ for real $\nu$ is the forward-scattering amplitude for pion-pion scattering, analytically continued to imaginary argument $\mathrm{i}y$. The multiplicity of a vector with $\vec{n}^2=1$ is $m(1)=6$. This number arises since only effects of pions with $\vec{n}^2=1$ are taken into account, which travel around the finite volume just once. 
The higher-order correction terms are exponentially suppressed by $\bar M L$, where $\bar M > \sqrt{2} m_\pi$. 
In leading order in the chiral expansion, the forward scattering amplitude is simply given by~\cite{Colangelo:2004sc}
\be
F(\nu) &=& -\frac{m_\pi^2}{f_\pi^2} + {\mathcal O}(m_\pi^4).
\ee
Using the lowest-order result for the pion scattering amplitude, L\"uscher's formula gives the same results as the direct calculation by Gasser and Leutwyler~\cite{Gasser:1986vb}.

Recent progress on the finite-volume effects in QCD has been made by combining L\"uscher's approach with higher-order calculations in ChPT for the $\pi\pi$-scattering amplitude~\cite{Bijnens:1995yn,Bijnens:1997vq,Bijnens:1998fm}, including two loops in infinite volume. It is based on the new insight that application of the L\"uscher result in combination with input from ChPT allows to obtain finite-volume effects in higher order in the chiral loop expansion, without having to perform difficult multi-loop calculations in finite volume. The price that one pays for using the asymptotic result is an error of $ {\mathcal O}(\mathrm{e}^{-\bar{M} L})$, since the L\"uscher formula \emph{by design} gives the leading-order term in the large-$L$ expansion~\cite{Colangelo:2003hf}.

\begin{figure}
\includegraphics[scale=0.55]{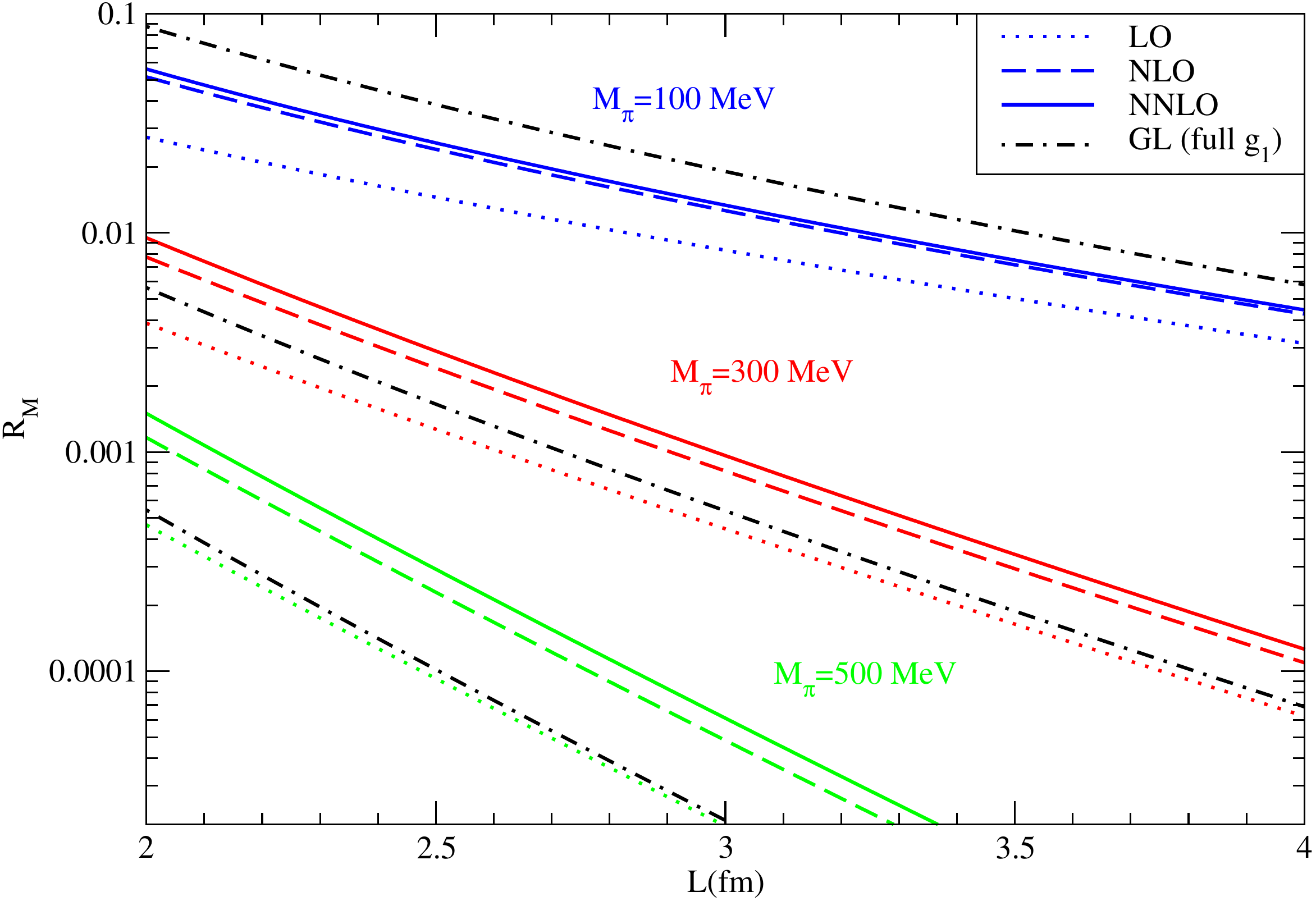}
\caption{Pion mass shift $R[m_\pi(L)] = \frac{m_\pi(L) - m_\pi(\infty)}{m_\pi(\infty)}$ in a finite volume $V=L^3$ as a function of $L$~\cite{Colangelo:2003hf}. Shown are the results obtained from L\"uscher's formula with input from ChPT  to Leading Order (LO), Next-(NLO) and Next-to-Next-to Leading Order (NNLO). For comparison, the full one-loop result from Gasser and Leutwyler (GL) \cite{Gasser:1986vb} is also given. Figure from G. Colangelo and S. Durr, Eur. Phys. J. C33, 543 (2004), Copyright (2004), reprinted with kind permission from Springer Science and Business Media.
\label{fig:RmpiChPT1}}
\end{figure}

The results from the calculation in  \cite{Colangelo:2003hf} for the relative shift in the pion mass $m_\pi(L)$ in finite volume 
\be
R[m_\pi(L)]&=& \frac{m_\pi(L) - m_\pi(\infty) }{m_\pi(\infty)}
\ee
 are shown in Fig.~\ref{fig:RmpiChPT1} as a function of the box size $L$, for pion masses $m_\pi(\infty)=100$ MeV, $300$ MeV, and $500$ MeV. For comparison, the result from Gasser and Leutwyler (GL) \cite{Gasser:1986vb} from the finite-volume one-loop calculation is also shown. 
The chiral expansion converges more quickly for small pion mass, which can be seen by comparing the results in next-to- and next-to-next-to-leading order (dashed and solid lines). The convergence of the finite-volume corrections behaves in the opposite way, since L\"uscher's result yields the leading term of a large-$L$ expansion, and corrections are suppressed more quickly for larger pion masses.
 
This can be seen clearly in Fig.~\ref{fig:RmpiChPT1}: For small pion mass, the difference between the results from L\"uscher's formula in leading order and from the exact one-loop calculation is rather large. For large pion mass, the exponentially suppressed corrections are small, and the exact finite-volume one-loop calculation and the leading-order result from L\"uscher's formula agree.  

The combination of L\"uscher's result with the ChPT-input can be significantly improved.
Generically, L\"uscher's formula has been derived by only taking pions into account that wrap around the lattice just once. 
The necessity of taking pions into account that wrap around the lattice more than once and to sum over all contributions was already discussed in \cite{Colangelo:2003hf} and also recognized in \cite{AliKhan:2003cu}.

L\"uscher's result can be re-derived, taking into account also the leading contribution from pions wrapping around the lattice more than once \cite{Colangelo:2005gd}. Summing the leading exponential contributions for all $\vec{n}$, one finds for the shift of the pion mass
\be
\lefteqn{m_\pi(L) - m_\pi(\infty) = }\nonumber\\
&& -\frac{1}{32\pi^2} \frac{1}{m_\pi L} \sum_{r=1}^\infty \frac{m(r)}{\sqrt{r}} \int_{-\infty}^\infty {\mathrm d}y F(\mathrm{i} y) \mathrm{e}^{-\sqrt{r(m_\pi^2+y^2)}L} + {\mathcal O}(\mathrm{e}^{-\bar M L}),\nn\el 
\ee
where $m(r)$ has the same meaning as in \eqref{eq:multiplicity}.
As observed in \cite{Colangelo:2005gd}, compared to the result with $r=\vec{n}^2=1$ only, this leads to a significant improvement, in particular for small pion masses. 

The re-summed L\"uscher formula has also more recently been applied to a calculation of the finite-volume mass shift of the nucleon and heavy mesons \cite{Colangelo:2010ba}, and to a calculation of finite-volume effects in twisted-mass lattice QCD \cite{Colangelo:2010cu}.

\begin{figure}
\includegraphics[scale=0.95]{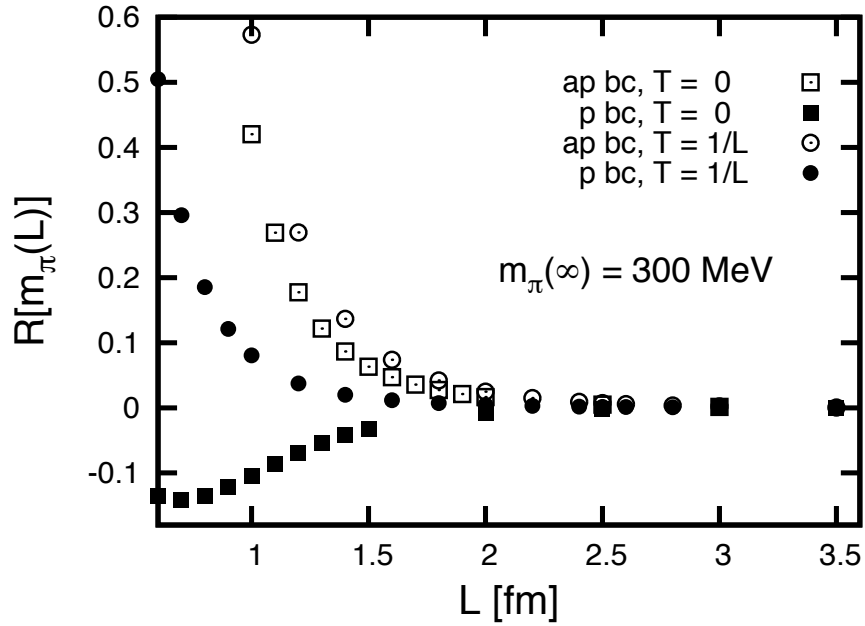}
\caption{Pion mass shift in finite volume $V\times 1/T= L^3 \times 1/T$ from the quark-meson model for different choices of the quark boundary conditions in the spatial directions \cite{Braun:2005gy}. (Open) solid symbols correspond to (anti-)periodic boundary conditions. Squares indicate results for the choice $T=0$, circles for $T=1/L$. All results are for $m_\pi(\infty) = 300$ MeV. For small $L$, the results for periodic boundary conditions lie significantly below those for anti-periodic boundary conditions, and exhibit a ``dip" for small $T$. Figure reprinted with permission from J.~Braun \emph{et al.}, Phys. Rev. D72, 034017 (2005). Copyright (2005) by the American Physical Society.}
\label{fig:qmmRmpiM300}
\end{figure}

The observation that additional low-energy constants in the finite-volume chiral Lagrangian can be excluded only for a specific choice of spatial boundary conditions for the quark fields is extremely important for the comparison of ChPT predictions to lattice QCD results. It is also at the heart of several phenomena that are observed in the investigations of the QCD finite-volume behavior with models for chiral symmetry breaking in~\cite{Braun:2004yk,Braun:2005gy,Braun:2005fj}. These models explicitly include quarks as degrees of freedom~\cite{Braun:2005gy}, and it is thus possible to assess the effect of different boundary conditions for the quarks in spatial directions.
The quark-meson-model is introduced in more detail in section~\ref{sec:qmm}.
The shift in pion mass and pion decay constant in a finite volume have been investigated in such a 
model~\cite{Braun:2004yk,Braun:2005gy,Braun:2005fj} and results for different choices of boundary conditions have been compared. The results for the mass shift for a pion with $m_\pi(\infty) = 300$ MeV can be seen in Fig.~\ref{fig:qmmRmpiM300}.
The results in the figure are for Euclidean volumes $V=L^4$ ($T=1/L$) and $V=L^3 \times \infty$ ($T=0$), for periodic and anti-periodic boundary conditions for quark fields in the spatial directions. It is obvious that the choice of boundary conditions has a significant effect on the results for the finite-volume mass shift. 
For periodic boundary conditions, the increase in the finite-volume pion mass with decreasing $L$ is significantly smaller than for anti-periodic boundary conditions. The effect is most pronounced for $T=0$, but appears for any aspect ratio of the four-dimensional Euclidean volume. For small $T$ and large pion masses, we even see a \emph{decrease} in the pion mass in finite volume. 
This is an effect of the fermionic zero mode that is present for periodic boundary conditions. For small values of the volume, the zero mode contributions to the chiral condensate 
$\langle \bar q q \rangle \sim \langle \sigma \rangle = \sigma_0$ 
are enhanced as $\sim 1/V$, and the pion mass 
$m_\pi^2 \sim \frac{m_c}{\sigma_0}$
 in the model becomes \emph{smaller} for a fixed value of the symmetry-breaking current quark mass $m_c$. (The Gell-Mann--Oakes--Renner relation $f_\pi^2 m_\pi^2 = m_c |\langle \bar q q \rangle | + \cdots$ holds, and in the quark-meson models both $f_\pi \sim \sigma_0$ and $\langle \bar q q \rangle \sim \sigma_0$).
In accordance with our expectation, for very small volumes the pion mass always becomes large when chiral symmetry is restored. 
This result is an indication that the choice of the quark boundary condition in QCD does indeed have an effect on the low-energy behavior, at least insofar as the results of the model are applicable to QCD. But are they?

\begin{figure}
\includegraphics[scale=0.55]{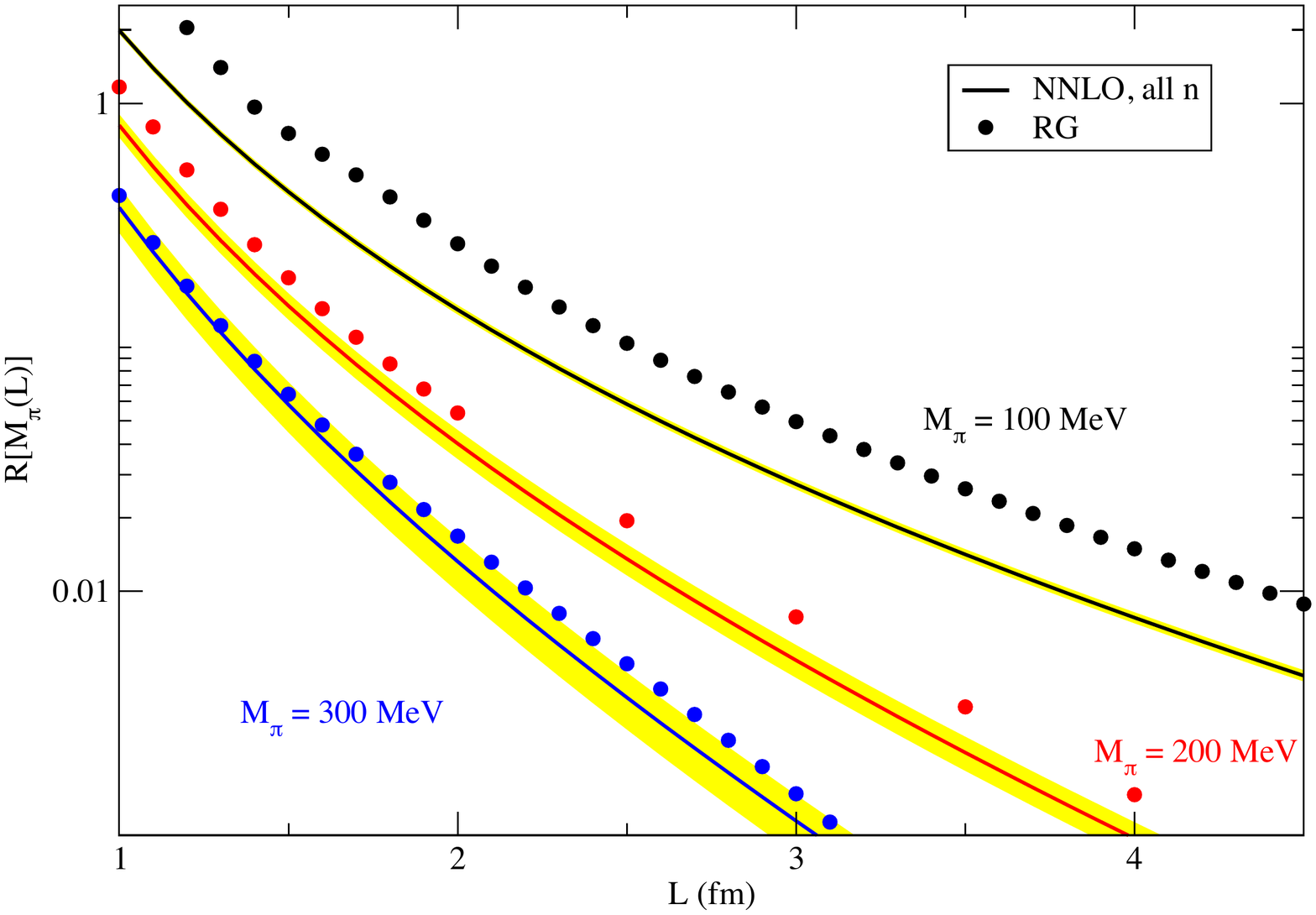}
\caption{Pion mass shift $R[m_\pi(L)] = \frac{m_\pi(L) - m_\pi(\infty)}{m_\pi(\infty)}$ in a finite volume $V = L^3$.  Results from the quark-meson model with \emph{anti-periodic} boundary conditions for quarks in the spatial directions obtained with Renormalization Group (RG) methods are compared to results from chiral perturbation theory. The chiral perturbation theory results are courtesy of G. Colangelo \emph{et al.} \cite{Colangelo:2003hf,Colangelo:2004xr,Colangelo:2005gd} and contain contributions to the L\"uscher formula from all $\vec{n}$ for the pions. The comparison below $L \approx 2- 3$ fm is for illustrative purposes only, as the ChPT description is presumed to reach its limit of applicability.}
\label{fig:qmmChPTcomparison}
\end{figure}

The quark-meson model in the form used here lacks important features of QCD: constituent quarks are not confined and can propagate over large distances, and the momentum dependence of the pion interactions is not fully implemented~\cite{Jungnickel:1997yu,Jendges:2006yk}. It is therefore not a quantitative theory of the low-energy interactions in QCD, and we have to see how far our insights carry. 
Of course we are still free to compare results to ChPT even on a quantitative level, and this will provide a  test for the applicability of these results to actual QCD. Such a comparison with the results from \cite{Colangelo:2005gd} is provided in Fig.~\ref{fig:qmmChPTcomparison}. 
We expect that only the results with anti-periodic quark boundary conditions are consistent with those from ChPT.
This is indeed the case: The results with periodic boundary conditions behave qualitatively different, while the results with anti-periodic quark boundary conditions behave qualitatively similar to those of ChPT. Thus only these latter results are shown in the quantitative comparison. This is in agreement with the arguments in \cite{Gasser:1987zq} which establish ChPT with the standard Lagrangian as the low-energy theory of QCD in a finite volume with quarks with \emph{anti-periodic} boundary conditions. 

We find that these results agree with those of the ChPT-calculation for large pion masses ($m_\pi(\infty)=300$ MeV) within errors. For small pion masses ($m_\pi(\infty)=100$ MeV), these results differ from those of ChPT by a constant factor, which appears as an offset in the logarithmic plot. Since the L\"uscher formula only provides the leading-order terms of the large-$L$ expansion, even including pions going around the lattice any number of times, the error is of order ${\mathcal O}(\mathrm{e}^{- \mathrm{const.}\,m_\pi L})$. Therefore convergence of L\"uscher's result is slower for smaller $m_\pi L$ and for smaller pion masses. The difference between the results seems consistent with such a slower convergence of this expansion for smaller pion masses.

We conclude that these results for the pion mass shift in a finite volume are compatible with ChPT for anti-periodic quark boundary conditions, but differ significantly for periodic quark boundary conditions. This is consistent with the arguments from \cite{Gasser:1987zq} and should caution against comparisons of ChPT to lattice results regardless of the quark boundary conditions in the simulation.  

\begin{figure}
\includegraphics[scale=0.95]{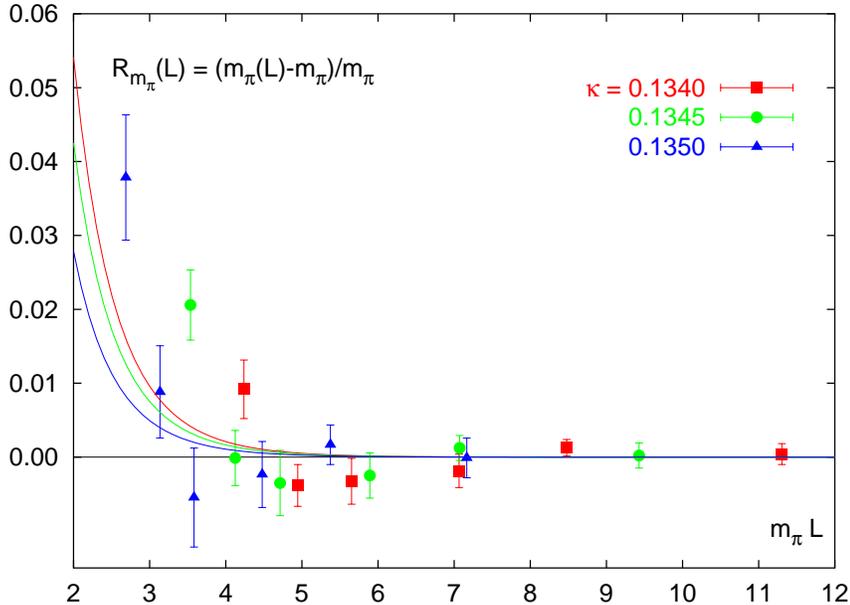}
\caption{Pion mass shift from the quenched lattice calculation \cite{Guagnelli:2004ww} with Wilson fermions and periodic boundary conditions for quarks in the spatial directions. The lattice spacing for the calculation is $a=0.079$ fm, the volume ranges from $0.9$ fm $\le L \le $ $2.5$ fm. The values for the pion masses corresponding to the hopping parameter values $\kappa$ are 
\cite{Guagnelli:2004ww,Garden:1999fg}: 
$\kappa = 0.1340$: $m_\pi(\infty) = 881$ MeV; 
$\kappa = 0.1345$:  $m_\pi(\infty) =735$ MeV;
$\kappa = 0.1350$:  $m_\pi(\infty) =559$ MeV.
The lines are the results from ChPT, including the two leading exponential terms from the result 
\cite{Gasser:1986vb}. Figure reprinted from M.~Guagnelli \emph{et al.}, Phys. Lett. B597, 216 (2004), Copyright (2004), with permission from Elsevier.
\label{fig:quenchedlatticecomparison}}
\end{figure}

A comparison of the ChPT-results based on the improved L\"uscher approach to lattice QCD results in both in quenched \cite{Guagnelli:2004ww} (see Fig.~\ref{fig:quenchedlatticecomparison}) and unquenched lattice calculations \cite{Orth:2005kq} indicate that the ChPT-results under-predict the observed mass shift. This might well be due to the simulation parameters still being outside of the domain of validity for the ChPT-calculation. 
However, in these calculations there is a deviation from the expected behavior even at the qualitative level: The pion mass in the finite volume is significantly \emph{smaller} than in infinite volume, i.e. the mass shift is \emph{negative}.
This is qualitatively very similar to the behavior that we observe in the model for periodic boundary conditions, although the results also cannot describe the lattice results quantitatively.

More recently, similar behavior has been observed in a calculation based on solutions of Dyson-Schwinger Equations (DSE) \cite{Luecker:2009bs}. The authors attribute this to possible quenching effects, since the feedback of pions is not completely taken into account.
This is not a contradiction to our interpretation, independently of possible quark effects:
One would indeed expect the pion mass to rise less quickly in a quenched calculation, compared to a calculation with full pion effects, since the main effect of pion fluctuations is the restoration of chiral symmetry. Thus, in a quenched calculation, the increase in the finite-volume pion mass with a decrease in the volume will be smaller.

Giusti {\it et al.} find that the ChPT formula predicts the correct exponential decrease for the mass shift with increasing volume, but that for small volumes ($\sim1.2$ fm) the theory underestimates the pre-factor by approximately one order of magnitude for the volumes and pion masses investigated in \cite{Giusti:2007hk}.

\begin{figure}
\includegraphics[scale=0.5]{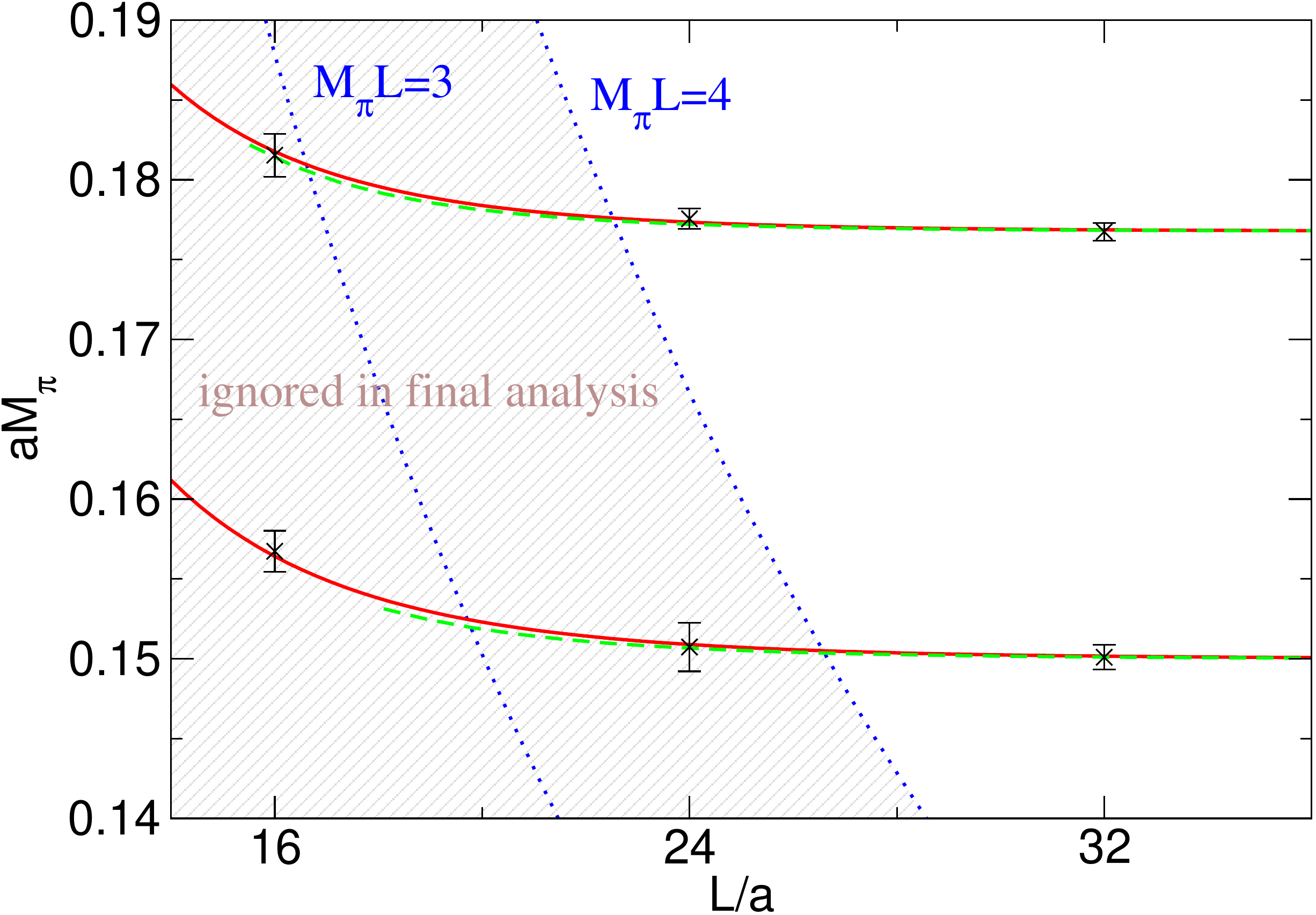}
\caption{Pion mass in a finite volume in lattice units $a M_\pi$ as a function of the volume size $L/a$ \cite{Durr:2010aw} ($a$ is the lattice spacing). The upper data set corresponds to a pion mass $M_\pi \simeq 300$~MeV, the lower set to a pion mass $M_\pi \simeq 250$~MeV.  The green dashed lines are the ChPT predictions for the pion mass from \cite{Colangelo:2005gd}. The solid red line is a fit to the data using the functional form of the ChPT result, multiplied by an overall constant as an additional free parameter. The blue dotted lines are given by the conditions $M_\pi L = 3$ and $M_\pi L = 4$ and denote boundaries for estimated magnitudes of finite-volume effects. Figure from S. Durr \emph{et al.}, JHEP 1108, 148 (2011), Copyright (2011),  reprinted with kind permission from Springer Science and Business Media.}
\label{fig:BMWfv}
\end{figure}
In \cite{Durr:2010aw}, details of a simulation at physical pion masses by the Budapest-Marseille-Wuppertal Collaboration \cite{Durr:2010vn} are given, and finite-volume effects are investigated in the course of estimating the simulation errors. Similar to the observations in~\cite{Giusti:2007hk}, the authors find that they can fit a wide range of results for the pion or other hadron masses in finite volume, provided they use the functional form of the ChPT-result with an overall adjustable prefactor. For the dedicated finite-volume investigation for the pion with results shown in Fig.~\ref{fig:BMWfv}, both such a fit and the ChPT result~\cite{Colangelo:2005gd} appear to agree with the lattice simulation results. 

In the review \cite{Fodor:2012gf}, very recent lattice simulation results are compared to the finite-volume predictions of ChPT for the finite-volume mass shift of the pion. It is found that most recent simulations are performed in regions of the $m_\pi$--$L$-space where the finite-volume effects are at most of the order of $1 \%$, according to the ChPT results \cite{Colangelo:2005gd}. Generally, the finite-volume effects for meson masses are negligible in the region with $m_\pi L >4$, while they can be quite sizable for $m_\pi L < 3$, see \cite{Fodor:2012gf}. 
For baryon masses, they can be substantially larger \cite{Colangelo:2010ba}. In these recent investigations, it appears that lattice simulations approach the region where finite-volume effects are exclusively due to pions. 

It appears likely that the possible presence and importance of additional terms $\sim 1/L$ in the chiral Lagrangian due to the broken permutation symmetry can only be decided in a dedicated comparison of simulations with different spatial boundary conditions for the quark fields.


\section{Models for dynamical chiral symmetry breaking}

\subsection{Quark-meson-model}
\label{sec:qmm}

In order to go beyond the low-energy effective description of QCD, it is very useful to employ models. For describing phase transitions in QCD, it is clearly necessary to include high-energy degrees of freedom, beyond those accounted for in a low-energy effective description.
The most satisfactory solution would be to solve QCD directly, including elementary fermionic and gluonic degrees of freedom, and the composite degrees of freedom arising at low energy. While a lot of progress has been achieved towards such a comprehensive solution of the theory, and brute-force computer simulations of this are possible as well, this approach is still difficult and not always the most advantageous. 

For one thing, a brute-force solution of the full theory, if possible, requires additional analysis to identify the physical mechanisms that underlie the observed phenomenology. While a full (numerical) solution contains all the relevant physics, it does not necessarily lead to an \emph{understanding} of the relevant mechanisms. This is where models can really shine, since they provide a restriction to only one particular set of operators and allow to explore the relevance of this set for a specific phenomenon. 
Additionally, models enable us to perform (simplified) calculations beyond those possible from a full solution of a theory, and therefore to explore its implications on a much wider scope.

In this spirit, the so-called chiral quark-meson model has been used to study the chiral phase transition both in infinite and in finite volume. The calculation has also been extended to finite baryon (or quark) chemical potential.
The model implements the chiral flavor symmetry of QCD. We concentrate on the case of two light flavors $N_{\mathrm{f}} = 2$. For the quark sector, the symmetry is implemented as an $SU(2)_{L} \times SU(2)_R$ symmetry.
In this case, we have assumed that the axial symmetry is broken explicitly by the anomaly and implement only the remaining invariance~\cite{Klevansky:1992qe}.
Conceptually, the quark-meson model can be obtained by bosonizing a purely fermionic Nambu--Jona-Lasinio model~\cite{Nambu:1961tp,Klevansky:1992qe}, which models spontaneous chiral symmetry breaking. 	While there are some conceptual differences, both implement the same mechanism for spontaneous breaking of chiral symmetry. They are reasonable models only reasonable below a particular momentum cutoff scale $\Lambda$, determined by typical hadronic scales. 
The model has been used in numerous studies of the low-energy phase of QCD and the chiral phase transition \cite{Jungnickel:1996aa,Jungnickel:1995fp,Berges:1997eu,Berges:1998sd,Schaefer:1999em,Braun:2003ii,Schaefer:2004en,Braun:2005fj,Braun:2010vd,Braun:2011iz,Pawlowski:2014zaa}.

The bare effective action of the quark-meson model at a large hadronic scale $\Lambda$ for $N_{\mathrm{f}}=2$ is given by
\begin{eqnarray} 
  \Gamma_{\Lambda}[\bar q,q,\phi]&=& \int d^{4}x \Big\{
  \bar{q} \left({\rm i}{\partial}\!\!\!\slash + 
    {\rm i}g(\sigma+i\vec{\tau}\cdot\vec{\pi}\gamma_{5}) + {\rm i} \gamma_0 \mu\right)q\nonumber \\ 
  && \qquad
  +\frac{1}{2} (\partial_{\mu}\phi)^{2}+U_{\Lambda}(\phi^2) - c \sigma \Big\} 
\label{eq:QM}
\end{eqnarray} 
with $\phi^{\mathrm{{T}}}=(\sigma,\vec{\pi})$. The quark fields are denoted by the Dirac spinors $\bar{q}$ and $q$.  
The meson sector in the two-flavor case ($N_{\mathrm{f}}=2$) is equivalent to a linear $\sigma$-model with an $O(4)$ symmetry. This symmetry is spontaneously broken to an $O(3)$ symmetry, with three emerging Goldstone modes $\vec{\pi}$. By convention, the meson field $\sigma$ acquires the finite expectation value $\sigma_0=\langle \sigma \rangle$ which signals the spontaneous breaking of chiral symmetry. The Yukawa interaction term between the elementary quark fields and the composite mesons is invariant under the chiral $SU(2) \times SU(2)$ symmetry, the coupling strength is characterized by the coupling $g$.

In nature, chiral symmetry is broken explicitly by the small masses of the light quarks. Because of this explicit symmetry breaking, the Goldstone bosons in the phase with broken chiral symmetry acquire finite masses. Since they remain light for light quarks, they determine long-range correlations and the infra-red behavior of the model.
In particular for investigations of finite-volume effects, the relation of the linear extent of the volume $L$ and the wavelength of the light fluctuations, determined by the mass of the Goldstone modes (pions) $m_\pi$, is of utmost importance~\cite{Braun:2004yk,Braun:2005gy}. In fact, in the absence of explicit symmetry breaking, the massless Goldstone fluctuations are going to restore the symmetry and lead to a vanishing of the expectation value of the field $\sigma$. This can be understood intuitively by picturing the Goldstone manifold: In the absence of explicit symmetry breaking, all points on the Goldstone manifold contribute with equal weight to the partition function. In a finite volume, the constant-field zero-mode contributions dominate the partition function if they are not damped by an explicit mass term \cite{Gasser:1987ah}. In order to find a phase with spontaneously broken symmetry, it is necessary to break the symmetry explicitly and make the Goldstone modes massive.

For these reasons, it is essential to include an explicit symmetry breaking term in the action. This can be done by including a quark mass term
\be
\bar{q} m_c q.
\ee
In these calculations, it is a choice to bosonize the current quark term at the UV scale, which leads to a term linear in $\sigma$ in the action. The symmetry-breaking parameter $c$ is related to the quark mass by the UV parameters of the model.
Different choices for the potential $U_\Lambda(\phi^2)$ at the UV scale are possible. The parameters of the potential are fixed to yield reasonable values for a set of observables ($f_\pi$, $m_\pi$, $m_\mathrm{q}$, $m_\sigma$) in the IR in the vacuum.
With an ansatz for the UV potential of the form
\be
U_\Lambda(\phi^2) &=& \frac{1}{2} m_\Lambda^2 \phi^2 + \frac{\lambda_\Lambda}{4} (\phi^2)^2 
\ee 
where the chiral $\mathrm{O}(4)$ symmetry is unbroken and the parameters $m_\Lambda$ and $\lambda_\Lambda$ characterize the potential, the parameter $c$ is given by
\be
c = \frac{m_c m_\Lambda^2}{g}
\ee

In these calculations a possible wave function renormalization is neglected. This is an approximation which has observable consequences in the critical behavior of the model, since it implies that the \emph{anomalous dimension} $\eta = 0$, which in turn leads to a modification of the critical exponents. Since the anomalous dimension for the $O(4)$ symmetry class in $d=3$ dimensions is small, this is not a serious limitation and justified for a first investigation of finite-size effects with the method.

Of the numerous calculations, in which the model has been used for QCD problems, we mention a few examples.
In~\cite{Berges:1997eu} thermodynamic quantities, the equation of state and the behavior at the phase transition were investigated with functional RG methods. 
In \cite{Berges:1998sd}, the chiral phase transition at finite density has been investigated, making use of a functional RG formalism, and a first-order transition at high baryon density is observed. 
In \cite{Schaefer:2004en}, the phase diagram of the model in the chemical-potential -- temperature plane has been calculated, using a non-perturbative renormalization group equation. 
The model has been used to study finite-volume effects on chiral symmetry breaking~\cite{Braun:2004yk,Braun:2005gy}, the chiral phase transition~\cite{Braun:2005fj,Braun:2011iz}, and to investigate scaling effects~\cite{Braun:2010vd}. 
In~\cite{Pawlowski:2014zaa}, the model has been studied using non-perturbative Renormalization Group methods and taking higher-order quark mesonic scattering processes into account. This approach goes beyond a local potential approximation, and takes wave function renormalizations for the fields as well as a renormalization of the Yukawa coupling into account. Among other quantities, the transition temperature and the curvature of the first-order transition line in the finite temperature - finite baryon chemical potential plane are investigated. However, since the meson wave function renormalization on the one hand tends to enhance mesonic fluctuations and thus to lower the transition temperature, and on the other hand a running Yukawa coupling enhances the strength of the chiral symmetry breaking and thus tends to increase the transition temperature, the effects counteract each other and the overall effect on the phase diagram is small. Results for the curvature of the transition line as a function of chemical potential are within $2\%$ of those obtained in~\cite{Braun:2011iz}.

\subsection{Polyakov-loop extended models}
An important \emph{caveat} regarding the NJL model or quark-meson models as its bosonized offspring is the absence of gauge field effects. Namely, there is no confinement of elementary quarks in the low-energy region. 
In a pure gauge theory with infinitely heavy quarks, the \emph{Polyakov loop} acts as an order parameter of quark confinement. The Polyakov loop $L(\vec{x}\,)$ is a Wilson loop in Euclidean time direction in compactified Euclidean space-time, at fixed spatial coordinate $\vec{x}$:
\be
L(\vec{x}\,) &=& \mathcal{P} \exp\Big\{ \mathrm{i} \int_0^{1/T} \mathrm{d} \tau A_4(\tau, \vec{x}) \Big\},
\ee 
where $\mathcal{P}$ denotes path ordering, and we define the expectation value 
\be
\Phi(\vec{x}) &=& \left\langle \frac{1}{N_\mathrm{c}} \mathrm{tr} L(\vec{x}\,)   \right\rangle
\ee
which acts as the order parameter. The Polyakov loop transforms non-trivially under an element of the $\mathrm{Z}(3)$ center symmetry of the gauge group $\mathrm{SU}(3)$;  in a high-temperature phase, where quarks are not confined, $\Phi \neq 0$ and the center symmetry is spontaneously broken, in the low-temperature phase with quark confinement $\Phi=0$ and the center symmetry is restored.

With regard to the free constituent quarks in an NJL model, the absence of confinement effects can be partially cured by introducing an effective potential for the Polyakov loop. By coupling the quarks to the Polyakov loop, and determining the expectation value from the effective potential, color non-singlet states of quarks can be suppressed in the low-temperature phase. This simulates quark confinement and removes thermodynamic effects of the constituent quarks at low temperature.

The basic idea of the Polyakov-loop extended models is to use an effective potential for the effective value of the Polyakov loop or the associated gauge field components~\cite{Fukushima:2003fw}.
While originally phenomenologically motivated~\cite{Meisinger:1995ih}, the potential has recently been determined by fitting to results from lattice gauge theory \cite{Fukushima:2003fw,Roessner:2006xn,Ratti:2006wg,Schaefer:2007pw,Fukushima:2008wg}, or by using functional RG results \cite{Braun:2007bx}.

For two light quark flavors, the Euclidean action of the Polyakov-loop extended NJL model is given by~\cite{Fukushima:2003fw,Roessner:2006xn,Ratti:2006wg}
\be
S[\bar\psi, \psi, \Phi]&=& \int_0^{1/T} \mathrm{d} \tau \int \mathrm{d}^3x \Bigg \{ \bar\psi (\mathrm{i} D \fslash+ \gamma_0 \hat \mu - \mathbf{m}) \psi 
 + G \Big[ (\bar \psi \psi)^2 + (\bar \psi \mathrm{i} \gamma_5 \vec{\tau} \psi)^2 \Big]  \Bigg\} \el
&& \quad -\frac{1}{T} \int \mathrm{d}^3x \, \mathcal{U}(\Phi,\Phi^*,T)
\ee
$\mathbf{m}=\mathrm{diag}(m_u, m_d)$ is the quark mass matrix, and $\hat\mu=\mathrm{diag}(\mu_u, \mu_d)$ is the quark chemical potential matrix.

Different parameterizations of such effective potentials are employed in these models~\cite{Roessner:2006xn,Ratti:2006wg,Schaefer:2007pw,Fukushima:2008wg,Schaefer:2009ui}. 
By construction, the effective potentials can only reproduce behavior at the physical expectation value of the Polyakov loop. For this reason, it is not possible to discriminate among the different possible parameterizations outside of the minimum of the potential. 

We use a parameterization according to \cite{Roessner:2006xn,Ratti:2006wg}, which is motivated by the structure of the Haar measure of the SU$(3)$ gauge group:
\be\label{eq:effpphi}
	\frac{\mathcal{U}(\Phi,\Phi^*,T)}{T^4}=-\frac{1}{2}a(T)\Phi^*\Phi+b(T)\ln[1-6\Phi^*\Phi+4(\Phi^{*3}+\Phi^3)-3(\Phi^*\Phi)^2].\el
\ee
The coefficients in this expression are temperature dependent  
\be
	a(T)=a_0+a_1\left(\frac{T_0}{T}\right)+a_2\left(\frac{T_0}{T}\right)^2 &\textrm{and}& b(T)=b_3\left(\frac{T_0}{T}\right)^3
\ee
and are adjusted to reproduce QCD lattice simulation results for the pressure, the entropy density, and the entropy density in a pure gauge theory.  

The model can be extended to include a non-local fermion interaction~\cite{Hell:2008cc,Hell:2009by,Hell:2011ic,Kashiwa:2011td}, in order to make contact with the effective scale-dependent interaction obtained from QCD.

The model can be bosonized in order to more easily include fluctuation effects of mesonic fields. This bosonized model is then know as the Polyakov-loop extended quark-meson (PQM) model~\cite{Schaefer:2007pw,Herbst:2010rf}.
For the PQM model in mean-field solution, the thermodynamics and fluctuation effects have been calculated in~\cite{Skokov:2010sf}.
 In \cite{Schaefer:2007pw}, the phase diagram of this Polyakov--quark-meson model at finite temperature and baryon chemical potential has been determined in a mean-field analysis. 
In \cite{Skokov:2010uh}, the phase diagram and quark density fluctuation effects have been explored with functional renormalization group equations. 
In \cite{Herbst:2010rf}, the phase structure of the model is again explored beyond the mean-field level, taking quark effects on the Polyakov loop potential into account, which lead to significant effects at large chemical potential. 

The interplay between the chiral and the deconfinement phase transitions has also recently been investigated in an RG framework for a purely fermionic model \cite{Braun:2011fw,Braun:2012zq}. By exploring the renormalization group running of the scalar fermion coupling, the authors were able to show how the onset of spontaneous chiral symmetry breaking is influenced by the presence of a finite expectation value of the Polyakov loop, signaling the onset of the deconfinement transition. They find that for a wide range of parameters $T_\chi \ge T_{\mathrm d}$, where the temperatures are the chiral phase transition and deconfinement phase transition temperatures, respectively. For physically meaningful choices of the parameters, there is a parameter window in which the two transition temperatures appear ``locked" to one another. A similar argument has been made earlier in a mean-field analysis \cite{Meisinger:1995ih} for the large-$N_c$ limit.

\section{Renormalization Group methods}
Renormalization Group (RG) methods are a powerful way to systematically account for the effects of quantum fluctuations and thermal fluctuations in a physical system.  
By systematically integrating out fluctuations in a defined momentum shell at a momentum scale $k$, one obtains an effective theory with couplings dependent on this scale. By evolving the theory from large to small momentum scales, and by following the \emph{flow} of the couplings, one obtains an effective theory which includes fluctuation effects. Under a change of the RG scale $k$ the values of the couplings change and approach their full effective values in the limit $k\to 0$.

In the context of critical behavior at phase transitions, RG methods have been essential for our understanding of scaling relations \cite{Wilson:1973jj,Wegner:1972ih}. At a critical point, the correlation length diverges for fluctuations associated with the order parameter of an order-disorder transitions. These long-range fluctuations need to be accounted for if we want to describe the scaling behavior in a quantitatively correct way.   
At the heart of the application to critical phenomena is the observation that a system at criticality has only one relevant length scale -- the correlation length --  
which diverges at the critical point itself. This implies that the system becomes invariant under a length rescaling at criticality, and that much of its physical response can be deduced from the behavior of the system under such a length rescaling. 
The consequence is the paramount importance of \emph{fixed points} in the RG flow of dimensionless couplings which govern the critical behavior.

While many of the initial successes of RG methods have been obtained in perturbative calculations \cite{Wilson:1973jj,Wilson:1971bg,Wilson:1971dc,Wegner:1972ih,Pisarski:1983ms}, recently non-perturbative functional formulations of the RG \cite{Polchinski:1983gv,Wetterich:1992yh,Liao:1992fm,Morris:1993qb,Liao:1994fp} have been applied to strongly interacting systems, which necessitate a non-perturbative treatment.
Examples include QCD with its strong gauge coupling \cite{Reuter:1993kw,Reuter:1997gx,Ellwanger:1995qf,Gies:2006wv,Braun:2007bx,Braun:2008pi,Marhauser:2008fz,Braun:2009gm,Braun:2010cy,Fister:2011uw,Fister:2013bh,Braun:2014ata,Mitter:2014wpa} or strongly interacting fermion systems \cite{Gies:2009da,Gies:2010st,Braun:2011pp,Braun:2010tt,Schmidt:2011zu,Janssen:2012pq} in condensed matter physics and for ultracold atoms \cite{Blaizot:2004qa,Diehl:2007ri,Diehl:2007th,Floerchinger:2008jf,Diehl:2009ma,Braun:2011uq}.
For recent reviews of functional RG methods see e.g. \cite{Litim:1998nf,Bagnuls:2000ae,Berges:2000ew,Polonyi:2001se,Delamotte:2003dw,Pawlowski:2005xe,Schaefer:2006sr,Gies:2006wv,Delamotte:2007pf,Sonoda:2007av,Rosten:2010vm,vonSmekal:2012vx,Strodthoff:2016dip}.

In the context of QCD, the effective degrees of freedom at large momentum scales (quarks and gluons) are different from those at small momentum scales (hadrons). In principle, RG methods allow us to transition from one set of fundamental to another set of composite degrees of freedom \cite{Gies:2001nw,Gies:2002kd}, and to describe a theory over a large range of momentum scales in a unified framework. The application to QCD \cite{Braun:2005uj,Gies:2006wv,Braun:2007bx,Braun:2008pi,Braun:2009gm, Mitter:2014wpa,Strodthoff:2016pxx} appears very promising. 

Such a description can also be used for models for chiral symmetry breaking across the symmetry-breaking scale. These models are then able to describe chiral symmetry restoration at finite temperature and in small, finite volumes. The RG approach provides a framework for a consistent treatment of the models over a very wide range of temperatures and volume sizes.

\subsection{Functional Renormalization Group and Wetterich equation} 
In the work considered in this review, mainly the RG flow equation in the formulation according to Wetterich
\cite{Wetterich:1992yh} is used, which is based on a scale-dependent effective action. In contrast to a formulation of the RG in terms of a renormalized \emph{Hamiltonian} of the system~\cite{Wilson:1971bg,Wilson:1971dh,Wilson:1973jj,Wegner:1972ih}, the central object in this case is the \emph{effective action} $\Gamma$, the generating functional of one-particle irreducible $n$-point correlation functions. 

A scale-dependent effective action is obtained by introducing a scale-dependent infrared cutoff term of the form
\be
\Delta S_k[\phi] = \frac{1}{2}\int \frac{\mathrm{d}^dq}{(2 \pi)^d} \phi(-q) R_k(q) \phi(q)
\ee
into the partition function. The function $R_k(q)$ is a momentum-dependent cutoff function with an intrinsic scale $k$. With the aid of this cutoff function, fluctuations in momentum space are then systematically integrated out and their effects are included in the renormalized action.
The scale-dependent or coarse-grained effective action $\Gamma_k$ contains fluctuation effects from the momentum modes with $q \gtrsim k$. 

Finally, from considering the change of the effective action under a lowering of the scale $k$, one obtains a one-loop flow equation for the effective action
\begin{equation} 
  \label{eq:Flow_EffAction} 
k  \partial_k \Gamma_{k} = \frac{1}{2} \STr\left\{  \left[\Gamma _{k}^{(2)} + R_k\right]^{-1} \left(k\partial_k
      R_k\right) \right\} \ .
\end{equation}
The expression which appears in the loop is the full effective propagator with the functional derivative $\Gamma_k^{(2)}[\phi](p, q) \equiv \frac{\delta^2}{\delta \phi(p) \delta \phi(q)} \Gamma_k[\phi]$. The cutoff function $R_k$ gives an effective mass to the IR modes and suppresses long-range fluctuations for finite values of $k$. 
The cutoff function has to satisfy several constraints~\cite{Berges:2000ew}:
\begin{enumerate}
\item In order to regulate IR fluctuations, it needs to act as an effective mass for small momenta $p^2<k^2$:
\begin{equation}\lim_{\frac{p^2}{k^2} \to 0} R_k(p^2) >0. \end{equation}
\item In order to ensure that one obtains the complete effective action in the limit $k \to 0$ when the cutoff vanishes, the cutoff function itself must vanish in this limit:
\begin{equation}\lim_{\frac{k^2}{p^2} \to 0} R_k(p^2) =0. \end{equation}
\item In order to recover the original classical action that defines the theory at either a finite, large ultraviolet cutoff scale $\Lambda$, or in the limit of $\Lambda \to \infty$, the cutoff function itself must diverge in this limit:
\begin{equation}\lim_{k \to \Lambda} R_k(p^2) \to \infty. \end{equation}
(This has the effect of quenching fluctuations around the expectation value of the fields, such that the bare, classical field $\varphi$ and the scale-dependent expectation value $\phi_k=\langle \varphi\rangle_k$ coincide for $k = \Lambda$.) 
\end{enumerate}
As a result, for a cutoff function that satisfies these requirements, the scale-dependent effective action $\Gamma_k$ interpolates between the classical action $S=\Gamma_{k \to \Lambda}$ and the full quantum effective action $\Gamma = \Gamma_{k \to 0}$.

The truncated effective action will converge towards the full effective action when an increasing number of coupling constants are included. The speed of convergence depends on the choice of truncation scheme. For scalar field theories, a reasonable starting point is an expansion in terms of a purely local potential. This is justified, since the non-canonical momentum dependence of the two-point function in a scalar field theory and the corresponding anomalous dimension $\eta$ is very small. This approximation does, however, introduce a systematic truncation error on the results. In light of the fact that finite-volume calculations were novel, this decision appears justified. A systematic improvement of the calculation is possible. 

A description in terms of purely \emph{local} couplings is a first approximation to the full result for the effective action. Since the effective action generates $n$-point correlation functions, 
only a calculation that takes actual correlations and hence the momentum-dependence of the $n$-point couplings into account can provide a complete picture of the theory. 

The derivative expansion \cite{Morris:1994ie,VonGersdorff:2000kp,Mazza:2001bp,Canet:2002gs,Canet:2003qd,Litim:2010tt} is such a systematic extension of the local expansion which includes momentum dependence of couplings, organized in powers of momenta or numbers of derivatives in position space representation. Recent progress in the implementation of momentum-dependent couplings has been made in \cite{Blaizot:2005xy,Blaizot:2005wd,Blaizot:2006vr,Benitez:2011xx} and in \cite{Tripolt:2013jra,Pawlowski:2014zaa,Pawlowski:2015mia,Strodthoff:2016pxx}.

Different solution methods have been employed to achieve global descriptions of the effective potential beyond an expansion around a local minimum:
In the grid approach, the effective potential and the couplings are discretized and described globally in the space of the fields \cite{Adams:1995cv,Schaefer:2004en,Schaefer:2006sr}.
In~ \cite{Bervillier:2007rc}, several different methods for high accuracy calculations are discussed, which also include solution methods for global descriptions of the potential.
More recently,  pseudo-spectral methods have been used as a high-accuracy method making use of Chebyshev polynomials as basis functions for a global description of the potential in field space~\cite{Borchardt:2015rxa,Borchardt:2016pif}, successfully benchmarked against known high-accuracy results, and among other applications employed to solve RG flows for $\mathrm{O}(N)$ models in $d=3$ and fractal $d=2.4$ dimensions. 

The convergence can be improved by \emph{optimizing} the RG flow through the choice of a particularly suitable cutoff function for a given truncation~\cite{Litim:2001up,Litim:2001fd,Litim:2000ci,Pawlowski:2005xe}. For this reason an optimized cutoff for the local potential approximation is used here.
For a calculation in $d$ dimensions at zero temperature, it makes sense to use a $d$-dimensional regulator function 
\be
R_{\mathrm{B}, k}(q) &=& q^2 r_{\mathrm{B}}(q^2/k^2)\quad\text{and}  \nn\\
R_{\mathrm{F}, k}(q) &=& q \,\fslash \, r_\mathrm{F}(q^2/k^2)
\label{eq:d-1optreg}
\ee
for bosons and fermions, respectively. The cutoff term for fermions is of the form \cite{Jungnickel:1995fp,Berges:2000ew}
$\Delta S_{\mathrm{F},k}[\bar \psi, \psi] = (2\pi)^{-d} \int \mathrm{d}^dq \bar\psi(-q) R_{\mathrm{F}, k}(q^2) \psi(q)$, where the fermion cutoff function must be of the form $R^2_{\mathrm{F}, k}(q^2) =  q^2 r_{\mathrm{F}}(q^2/k^2)$~\cite{Litim:2001up}.

For calculations at finite temperature, it is sufficient to use a regulator function that depends only on the \emph{spatial} momenta $R_k(\vec{p}^{\;2})$ \cite{Litim:2006ag,Blaizot:2006rj}, i.e.  a $d-1$-dimensional regulator for a $d$-dimensional theory.  This has the advantage that the Matsubara frequency sum can still be performed analytically.
Therefore it is sensible to use the following regulator functions for calculations in $d=3+1$ space-time dimensions \cite{Litim:2001up,Litim:2006ag,Blaizot:2006rj} at finite temperature: 
\begin{eqnarray} 
R_{\mathrm{B}, k} (p_0,\vec{p}\,) &=&\vec{p}^{\,2} r_{\mathrm{B}} (\vec{p}^{\,2}/k^2)\quad\text{and} \nonumber \\
R_{\mathrm{F}, k} (p_0,\vec{p}\,) &=&  \vec{p}\, \fslash \,r_{\mathrm{F}}(\vec{p}^{\,2}/k^2)\ .
\end{eqnarray}
\label{eq:doptreg}
The shape functions $r_{\mathrm{B}} (x)$ and $r_{\mathrm{F}} (x)$ are given
explicitly by
\begin{eqnarray} 
  r_{\mathrm{B}} (x)&=&\left(\frac{1}{x} - 1\right)\Theta (1-x)\quad \text{and}\nonumber \\
  r_{\mathrm{F}} (x)&=&\left(\frac{1}{\sqrt{x}}-1\right)\Theta (1-x)\ .
\end{eqnarray}

Different calculation strategies have been used, depending on the problem. For calculations with the quark-meson model, initially model parameters were fixed in infinite volume, at vanishing temperature, to reproduce several physical observables. 
Calculations at finite temperature and in finite volume were then performed with the same initial parameters, and thermodynamic effects and effects of the finite volume were estimated. In each set of calculations, first an appropriate regulator function for infinite volume and zero temperature was chosen, and the finite-temperature and finite-volume calculations were then performed with the corresponding regulator function for this situation.

In the case of $\mathrm{O}(N)$-models, the transition between the ordered phase with spontaneously broken symmetry and the disordered phase with restored symmetry can be controlled by tuning the initial values of the couplings in the theory. For an investigation of the scaling behavior in $d=3$ dimensions, it is sufficient to treat the model in a $d=3$-dimensional framework and to use the relevant coupling that controls the transition as a ``temperature".   

In the following, we give an overview over the RG flow equations in infinite and finite volume that have been used in different studies. 

\paragraph{Infinite Volume}
Flow equations in infinite volume can be derived in different cutoff schemes, depending on the goal of the calculation. 
We first consider a regularization in $d$ dimensions at $T=0$ with a $d$-dimensional regulator. In this case, the flow equation for the quark-meson model becomes
\be
k \partial_k U_k(\phi^2) &=& \frac{k^{d+2}}{(4 \pi)^{d/2}} \frac{1}{\Gamma(\frac{d}{2}+1)} \left\{ 
\frac{N_{\mathrm{f}}^2-1}{E_\pi^2} + \frac{1}{E_\sigma^2}-\frac{4 N_{\mathrm{f}} N_{\mathrm{c}}}{E_\mathrm{q}^2} \right\}
\ee
where the effective energies are given by
\be
E_\pi = \sqrt{k^2+M_\pi^2}, \;\; E_\sigma = \sqrt{k^2+M_\sigma^2}, \;\; \mathrm{and} \;\; E_\mathrm{q} = \sqrt{k^2+M_\mathrm{q} ^2},
\ee
and the scale-dependent particle masses are
\be
M_\mathrm{q}^2= g^2 \phi^2, \;\; M_\pi^2 = 2 U_k^\prime(\phi^2), \;\; M_\sigma^2 = 2 U_k^\prime(\phi^2)+ 4 \phi^2 U_k^{\prime\prime}(\phi^2). 
\ee
We will use this notation for the effective energies and the effective particle masses also in the following. 
Specifically for $d=4$ Euclidean dimensions and with a $d=4$-dimensional cutoff function, one finds
\be
k \partial_k U_k(\phi^2) &=& \frac{k^{6}}{32 \pi^2}  \left\{ 
\frac{N_{\mathrm{f}}^2-1}{E_\pi^2} + \frac{1}{E_\sigma^2}-\frac{4 N_{\mathrm{f}} N_{\mathrm{c}}}{E_\mathrm{q}^2} \right\}.
\label{eq:inf_qmm_flow}
\ee
This flow equation can be used to investigate quantum effects, the mechanism of chiral symmetry breaking, and meson dynamics at $T=0$ \cite{Jungnickel:1995fp,Jendges:2006yk} with regard to chiral perturbation theory.  
It can also serve as a starting point for the investigation of finite-volume effects at $T=0$, where the parameters of the model are fixed from this flow equation and the results from a calculation in finite spatial volume $V=L^3$ are then a prediction \cite{Braun:2004yk,Braun:2005gy}. 

For the $\mathrm{O}(N)$ model in $d=3$ dimensions, our starting point for the investigation of the finite-volume effects is a flow with a $d=3$-dimensional regulator, where the flow equation is given by
  \begin{eqnarray}
    \label{eq:infv_o_n_flow_equation}
  k \partial_k U_k(\phi^2)
    &=& \frac{k^{5} }{12 \pi^2}\Bigg\{ \frac{N-1}{E_\pi^2} 
    +\frac{1}{E_\sigma^2}  \Bigg \}.
  \end{eqnarray}

When we want to consider flow equations in infinite volume in $d$ Euclidean dimensions at finite temperature in the Matsubara formalism, it is sufficient to regularize the theory in $d-1$ spatial dimensions. We then use a cutoff function for the spatial momenta only, as already outlined above. The summation over the Matsubara frequencies factorizes in this case and can be carried out explicitly.
In a modeling strategy, in order to determine the model parameters it is nonetheless necessary to also calculate the flow in the same cutoff scheme at $T=0$. For this reason, we also require flow equations with a $d-1$-dimensional cutoff at $T=0$.

The general result for the flow equation of the quark-meson model in $d$ dimensions at $T=0$ with a $d-1$-dimensional cutoff function for the spatial momenta only is given by
\be
k \partial_k U_k(\phi^2) &=& \frac{k^{d+1}}{(4 \pi)^{(d-1)/2}}\frac{1}{2\, \Gamma(\frac{d+1}{2} )}   \left\{-\frac{4 N_{\mathrm{f}} N_{\mathrm{c}}}{E_\mathrm{q}} + \frac{N_{\mathrm{f}}^2-1}{E_\pi} + \frac{1}{E_\sigma}\right\}.
\ee
Specifically for the flow equations in infinite volume for $d=3+1$ space-time dimensions at $T=0$ with a $d=3$-regulator for the spatial momenta only  the flow equation becomes
\be
k \partial_k U_k(\phi^2) &=& \frac{k^5}{12 \pi^2}  \left\{-\frac{4 N_{\mathrm{f}} N_{\mathrm{c}}}{E_\mathrm{q}} + \frac{N_{\mathrm{f}}^2-1}{E_\pi} + \frac{1}{E_\sigma}\right\}.
\ee
We solve this flow in order to fix the model parameters at $T=0$ and $L \to \infty$, and can then use the model to investigate effects at finite $T$ and $L$. 

Extending this to finite temperature and including a chemical potential for the quarks, which carry a conserved baryonic charge, one finds in $d=3+1$ Euclidean dimensions and with a $d=3$-dimensional cutoff function the quark-meson model flow equation
  \begin{eqnarray}
    \label{eq:ft_flow_equation}
  k \partial_k U_k(\phi^2)
    &=& \frac{k^5}{6 \pi^2} \Bigg[ \frac{N_{\mathrm{f}}^2-1}{E_\pi} 
    \left( \frac{1}{2}+n_B(E_\pi) \right)
    +\frac{1}{E_\sigma} \left(
      \frac{1}{2}+n_B(E_\sigma)\right) \nn\\
   && -\frac{2 N_\mathrm{c} N_\mathrm{f}}{E_q} \Big( 1-n_F(E_q,\mu)  - n_F(E_q,-\mu)
    \Big) \Bigg] \, ,
  \end{eqnarray}
where the distribution functions have the usual definitions
\be
n_{\mathrm{B}} (E) = \frac{1}{\mathrm{e}^{\beta E}-1}, \;\; n_{\mathrm{F}}(E, \mu) = \frac{1}{\mathrm{e}^{\beta(E-\mu)} +1}.
\ee
One recovers the $T=0$-limit at finite chemical potential from this expression~\cite{Braun:2003ii,Schaefer:2004en}:
  \begin{eqnarray}
    k \partial_k U_k(\phi^2) &=& \frac{k^5}{12 \pi^2}\left [ \frac{N_{\mathrm{f}}^2-1}{E_\pi}   +\frac{1}{E_\sigma}  -\frac{4 N_\mathrm{c} N_\mathrm{f}}{E_q} \Theta(E_q - \mu)  \right]
  \end{eqnarray}

\paragraph{Finite Volume}
For finite volume $V=L^d$, the derivation of the RG flow equations is essentially unchanged. The $d$-dimensional integral over all spatial loop momenta in the trace of the functional one-loop flow equation is replaced by a sum over discrete momentum modes:
\be
\int_{-\infty}^\infty \mathrm{d} p_1 \cdots \int_{-\infty}^\infty  \mathrm{d} p_{d} \longrightarrow \left(\frac{2 \pi}{L}\right)^{d}\sum_{n_1=-\infty}^\infty \cdots \sum_{n_{d}=-\infty}^\infty.
\ee
Depending on the boundary conditions for the spatial directions, the discrete momenta are
\be
\vec{p}_{\mathrm{p}}^{\,2} = \frac{4\pi^2}{L^2} \sum_{i=1}^{d} n_i^2
\label{eq:fv_mom_p}
\ee
for periodic (p) boundary conditions, and 
\be
\vec{p}_{\mathrm{ap}}^{\,2} = \frac{4\pi^2}{L^2} \sum_{i=1}^{d} \left(n_i+\frac{1}{2}\right)^2
\label{eq:fv_mom_ap}
\ee
for anti-periodic (ap) boundary conditions.

In an $\mathrm{O}(N)$-model, the transition between ordered and disordered phase can be controlled by the initial values of the relevant coupling, and thus we can perform investigations of the finite-size scaling behavior in $d=3$ in this model in a finite spatial volume $V=L^3$. Corresponding to eq.~\eqref{eq:infv_o_n_flow_equation}, the flow equation in finite $d=3$-volume with a $d=3$-regulator is
  \begin{eqnarray}
    \label{eq:fv_o_n_flow_equation}
  k \partial_k U_k(\phi^2)
    &=& \frac{k^{d+2} }{2} {\mathcal B}_{\text{p}}(kL) \Bigg[ \frac{N-1}{E_\pi^2} 
    +\frac{1}{E_\sigma^2}  \Bigg ],
  \end{eqnarray}
where ${\mathcal B}_{\text{p}}(kL)$ is a mode counting function for the nodes of a unit grid inside a $d=3$-dimensional sphere with radius $kL$ (see below). 

For the quark-meson model, we once again consider a $d=3+1$-dimensional model in a finite Euclidean volume $V=L^{3} \times 1/T$. A $d=3$ regulator function again allows to sum the Matsubara frequencies, and we include a chemical potential for the quark fields for completeness:
  \begin{eqnarray}
    \label{eq:fv_ft_flow_equation}
  k \partial_k U_k(\phi^2)
    &=& k^5 \Bigg[ \frac{N_{\mathrm{f}}^2-1}{E_\pi} 
    \left( \frac{1}{2}+n_B(E_\pi) \right){\mathcal B}_{\text{p}}(kL) \nn \\
    && \quad\quad\quad
    +\frac{1}{E_\sigma} \left(
      \frac{1}{2}+n_B(E_\sigma)\right){\mathcal
      B}_{\text{p}}(kL) \nn\\
   && -\frac{2 N_\mathrm{c} N_\mathrm{f}}{E_q} \Big( 1-n_F(E_q,\mu) -  n_F(E_q,-\mu)
    \Big){\mathcal B}_{\text{l}}(kL) \Bigg]  .
  \end{eqnarray}
While only periodic boundary conditions for the meson fields appear sensible, for the quark fields both periodic and anti-periodic boundary conditions are a reasonable choice. Consequently two different mode counting functions are needed. 
The definition of the mode-counting functions  ${\mathcal B}_{\text{l}}$ with $\mathrm{l} \in \{ \mathrm{p}, \mathrm{ap} \}$ is: 
\begin{equation} 
  {\mathcal B}_\text{l}(kL)=\frac{1}{(kL)^3} \sum_{\vec{n} \in \mathbb{Z}^3} 
  \Theta\!\left( (kL)^2    
    - \vec{p}_{\text{l}}^{\,2}L^2\right)\,,
\end{equation} 
where the momenta $\vec{p}^{\,2}_\mathrm{l}$ have been defined in eqns.~\eqref{eq:fv_mom_p} and \eqref{eq:fv_mom_ap}.
For small volumes, the boundary conditions and the presence of a zero-mode for periodic and the absence of such a mode for anti-periodic boundary conditions become important. Hence for small $kL$, we find 
\be
\lim_{kL \to 0}  {\mathcal B}_\text{p}(kL) \sim \frac{1}{(kL)^3}
\ee
for periodic and
\be
\lim_{kL \to 0}  {\mathcal B}_\text{ap}(kL) =0
\ee
for anti-periodic boundary conditions~\cite{Braun:2010vd,Braun:2011iz}. 
Asymptotically, the mode counting functions behave for large volumes $V=L^3$ in the limit $L\to \infty$ for fixed $k$ and for both $\text{l}=\mathrm{p}$ and $\text{l}=\mathrm{ap}$ as
\be
\lim_{kL \to \infty}  {\mathcal B}_\text{l}(kL) = \frac{1}{6\pi^2},
\ee
i.e. the number of modes grows like the volume of a $3$-dimensional sphere for $kL \gg 1$.
Using these relations for the large-volume limit, one recovers from eq.~\eqref{eq:fv_ft_flow_equation} the flow equations for infinite volume~\cite{Braun:2003ii,Schaefer:2004en,Stokic:2009uv} for the quark meson model.

For completeness, we mention that 
applications of the finite-volume formalism to non-relativistic fermions at finite density -- such as ultracold atoms in a finite trap -- require a modification in the argument of the cutoff functions:  At finite density, the momentum shell renormalization is performed towards the Fermi surface of the system, and appropriate cutoff functions with a dependence on the chemical potential are chosen~\cite{Diehl:2009ma}. The finite-volume case is discussed in~\cite{Braun:2011uq}. 
For our purposes, it is sufficient to restrict ourselves to relativistic systems or to scalar theories without conserved charges and hence without finite chemical potential for the scalar fields. 
More recently,  the dimensional crossover from three to two dimensions for a confined system of non-relativistic bosons in a trap was studied in~\cite{Lammers:2016xxx} with finite-volume functional RG methods. For this purpose the extent of one spatial dimension of the system was varied and the effect on the phase transition temperature was studied. In addition, $s$-wave scattering lengths are studied in a T-matrix formalism and related between two and three dimensions.

An additional application of functional renormalization group methods in a finite geometry is the calculation of forces from the Casimir effect, i.e. the forces which appear in a system with (quantum) fluctuations between macroscopic confining boundaries limiting the fluctuation spectrum. The Casimir forces for $\mathrm{O}(N)$ models with $N=1,2$ and dimension between two and three are calculated in ~\cite{Jakubczyk:2012iza} with functional RG methods for a $d$-dimensional system with one dimension limited to a finite extent $L$.

The general requirements for regulators in finite-volume FRG calculations have recently been explored in \cite{Fister:2015eca}, using as an example a $\phi^4$-theory with a single scalar field ($N=1$) in the Ising universality class. In addition to the optimized regulator function discussed above, a family of exponential cutoff functions and a sharp cutoff function are studied in finite volume. The behavior of the free energy density and the pressure function in the infinite-volume limit are studied.
Similar to the discussion in this review, the interplay of length scales and temperature scale are discussed in detail.  The question of a vanishing condensate for such a model in finite volume and the closely connected intricate question of the convexity of the potential are also explored. The mechanism for the vanishing of the condensate for an Ising model and in the absence of Goldstone modes is demonstrated in the framework of the FRG equations.

\subsection{Proper-Time Renormalization Group}

Some of the earlier work for finite volumes has been performed in the framework of the \emph{Proper-Time Renormalization Group} (PTRG) \cite{Braun:2004yk,Braun:2005gy,Braun:2005fj,Braun:2008sg}. The method is based on an \emph{ad-hoc} regularization of the one-loop effective action, which is then promoted to a renormalization scale-dependent effective action. In contrast to the rigorous field-theoretical derivation of the flow for the effective action in Wetterich's scheme, there is only a \emph{``heuristic"}~\cite{Litim:2001hk} derivation of the RG flow of the effective action in the PTRG scheme. In special cases, the resulting flow equations are identical to those from an exact derivation in Wetterich's scheme.

The method goes back to work in~\cite{Liao:1994fp} using Schwinger's proper-time operator regularization method, and has been extended and used in numerous applications \cite{Spitzenberg:2002tq,Braun:2004yk,Braun:2005gy,Braun:2005fj,Jendges:2006yk,Schaefer:1999em,Braun:2003ii,Schaefer:2006sr,Mazza:2001bp,Meyer:2001zp,Zappala:2002nx,Papp:1999he,Bohr:2000gp}.

For a scalar theory with an action of the form
\be
S_{\mathrm{cl}}[\phi] &=& \int \mathrm{d}^d x \,\left\{ \frac{1}{2} (\partial_\mu \phi)^2 + U(\phi) \right\}, 
\ee
the corresponding one-loop effective action is given by 
\be
\Gamma^{\mathrm{1-loop}}[\phi] &=& S_{\mathrm{cl}}[\phi] +\frac{1}{2} \Tr \log S^{(2)}_{\mathrm{cl}}[\phi]
\ee
where $S_{\mathrm{cl}}^{(2)}[\phi] = \frac{\delta^2}{\delta \phi \delta \phi} S_{\mathrm{cl}}[\phi] $ is the second variation of the action. This expression can be rewritten in terms of a Schwinger proper-time integral as
\be
\Gamma^{\mathrm{1-loop}}[\phi] &=& S_{\mathrm{cl}}[\phi] - \frac{1}{2} \int_0^\infty \mathrm{d} \tau \frac{1}{\tau} \mathrm{Tr} \exp \left(-\tau S_{\mathrm{cl}}^{(2)}\right).
\ee
This can be used for an UV regularization for a divergent loop contribution by introducing an appropriate cutoff function in the integral \cite{Oleszczuk:1994st}. 
Beyond this, it has been recognized in \cite{Liao:1994fp} that the same formalism can be used to also include an infrared cutoff $k$ in the cutoff function. From this one can obtain a flow equation for a scale-dependent effective action 
\be
k \partial_k \Gamma_k[\phi] &=& -\frac{1}{2} \int_0^\infty \mathrm{d} \tau \frac{1}{\tau}\left(k \partial_k f_k(\tau)\right) \mathrm{Tr} \exp\left(- \tau \Gamma_k^{(2)}\right).
\ee
Here the derivative $\Gamma_k^{(2)}[\phi] = \frac{\delta^2}{\delta \phi \delta \phi} \Gamma_k[\phi]$ of a scale-dependent effective action replaces the derivative of the constant classical action $S_{\mathrm{cl}}[\phi]$ in the original equation and turns it into a functional flow equation for the effective action. In a sense, this renormalization-group improved equation is a self-consistency equation for the change in the functional $\Gamma_k$ to be solved at the IR scale $k$. 

The cutoff function has to satisfy several constraints \cite{Litim:2001hk}. It must ensure that for the infrared scale $k \to 0$, effects of the cutoff are removed; and that for $k \to \Lambda$ at the UV cutoff scale, the original (classical) theory is recovered. 
A suitable choice for the cutoff function is \cite{Liao:1994fp}
\be
f_{a}(\tau k^2) &=& \frac{\Gamma(a+1, \tau k^2)}{\Gamma(a+1)}
\ee
where $\Gamma(a+1, x) = \int_0^x \mathrm{d} s \, s^a \mathrm{e}^{-s}$ is the incomplete $\Gamma$-function. The derivative with regard to the cutoff scale $k$, which appears in the flow equation, is
\be
k \partial_k f_{a}(\tau k^2) &=& -\frac{2}{\Gamma(a+1)} \left(\tau k^2\right)^{a+1} \exp(-\tau k^2)
\label{eq:PTRGcutoff}.
\ee

Since the proper-time RG is based on a particular choice of regularization of a one-loop effective action, the results depend on this regularization. To some extent, this is common to all functional RG methods: The flow is dependent on the choice of regularization, but the result in the deep infrared limit is independent of this choice after all quantum fluctuations have been integrated out. Such an RG flow is called an \emph{exact} flow \cite{Litim:2002xm}. In practice, there is always a residual cutoff dependence when the RG flow equations are truncated in any expansion scheme.

The proper-time RG in an infinite-volume application has been shown to be exact in this sense for a particular choice of regulator function \cite{Litim:2001hk}. For this choice, it is equivalent to the flow obtained from the Wetterich equation, making use of an optimized regulator function for the local approximation \cite{Litim:2001up}.
The relation of the PTRG to exact RG flows and the conditions under which the PTRG flow becomes exact, as well as the connection to perturbation theory have been further explored and clarified in \cite{Litim:2002xm}. 

For the form of the cutoff function given above in eq.~\eqref{eq:PTRGcutoff}, the choice $a=\frac{d}{2}$ reproduces the exact Wetterich flow with the optimized Litim-regulator exactly in infinite volume \cite{Litim:2001hk}.

\paragraph{Infinite Volume}

The PTRG flow for the quark-meson model with the general cutoff function with parameter $a$ in $d$ space-time dimensions is 
\be
k\partial_k U_k(\phi^2)&=& \frac{k^{2(a+1)} }{(4\pi)^{d/2}} \frac{\Gamma(a+1-d/2)}{\Gamma(a+1)}\left\{  \frac{N_\mathrm{f}^2-1}{E_\pi^{2(a+1-d/2)}} + \frac{1}{E_\sigma^{2(a+1-d/2)}} \right. \el
&&  \left. -\frac{4 N_\mathrm{c} N_\mathrm{f}}{E_\mathrm{q}^{2(a+1-d/2)}}
 \right\}
\ee
For $a=d/2$ or $a=2$ at zero temperature for $d=4$ space-time dimensions, this reduces to the flow equation~\eqref{eq:inf_qmm_flow} from the Wetterich scheme with optimized cutoff. The corresponding expression for the $\mathrm{O}(N)$-model is
\be
k\partial_k U_k(\phi^2)&=& \frac{k^{2(a+1)} }{(4\pi)^{d/2}} \frac{\Gamma(a+1-d/2)}{\Gamma(a+1)}\left\{  \frac{N-1}{E_\pi^{2(a+1-d/2)}} + \frac{1}{E_\sigma^{2(a+1-d/2)}}  \right\}.\el
\label{eq:on_ptrg_flow_iv}
\ee

In the PTRG method, all space-time dimensions are included in the regularization procedure. This is a necessary consequence of the proper-time regularization, which is symmetry-preserving by construction and does not break Lorentz or Euclidean invariance. It is therefore not possible to implement a cutoff scheme in which only the spatial dimensions are regularized, and the Euclidean time dimension is left unaffected. However, for a specific choice of the cutoff parameter, the Matsubara sums in the proper-time RG flow equations can be performed analytically, and one recovers a flow equation coincident with one from a three-dimensional cutoff scheme in the exact Wetterich RG formulation.

As shown in \cite{Braun:2003ii,Schaefer:2004en}, for a choice of $a=3/2$ in $d=3+1$ dimensions, the sums over Matsubara frequencies yield after analytic evaluation the flow equation in the form
\be
k \partial_k U_k(\phi^2)&=& \frac{k^5}{6 \pi^2} \left\{\frac{N_\mathrm{f}^2-1}{E_\pi} \left(\frac{1}{2} +n_{\mathrm{B}}(E_\pi) \right) + \frac{1}{E_\sigma} \left(\frac{1}{2} +n_{\mathrm{B}}(E_\sigma) \right) \right. \el
&& \left.- \frac{2 N_{\mathrm{c}}N_{\mathrm{f}}}{E_\mathrm{q}} (1- n_{\mathrm{F}}(E_\mathrm{q}, \mu) -n_{\mathrm{F}}(E_{\mathrm{q}}, -\mu)  )\right\}.
\ee
For this choice, the flow equation coincides with the exact equation \eqref{eq:ft_flow_equation}  with $d=3$-dimensional optimized cutoff function.

\paragraph{Finite Volume}
In a finite volume, the general derivation of the RG flow equations is unchanged. The volume size $L$ appears as an additional relevant coupling or relevant scale parameter in the flow equations. 
From a practical point of view, once again the volume integration over a continuous momentum variable $\vec{p}$ is replaced by a summation over discrete momentum modes $\vec{p}_{\mathrm{p}}^{\,2} = \frac{4\pi^2}{L^2} \sum_{i=1}^d n_i^2$ (periodic (p) boundary conditions) or  $\vec{p}_{\mathrm{ap}}^{\,2} = \frac{4\pi^2}{L^2} \sum_{i=1}^d \left(n_i+\frac{1}{2}\right)^2$ 
(anti-periodic (ap) boundary conditions). While the flow equations for an appropriate choice of cutoff function again reduce to those of Wetterich's RG formulation with Litim's optimized cutoff in the infinite volume, this equivalence no longer holds in the finite-volume case. To our knowledge, it is not possible to obtain an exact flow from the finite-volume PTRG flow. 

For an $\mathrm{O}(N)$-model, the flow equations in an infinite $d$-dimensional volume and in a finite volume $V=L^d$ can be written in very compact form in terms of threshold functions  $\ell_{a+1}^{(d)}(L, \omega)$ for finite volume and $\ell_{a+1}^{(d)}(\omega)$ for infinite volume. The expressions for the threshold functions are given in the appendix.
In a finite volume $V=L^d$, the PTRG flow for the $\mathrm{O}(N)$ model is given by the equation
\be
k \partial_k U_k(\phi^2) &=& (k^2)^{a+1}\left\{ (N-1) \ell_{a+1}^{(d)}(L, E_\pi^2)+  \ell_{a+1}^{(d)}(L, E_\sigma^2)\right\}
\ee
where the threshold functions are now dependent on the volume size $L$. 
In the infinite-volume limit $L \to \infty$, this reduces to the infinite-volume flow equation~\eqref{eq:on_ptrg_flow_iv}
\be
k \partial_k U_k(\phi^2) &=& (k^2)^{a+1}\left\{ (N-1) \ell_{a+1}^{(d)}(E_\pi^2)+ \ell_{a+1}^{(d)}(E_\sigma^2)\right\}
\ee
written here in terms of the infinite-volume threshold function
\be
\ell_{a+1}^{(d)}(\omega)&=& \frac{1}{(4\pi)^{d/2}} \frac{\Gamma(a+1-d/2)}{\Gamma(a+1)} \frac{1}{\omega^{(a+1-d/2)}}
\ee
to which the finite-volume threshold functions reduce in this limit. For the analysis of scaling behavior, it is again sufficient to control the phase transition by varying the initial values of the relevant coupling. 

It is slightly more involved to now also implement a finite temperature in a field-theoretical sense as an additional, compactified $d^{\,\mathrm{th}}$ dimension. In this case, the cutoff functions will also depend on the ratio of the Euclidean time-dimension $\beta=1/T$ and the size of the spatial volume extent $V=L^{d-1}$ in the form of a new parameter $t=TL$ \cite{Braun:2005fj}.
The threshold functions $ \ell_{a+1}^{(\mathrm{F,B},\mathrm{ap,p}),(d)}(TL, L, \omega)$ that appear in this case depend on the choice of boundary conditions in the spatial directions, the volume size $L$ and the ratio $t = T L$. Since the boundary conditions in the Euclidean time direction are fixed, they also differ for bosons and fermions. 

The flow equation for the effective potential of the quark-meson model in a finite Euclidean volume $V=\frac{1}{T} \times L^{d-1}$ is in terms of these threshold functions
\be
k \partial_k U_k(\phi^2) &=& (k^2)^{a+1}\left\{  (N_\mathrm{f}^2-1) \ell_{a+1}^{(\mathrm{B}, \mathrm{p}),(d)}(TL, L, E_\pi^2)+\ell_{a+1}^{(\mathrm{B},\mathrm{p}),(d)}(TL, L, E_\sigma^2)\right. \el
&&\quad\quad\quad\quad\quad  \left. - 4 N_\mathrm{f} N_\mathrm{c} \ell_{a+1}^{(\mathrm{F}, \mathrm{l}),(d)}(TL, L, E_\mathrm{q}^2)
\right\}
\ee
where $\mathrm{l} \in \{\mathrm{p}, \mathrm{ap} \}$, depending on the choice for the quark boundary conditions in the spatial directions.
For $d=3+1$ dimensions, the choice $a=d/2=2$ is closest to the optimal choice in the rigorous Wetterich scheme. We note again that because of the inevitable mixing of space-time dimensions in this cutoff scheme, there is no direct analogue to the cutoff choice with only a spatial-momentum cutoff that is possible in that scheme.

A drawback of the PTRG method is the difficulty of keeping track of external momenta in correlation functions. While it is not impossible to extend the method beyond an expansion in purely local correlations \cite{Bonanno:2000yp,Bohr:2000gp,Jendges:2006yk}, it is very difficult to include the dependence of vertex functions and correlation functions on external momenta. This restricts the usefulness of the method beyond the local expansion, as it is difficult to describe systems where the momentum dependence of interaction vertices becomes essential and a purely local expansion is not sufficient.  

PTRG cutoff functions have been used in \cite{Braun:2004yk,Braun:2005gy,Braun:2005fj} for finite-volume calculations, and a comparison between the PTRG cutoff scheme and the one inherent in the Wetterich equation in~\cite{Braun:2008sg} has been used for a systematic estimate of the cutoff effects in finite volume between these two cutoff schemes. Such a comparison allows to estimate the size of truncation effects and gives a bound on the error in the calculation.
For more recent calculations~\cite{Braun:2010vd,Braun:2011iz,Tripolt:2013zfa,Springer:2015kxa}, the equations obtained in the formally exact Wetterich scheme have been used exclusively.

\section{Critical behavior and the Renormalization Group}
Phase transitions in physical systems are the primary examples of \emph{critical behavior}. Our understanding of \emph{continuous} or \emph{second-order} phase transitions would be woefully inadequate without understanding the effects of \emph{critical fluctuations}. 
Phase transitions are characterized by non-analytic behavior in thermodynamic observables, stemming from a non-analyticity in the free energy density of the system. In the case of continuous phase transitions, these non-analyticities are closely linked with a diverging correlation length at the transition. 

Because of the diverging correlation length at the critical point the system becomes essentially \emph{scale-free} and the behavior in the vicinity of this point is governed by power laws. This had been observed empirically long before the renormalization group provided a thorough understanding of the phenomenon~\cite{Widom:1965xx,Griffiths:1967xx}. 
In fact, the existence of power-law scaling behavior as such can be deduced from the assumption of a diverging correlation length~\cite{Wilson:1973jj}.

A consequence of the dominance of the long-range fluctuations over the physics of a system at a critical point is \emph{universality} of the observed scaling behavior: Since long-range infrared fluctuations determine the behavior, the microscopic details of a system become unimportant and the scaling behavior depends only on the dimensionality and the relevant symmetries of the system. 

The advent of renormalization group methods has ultimately provided us with a method to actually calculate the numerical values of critical exponents directly, and to also calculate the universal  scaling functions which characterize the observables in the scaling region around a critical point.
For this purpose, perturbative methods have been first employed \cite{Brezin:1972fb,Brezin:1973xx,Wallace:1975vi,Guida:1998bx}, but also non-perturbative renormalization group methods have been used to calculate critical exponents and to analyze scaling behavior \cite{Tetradis:1993ts,Berges:1995mw,Berges:1997eu,VonGersdorff:2000kp,Bohr:2000gp}. The correct determination of critical exponents remains an important test of RG methods \cite{Bervillier:2007rc,Litim:2002cf, Canet:2002gs,Canet:2003qd}.
Because critical behavior is dominated by the long-range fluctuation effects, it is extraordinarily sensitive towards finite-volume effects. If the system close to criticality is placed in a finite box, the scaling behavior is immediately altered, and actual critical behavior can only be recovered in the limit of the box size $L \to \infty$. This poses considerable problems for numerical simulations on finite, discrete lattices, where critical behavior is influenced by the finite simulation volume. The disadvantage of the small volume can only be overcome by understanding the finite-volume effects, and with a theory of finite-size scaling, it can actually be turned to an advantage: Finite-size scaling analysis can be used as an additional tool  and provide more information about the scaling behavior by comparing to finite-size scaling functions.  

\subsection{Scaling behavior in infinite volume}
In the vicinity of a critical point, such as a continuous or second-order phase transition, the dynamics of the system are dominated by the critical long-range fluctuations. Renormalization group arguments show that the singular part of the free energy density of the system satisfies to leading order the scaling relation  
\be
f_{\mathrm{s}}(t, h)&=& \ell^{-d} f_{\mathrm{s}}(t \ell^{y_t}, h \ell^{y_h})
\label{eq:fscaling}
\ee
under a length rescaling by the dimensionless rescaling factor $\ell$, which can be arbitrarily chosen.  
The parameters $t=(T-T_\mathrm{c})/T_0$ and $h=H/H_0$ are the reduced temperature, measured from the critical value $T_\mathrm{c}$, and the strength of the external symmetry-breaking field. The normalization constants $T_0$ and $H_0$ depend on the microscopic details of the system. After normalization, the results describe the universal critical behavior of a system independent of the UV structure.

For our purposes, it is sufficient to consider only two relevant couplings, temperature $t$ and symmetry-breaking field $h$, introduced above. Associated with these couplings are two exponents  $y_t$ and $y_h$, which specify all critical exponents for the scaling behavior, 
\be
y_t = \frac{1}{\nu}, \quad y_h = \frac{\beta \delta}{\nu},
\ee
when taken together with the scaling laws $\gamma= \beta (\delta-1)$ and $\gamma = (2-\eta) \nu$.

As a consequence of the scaling relations, observables of the system can now be expressed in terms of power-law behavior and the universal scaling functions. For our model, the observables we consider are the order parameter, here identified with the pion decay constant $M \equiv f_\pi$, and the susceptibilities $\chi_\pi$ for the Goldstone modes and $\chi_\sigma$, transverse and longitudinal with respect to the direction of the external field $h$.  By choosing the arbitrary rescaling factor $\ell$ such that either $t\ell^{y_t}=1$ or that $h \ell^{y_h}=1$, the free energy density becomes a function of only a single scaling variable, combining $t$ and $h$, with an additional explicit dependence on either $t$ or $h$. It follows immediately that any thermodynamic observables which can be expressed in terms of derivatives of $f(t, h)$ with respect to the two couplings can also be expressed in terms of such scaling functions~\cite{Widom:1965xx}.

For the order parameter, the scaling relation is given by
\be
M(t, h)&=& h^{1/\delta} f_M(z), \quad z = t/h^{1/(\beta\delta)},
\ee
where $z$ is the scaling variable, and $f_M(z)$ is the scaling function, normalized to $f_M(0)=1$. For small values of $h$ and $t<0$, the scaling function behaves asymptotically as $f_M(z) \simeq (-z)^\beta$ for large values of $-z \to \infty$. The two normalization constants $T_0$ and $H_0$ are determined from these two conditions, such that
\be 
M(h)= h^{1/\delta}
\label{eq:Hnormal}
\ee for $t=0$ and 
\be
M(t) = (-t)^\beta
\label{eq:Tnormal}
\ee for $h=0$ and $t<0$.

In the calculations presented in the following, a local potential approximation has been used, the leading order of an expansion in powers of derivatives, where the anomalous dimension vanishes, $\eta=0$. In this case, the static susceptibilities in the model are directly related to the masses of the mesons according to
\be
\chi_\pi = \frac{1}{M_\pi^2} \quad \mathrm{and} \quad \chi_\sigma = \frac{1}{M_\sigma^2}
\ee 
for the transverse and the radial modes. Both in an $\mathrm{O}(N)$-model and the quark-meson model, the susceptibility for the transverse mode is directly related to the order parameter $M= \langle \sigma \rangle = \sigma_0$:
\be
\chi_\pi &=& \frac{M}{H} = \frac{\sigma_0}{H}.
\ee
Thus the transverse susceptibility does not provide any additional information beyond that already contained in the order parameter $M$, and we do not need to consider it separately. 
The scaling function $f_\chi(z)$ of the longitudinal susceptibility $\chi_\sigma = \frac{\partial M}{\partial H}$ is closely related to the scaling function $f_M(z)$ of the order parameter and its derivative:
\be
\chi_\sigma &=& \frac{h^{1-1/\delta}}{H_0} f_\chi(z) = \frac{h^{1-1/\delta}}{H_0}\frac{1}{\delta} \left\{f_M(z) -\frac{z}{\beta} f^\prime_M(z)\right\}.
\ee
It has been verified that this relation holds between the scaling functions in the functional RG calculation \cite{Braun:2007td}, which is a highly non-trivial test of the method, since it requires that both the order parameter and the higher-order four-point coupling correctly include the relevant fluctuations. 

The calculation of scaling functions and critical exponents has been a very important application of the Renormalization Group.
First determinations of the scaling function for the order parameter or the magnetic equation of state have been obtained from perturbative RG methods, such as the $\varepsilon$-expansion \cite{Brezin:1972fb,Brezin:1973xx,Wallace:1975vi}.

Much of the work in the determination of critical exponents and scaling functions has been done with the simulation of $\mathrm{O}(N)$ lattice spin models. 
Critical exponents for the $\mathrm{O}(4)$-model were obtained in \cite{Kanaya:1994qe} to a high accuracy in such a spin model calculation.
Finite-size effects and scaling corrections in the determination of critical exponents in such models were corrected in~\cite{Ballesteros:1996bd,Hasenbusch:2000ph}.
A first determination of the $\mathrm{O}(4)$ scaling functions in $d=3$ dimensions from a lattice spin model has been done in~ \cite{Toussaint:1996qr} for the purpose of analyzing the chiral phase transition for QCD.
Goldstone mode effects  near the coexistence line (the first-order transition line at $H=0$) were investigated and a determination of the scaling functions was completed by Engels {\it et al.} for the 
$\mathrm{O}(4)$-~\cite{Engels:1999wf}  and $\mathrm{O}(2)$-models~\cite{Engels:2000xw}.
More recently, the transverse and longitudinal correlation lengths and their scaling functions in the O(4) lattice spin model have been investigated in~\cite{Engels:2003nq,Engels:2009tv}.
A representation of the $3d$ scaling functions for the $\mathrm{O}(4)$ model has been determined by Engels and Karsch in \cite{Engels:2011km} by re-analyzing data from the high precision spin model lattice simulations in large volumes in \cite{Engels:2009tv}. The volumes are sufficiently large that all finite-volume effects should have disappeared. The high precision and the parameterization of the scaling functions make these results very suitable for a scaling analysis of lattice QCD data.

Field-theoretical results for the $\mathrm{O}(4)$ scaling function for the order parameter were obtained in \cite{ParisenToldin:2003hq} from a small-field expansion, making use of the known analytical constraints on the equation of state.

Functional RG methods have been applied to the problem as well.
The observation of critical scaling behavior in a quark-meson model with $N_\mathrm{f}=2$ flavors and determination of the $\mathrm{O}(4)$ scaling function with the Wetterich equation were achieved in~\cite{Berges:1997eu}.
In a study using the functional PTRG for an $\mathrm{O}(N)$-model \cite{Bohr:2000gp}, critical exponents for different values of $N$ were obtained and the scaling behavior in the temperature dependence in the absence of explicit symmetry breaking was investigated.

In our own work, we have used the Wetterich equation to calculate the scaling functions for the O$(4)$ model in $d=3$ in a local potential approximation. This provided primarily a starting point and a reference for a scaling analysis in a finite volume. As a secondary goal, we also investigated the scaling behavior for large symmetry-breaking fields and determine the violations of scaling behavior on scales that are presumably relevant for the analysis of QCD lattice results. 

As mentioned above, a first study of the $\mathrm{O}(4)$ scaling behavior in an $N_\mathrm{f}=2$ quark-meson model with the non-perturbative RG was done in~\cite{Berges:1997eu}.
Also in~\cite{Stokic:2009uv} a scaling analysis of the critical behavior of an $N_\mathrm{f}=2$ quark-meson model was performed with the aid of the functional RG. The emphasis of these studies was on the attempt to observe the scaling behavior, and deviations were not investigated. 

In~\cite{Braun:2010vd}, a scaling analysis for the quark-meson model, in infinite as well as finite volume, has been performed. 
The goal was to advance to realistic pion masses and to map out the scaling region and investigate scaling deviations as they might occur in lattice simulation data. 

The scaling analysis for a theory in the O(4) universality class in $d=3$ dimensions is here also performed for a model in $d=4$ Euclidean dimensions, where the temperature is given by the extent of the volume in the Euclidean time direction. Consequently, scaling in the $d=3$ universality class will only be observed when the system satisfies the conditions for an effective dimensional reduction. The 
question of the applicability of dimensional reduction to the chiral phase transition was also raised in~\cite{Berges:1997eu}.

This is very similar to the situation that one actually encounters in the scaling analysis of lattice QCD results.  If one assumes a chiral transition temperature of $T_\mathrm{c} \approx 150 $ MeV and a pion mass of $m_\pi \approx 140$ MeV, it is \emph{a priori} not at all obvious that there are no critical fluctuations propagating in the Euclidean time direction. 

We find both in the O(4)-model study and the quark-meson model study that pion masses have to be fairly small in order to observe scaling behavior without any large corrections to scaling. While it is difficult to compare universal results on an absolute scale between different physical systems, the scales in the quark-meson model are similar enough to those in QCD to allow some conclusions. In particular, the transition temperature with $T_\mathrm{c} = 150$~MeV is similar to the one found on the lattice \cite{Aoki:2006br,Cheng:2006qk,Bazavov:2009zn,Aoki:2009sc,Borsanyi:2010bp,Bazavov:2011nk}. 

\begin{figure}
\begin{center}
\hspace*{-5mm}\includegraphics[scale=0.85, clip=true]{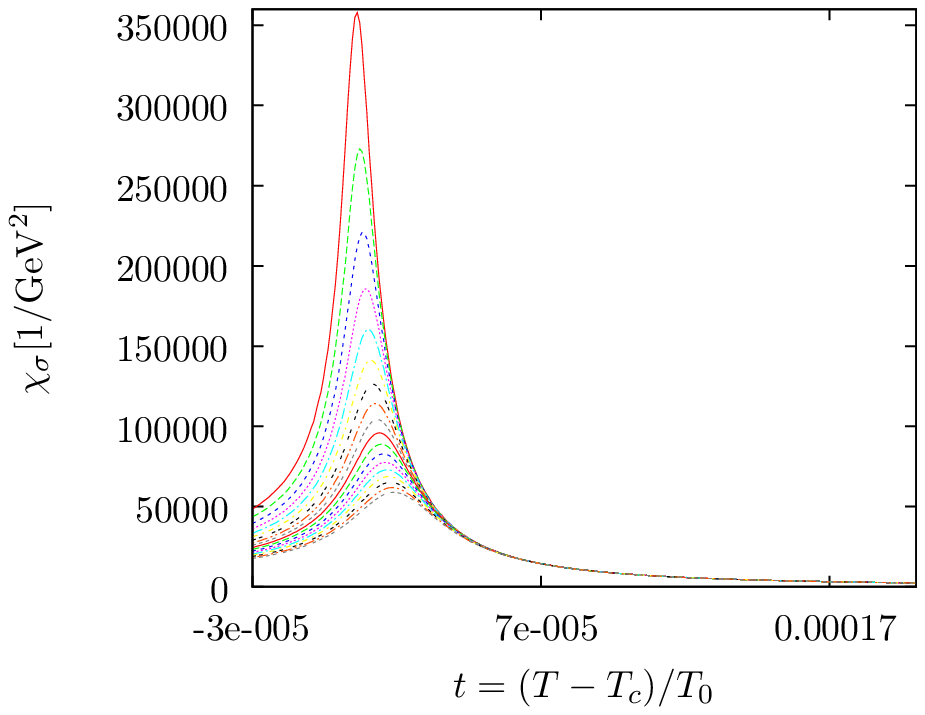}\\
\includegraphics[scale=0.8, clip=true]{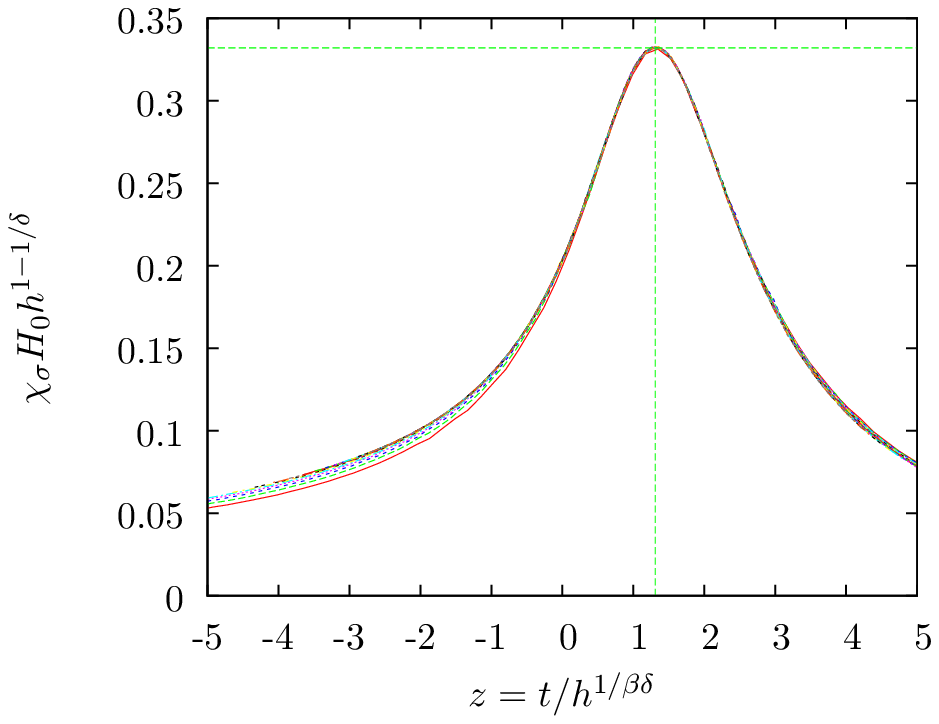}
\end{center}
\caption{Chiral (longitudinal) susceptibility $\chi_\sigma$ from the quark-meson model in the pion mass range $M_\pi \in [0.2, 0.9]$ MeV~\cite{Braun:2010vd}. The critical temperature in the model is  
$T_\mathrm{c} = 145$ MeV. Shown is the susceptibility $\chi_\sigma$ as a function of the reduced temperature $t$ (first panel), and the rescaled susceptibility $H_0 h^{1-1/\delta} \chi_\sigma$ as a function of the scaling variable $z=t/h^{1/(\beta\delta)}$. Scaling deviations are small only for these pion masses which are small compared to the other scales in the model. Figures from J.~Braun \emph{et al.}, Eur. Phys. J. C71, 1576 (2011), Copyright (2011), reprinted with kind permission from Springer Science and Business Media.}
\label{fig:qmmsuscsmall}
\end{figure}

\begin{figure}
\begin{center}
\includegraphics[scale=0.8, clip=true]{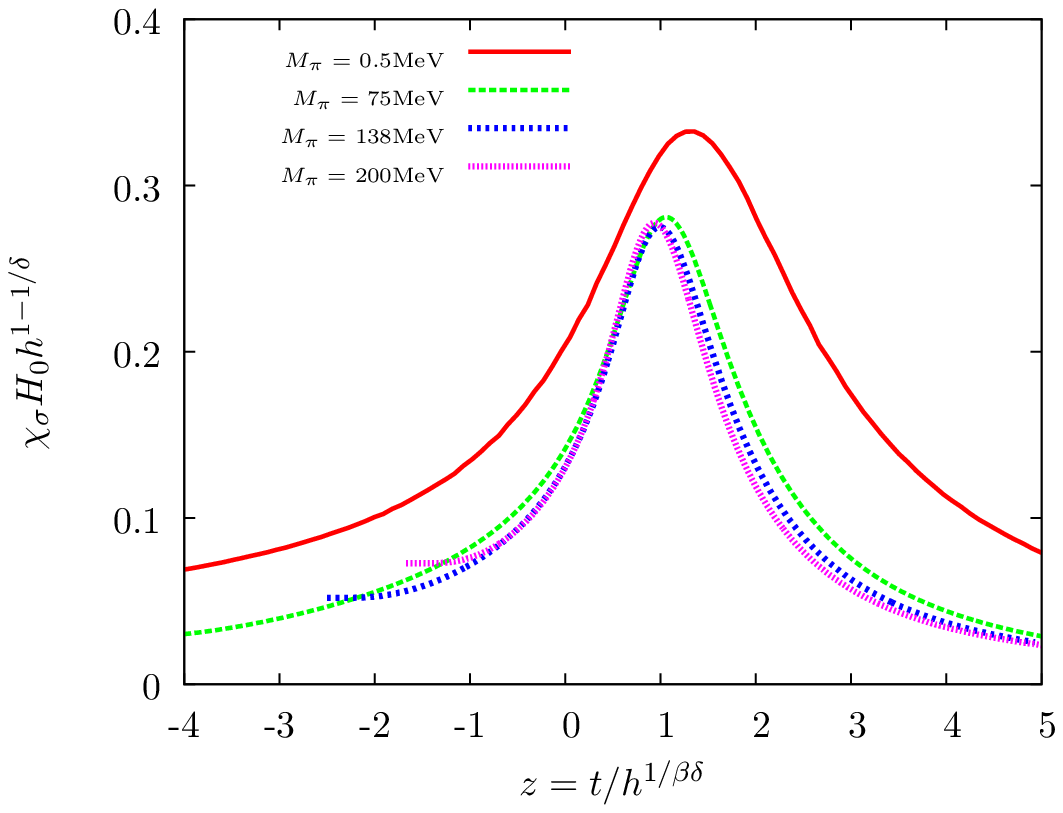}
\end{center}
\caption{Rescaled longitudinal susceptibility $H_0 h^{1-1/\delta} \chi_\sigma$ as a function of the scaling variable $z=t/h^{1/(\beta\delta)}$~\cite{Braun:2010vd}. Scaling deviations become large for pion masses that are realistically large and of the same size as those in current lattice simulations. Figure from J.~Braun \emph{et al.}, Eur.Phys.J. C71, 1576 (2011), Copyright (2011), reprinted with kind permission from Springer Science and Business Media.}
\label{fig:qmmsusclarge}
\end{figure}

In Fig.~\ref{fig:qmmsuscsmall}, the scaling behavior of the longitudinal susceptibility from the quark-meson model is shown for pion masses in the range of $0.2\;\mathrm{MeV}\; \le M_\pi \le 0.9$ MeV, i.e. considerably smaller than the one accessible in current lattice simulations. In this case, the scaling behavior is almost perfect and deviations are negligible.
In contrast, in Fig.~\ref{fig:qmmsusclarge} the results for pion masses of $M_\pi=75$ MeV and $M_\pi = 200$ MeV are shown, which are used in current simulations \cite{Ejiri:2009ac,Bazavov:2012jaa}. In this case, scaling violations are rather large, and the rescaled results differ significantly from the actual scaling functions. Note that there is an apparent scaling behavior if only results in a small interval of pion masses are taken into account. 

Many of the results for scaling functions obtained with different methods have been used in the analysis of the scaling behavior of lattice QCD data and have proven to be very useful. While the emphasis in earlier calculations was on the determination of the then-unknown scaling functions and the observation of scaling behavior in the first place, our goal has been to investigate the conditions under which these results can be reasonably compared to those of lattice QCD simulations. Since the quark masses and therefore the amount of chiral symmetry breaking was large in early lattice simulation studies of scaling at the phase transition~\cite{Iwasaki:1996ya,Aoki:1998wg}, it appears likely that the effect of large quark masses led to significant corrections to the scaling behavior. 
Remarkably, while of course also affected by large quark masses, an analysis of finite-size scaling behavior might have been easier in the relatively small lattice volumes of early calculations, but was not attempted. For more recent scaling studies \cite{Ejiri:2009ac}, where pion masses are even decreased below their physical values, an analysis of finite-volume effects will be ever more relevant.

\subsection{Finite-Size Scaling}
Any study of phase transition behavior in a system by means of a numerical simulation in a finite volume is confronted with the problem that, strictly speaking, phase transitions can only take place in infinite-volume systems: Only in the thermodynamic limit can the partition function develop non-analytic behavior. From a theoretical point of view, this is a pre-requisite of the discontinuous or singular behavior in thermodynamic observables which characterizes a phase transition. 
Finite-size effects are not solely a complication for finite-volume simulations of a physical system, but can actually serve as an additional analytical tool for the analysis of critical phenomena.
They can be used in a systematic fashion to extract additional information from a finite-volume system close to criticality. Putting a system into a finite volume always introduces an additional relevant variable into the system: Only in the limit $L \to \infty$ of infinitely large volume size $L$ does the system actually reach criticality. For any finite value of $L$, the system is necessarily away from the critical point. 

This can be also understood easily by considering the correlation length associated with the order in the system. The correlation length obviously cannot exceed the system size, and this imposes an absolute limit on its magnitude. In particular, it can never diverge and the finite-size system cannot reach the critical point. The correlation length can only become infinite again in the thermodynamic limit.

This consideration leads one directly to Fisher's finite-size scaling hypothesis \cite{Fisher:1971ks}: It ought to be possible to describe finite-size effects in the vicinity of the critical point entirely in terms of the correlation length and the system size. According to Fisher's hypothesis, the finite-size scaling should in fact only depend on the ratio of these length scales.  

For example, applying Fisher's scaling hypothesis to the order parameter $M$, one expects that the ratio between its values in finite and infinite volume is given by a function which depends only on the ratio of the infinite-volume correlation length $\xi(t, h, L \to \infty)$ to the volume size $L$:
\be
\frac{M(t, h, L)}{M(t, h, L \to \infty)} &=& \mathcal{F}\left( \frac{\xi(t, h, L \to \infty)}{L}\right).
\ee
From this hypothesis, taken together with the known power-law scaling behavior of thermodynamic quantities from infinite volume, one can already deduce the form of finite-volume scaling functions.
Since the infinite-volume correlation length scales as $\xi \sim t^{-\nu}$ with the temperature in the absence of the field $h$, in order to keep the ratio $\xi/L$ constant under a length rescaling of $L$ requires to vary $t$ in such a way to keep in turn $tL^{1/\nu}$ constant. If we also want to make contact with infinite-volume scaling, we need to keep the scaling variable $z=t/h^{1/(\beta\delta)}$ constant, and hence need to keep $ h L^{\beta \delta/\nu}$ constant. This already suggests which variables will be useful for a finite-size scaling analysis.  

More generally, the volume size $L$ must be treated as an additional relevant coupling at a critical point, since only in the limit $L \to \infty$ the critical point can be attained. Including $L$ in the scaling relation \eqref{eq:fscaling}, the singular part of the free energy of the system behaves as
\be
f_s(t, h, L)&=& \ell^{-d} f_s(t\ell^{y_t}, h \ell^{y_h}, L \ell^{-1}). 
\ee
The rescaling factor $\ell$ can now be chosen to keep one argument constant, so that the free energy density becomes a function of only two scaling variables. In order to obtain the infinite-volume scaling behavior for $L \to \infty$, it is advantageous to choose the scaling variable $z$ as one of these two variables. The leading-order behavior for the order parameter $M$ can then be expressed as
\be
M(t, h, L) &=& L^{-\beta/\nu} Q_M(z, h L^{\beta\delta/\nu}),
\ee
i.e. in terms of a finite-size scaling function and the new scaling variable $h L^{\beta\delta/\nu}$.

Asymptotically for $L \to \infty$, we expect to recover the infinite-volume scaling behavior from the finite-size scaling function. Therefore
\be
\lim_{x \to \infty}Q_M(z, x) = x^{1/\delta}  f_M(z)  \quad \mathrm{with} \quad x =h L^{\beta\delta/\nu} .
\ee  

A similar relation can be derived for the longitudinal susceptibility. In leading order, one finds
\be
\chi_\sigma(t, h, L) &=& L^{\gamma/\nu} Q_\chi(z, h L^{\beta\delta/\nu}).
\ee
For very small volumes or very small symmetry breaking, the finite-size scaling function for the susceptibility becomes almost constant as a function of the scaling variable $h L^{\beta\delta/\nu}$. Since $\gamma/\nu = (2 -\eta)$ because of the scaling relations between the critical exponents and $\eta \approx 0$ for O(N) models, 
one finds the canonical finite-volume scaling behavior $\chi_\sigma \sim L^2$ for the susceptibility, as expected from basic dimensional arguments. 

Since the system size $L$ is a dimensionful length scale, it cannot be a universal quantity. The absolute length scale will be sensitive to the microscopic details of the theory or model considered. For this reason, the finite-size scaling functions are not universal as long as they still depend on  a dimensionful $L$. We obtain truly universal results only if we introduce a non-universal normalization factor $L_0$, in analogy to $H_0$ and $T_0$, and parametrize results in terms of the dimensionless size variable $l = L/L_0$. 

A possible choice for the normalization constant $L_0$ is provided by the correlation length, which behaves for $h =0$ and $L \to \infty$ as 
\be
\xi(t) &=& {L_0} t^{-\nu} \\
M_\sigma(t)&=& \frac{1}{L_0} t^\nu.
\ee
Then the ratio
\be
\frac{L}{\xi(t, L \to \infty)} &=& M_\sigma(t) L = \frac{L}{L_0} t^\nu = l t^\nu
\label{eq:lengthnormal}
\ee
becomes a truly scale-independent variable which can be compared directly between different systems. 
We note that the necessity to normalize the length scale to a universal, dimensionless quantity has not always been observed in comparisons of e.g. spin model and lattice QCD results. However, this is well recognized: In the lattice investigation of $\mathrm{O}(2)$ scaling behavior for QCD with staggered fermions~\cite{Kogut:2006gt}, for example, careful attention has been paid to compare only results with the same dimensionless ratio $\xi/L$ for spin systems and QCD.

\begin{figure}
\begin{center}
\includegraphics[scale=0.8, clip=true]{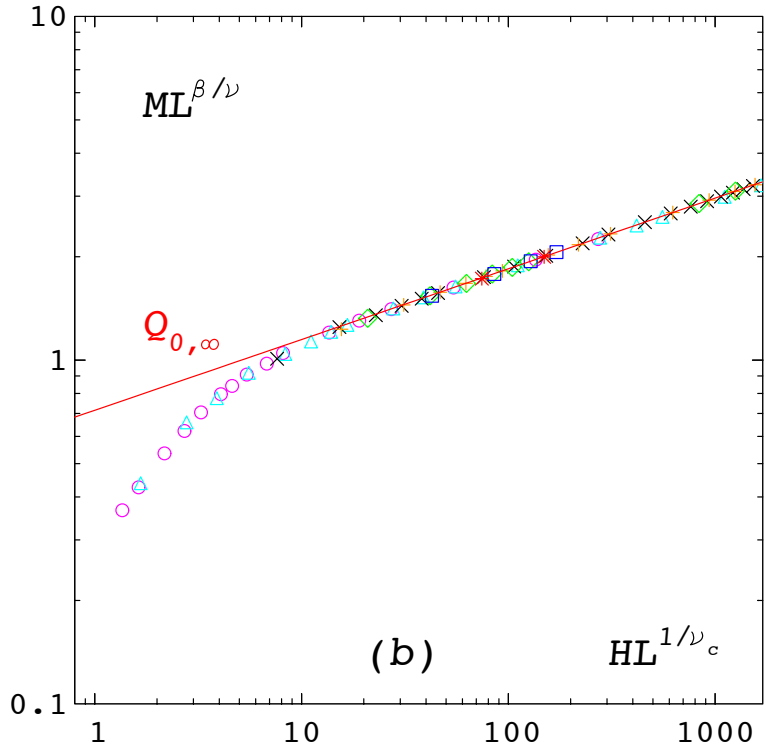}
\end{center}
\caption{Finite-size scaled data for the order parameter $M$ from an O(4) spin model lattice simulation~\cite{Engels:2001bq}. The simulation was performed at the critical temperature $T=T_\mathrm{c}$, i.e. at $t=0$, $z=0$. The line indicates the asymptotic behavior $Q_{0, \infty}\sim (HL^{1/\nu_\mathrm{c}})^{1/\delta}$ of the finite-size scaling function $Q_M(z, hL^{1/\nu_\mathrm{c}})$ for $z=0$, $L \to \infty$, where  $\nu_\mathrm{c} = \nu/(\beta \delta)$. Figure reprinted from J. Engels \emph{et al.}, Phys. Lett. B514, 299 (2001), Copyright (2001), with permission from Elsevier.}
\label{fig:mendeso4fss}
\end{figure}

Pioneering work on the effects of Goldstone modes and the scaling behavior of different observables in a finite volume for O(2) and O(4) spin models was done in \cite{Engels:1999wf,Engels:2000xw,Engels:2001bq}.
The finite-size scaling functions from O(2) and O(4) spin model lattice simulations in $d=3$ were first determined in~\cite{Engels:2001bq} by Engels {\it et al.}, where also a first attempt at a finite-size scaling analysis of QCD lattice simulation data was made. The results for the scaling function from this simulation at the critical temperature $T=T_\mathrm{c}$ ($z=0$) are shown in Fig.~\ref{fig:mendeso4fss}.
For large values of the symmetry-breaking parameter $H$ and short correlation length, the behavior of the system approaches the infinite-volume behavior $H^{1/\delta}$ asymptotically, which is reflected in the behavior of the finite-size scaling function. For small symmetry breaking parameter, the correlation length increases and finite-size effects become apparent in the deviation of the results from the asymptotic behavior. This is the region which is most interesting for studying finite-size scaling effects. 
Unfortunately the original investigation in~\cite{Engels:2001bq} only provided relatively few data points in this region. 

In a recent update of this work by Engels and Karsch~\cite{Engels:2014bra}, which extends the infinite-volume investigation~\cite{Engels:2011km} to finite volumes, the authors calculate finite-size scaling functions for the order parameter and the susceptibility for the O(4) spin model in $d=3$ from high precision lattice simulations. Polynomial parameterizations of the finite-size scaling functions are given, and the normalization conditions for the length scale are discussed. The analysis is extended over a much larger range of $z$-values, compared to the first investigation in~\cite{Engels:2001bq}. A new parameterization for the finite-size scaling functions, which allows to easily make direct contact with the scaling functions in the thermodynamic (infinite-volume) limit, is also introduced. The results will likely be extremely valuable for the finite-size scaling analysis of lattice QCD data.  

in~\cite{Braun:2008sg}, the finite-size scaling behavior in the $\mathrm{O}(4)$ model with functional Renormalization Group methods has been investigated, using both the Wetterich equation and the PTRG cutoff scheme. A significant advantage of the RG approach above a model simulation is the wide range of possible volume sizes and values of the symmetry-breaking parameter (and hence the wide range of masses for the Goldstone modes). This makes it possible to explore the scaling region in detail, and to investigate both asymptotic behavior and deviations from or corrections to the scaling behavior.
The investigation was restricted to an expansion of the effective action for the $\mathrm{O}(4)$ model in purely local couplings. This means that the anomalous dimension vanishes, $\eta = 0$, and there is a small, systematic error in the values for the critical exponents. 
It was possible to obtain results for the scaling functions for the order parameter and the longitudinal susceptibility over a wide range of volumes and values of the symmetry-breaking field. The corrections to the scaling behavior due to large symmetry breaking agree with the predictions of a general RG analysis of the scaling behavior.   

\begin{figure}
\vspace*{-10mm}
\begin{center}
\includegraphics[scale=0.35, clip=true]{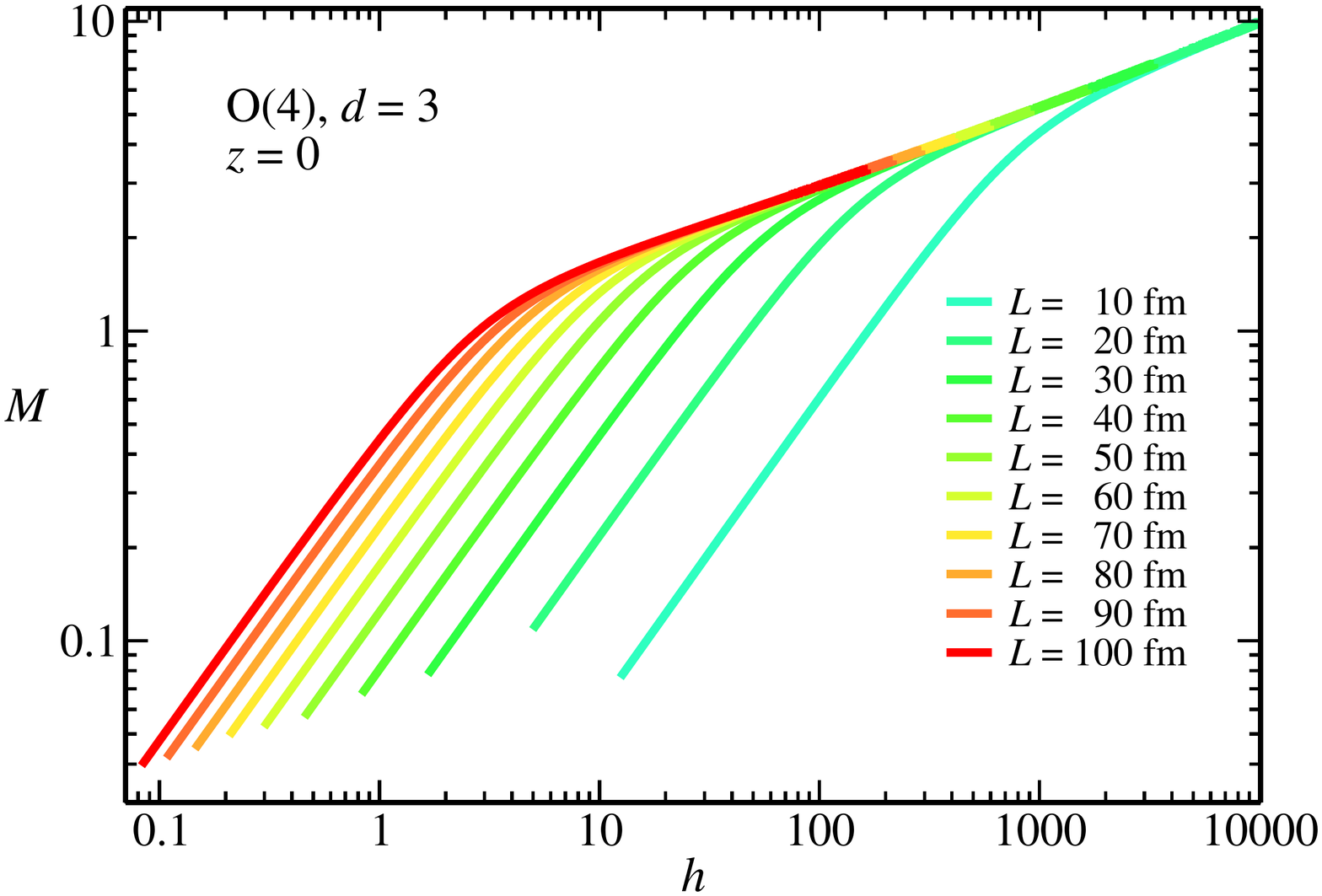}\\
\includegraphics[scale=0.36, clip=true]{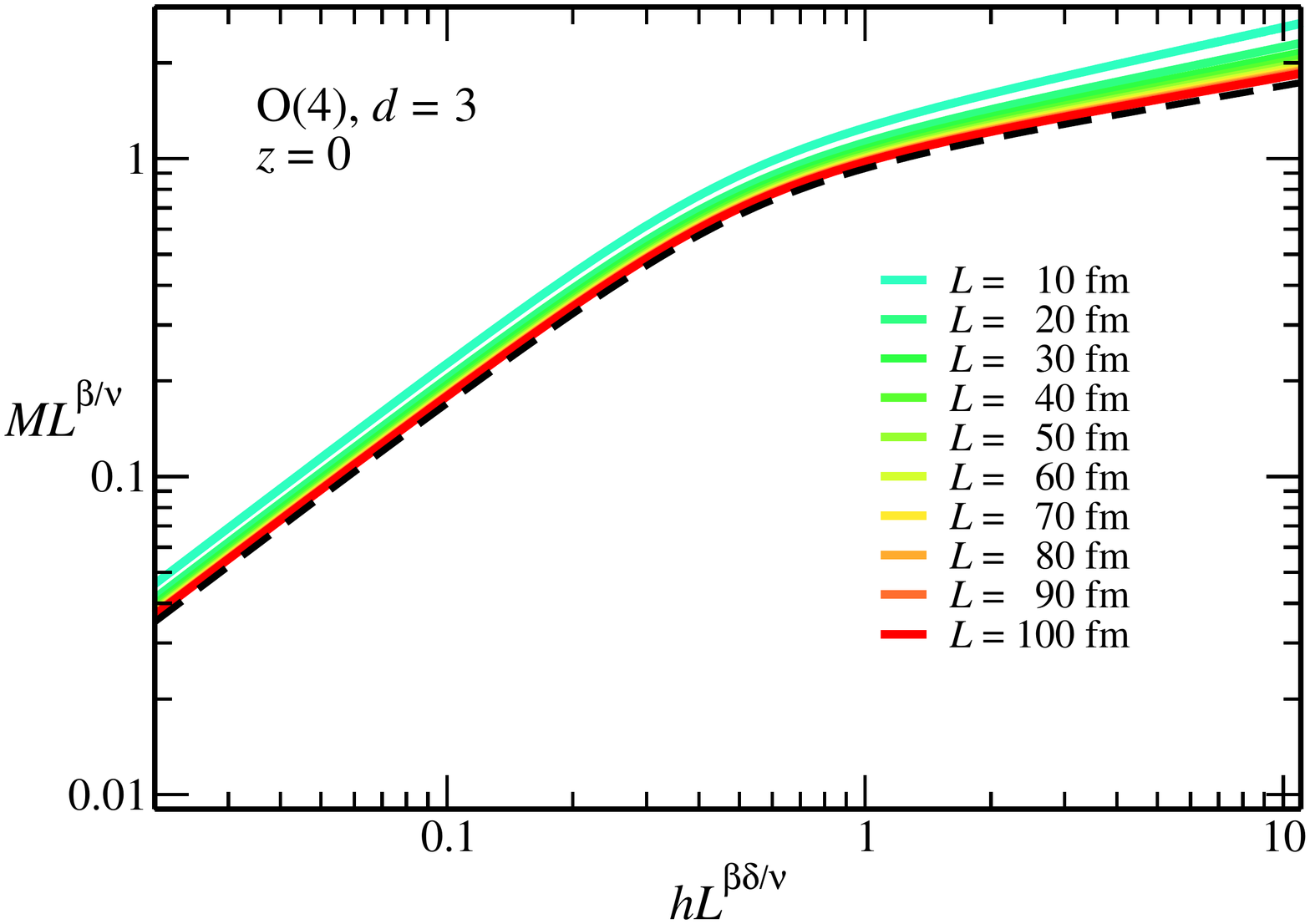}
\vspace*{-10mm}
\end{center}
\caption{Finite-size scaling behavior of the order parameter $M$ at the critical temperature, $z=0$, for the $\mathrm{O}(4)$ model in $d=3$ from \cite{Braun:2008sg}.
The first panel shows the order parameter $M$ as a function of $h$ for different volume sizes,  and the second panel the finite-size scaled order parameter $L^{\beta/\nu} M$ as a function of  $h L^{\beta \delta/\nu}$.
For large values of $h$, where the correlation length is small, the unscaled results for different volume sizes all converge towards the same infinite-volume limit (first panel).  The black dashed line is the scaling function after accounting for scaling corrections (second panel). Figure from J. Braun and B. Klein, Eur. Phys. J. C63, 443 (2009), Copyright (2009),  reprinted with kind permission from Springer Science and Business Media.}
\label{fig:o4-fss}
\end{figure}

Exemplary results for the scaling function for the order parameter $M=\langle \sigma \rangle$ are shown in Fig.~\ref{fig:o4-fss} for the system at the critical temperature ($T=T_\mathrm{c}$ or $z=0$ for the scaling variable). From the unscaled results for $M$ as a function of the symmetry-breaking field $h$, one observes that the curves for different volume size $L$ deviate from the asymptotic infinite-volume result at different values of $h$: The symmetry-breaking parameter controls the correlation length, large values of $h$ corresponding to large fluctuation masses and short correlation lengths. Hence the deviations from infinite-volume behavior appear at small values of $h$ for large volume size $L$, where the correlation length $\xi$ is also large. The rescaled results for $ML^{\beta/\nu}$ vs. $hL^{\beta \delta/\nu}$ collapse nicely onto the scaling function. The expected deviations appear for large values of $h$. 
Parameterizations are provided for the scaling functions for order parameter and longitudinal susceptibility for a wide range of values of the infinite-volume scaling variable $z$.

For a direct comparison with QCD lattice simulation data, a determination of the length renormalization constant as in Eq.~\eqref{eq:lengthnormal} for the lattice results is necessary, which in turn requires a measurement of the correlation length on the lattice. Even though this is difficult to do, using the pion mass as an estimate, these results suggest that current lattice simulations are in the asymptotic scaling region and finite-size scaling effects are small. This is consistent with the observations made in~\cite{Engels:2001bq} for older lattice simulation results.

\begin{figure}
\begin{center}
\includegraphics[scale=0.5, clip=true]{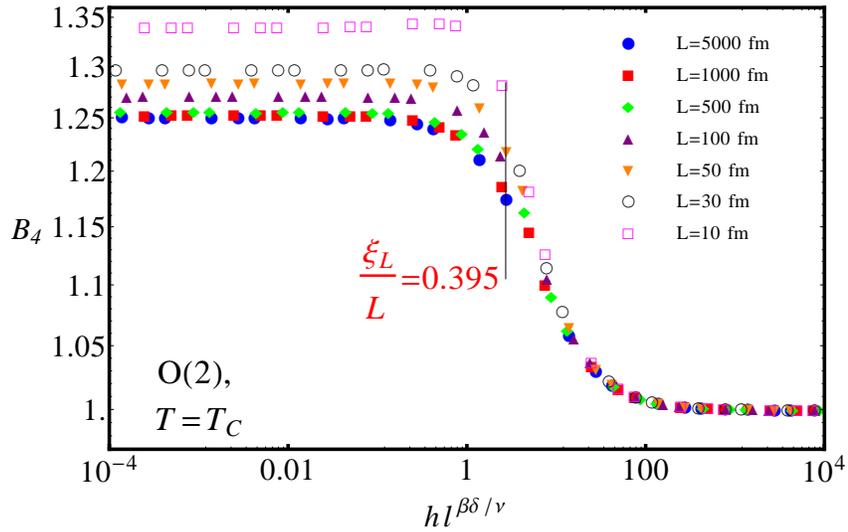}
\end{center}
\caption{$4^{\mathrm{th}}$ order Binder cumulant $B_4(T_c)$ for the $\mathrm{O}(2)$ model in $d=3$ at the critical temperature as a function of the symmetry-breaking field $h$, rescaled with the volume, from~\cite{Springer:2015kxa}. The universal value of $\xi_L/L$ at the onset of the finite-size scaling region for the order parameter is indicated in the plot. For small symmetry-breaking field $h$, the correlation length is large and the system exhibits finite-size scaling behavior. For large values of $h$, the correlation length is much smaller than the system extent, the system is in the ordered phase and $B_4 \to 1$. In the finite-size scaling region, only the results $B_4(T_c, L)$ for large volumes, $L>500~\mathrm{fm}$, approach the universal value $B_4(T_c)=1.2491(39)$. Non-universal finite-volume corrections are significant for the smaller volume sizes, $L<100~\mathrm{fm}$.}
\label{fig:BinderO2}
\end{figure}

\begin{figure}
\begin{center}
\includegraphics[scale=0.5, clip=true]{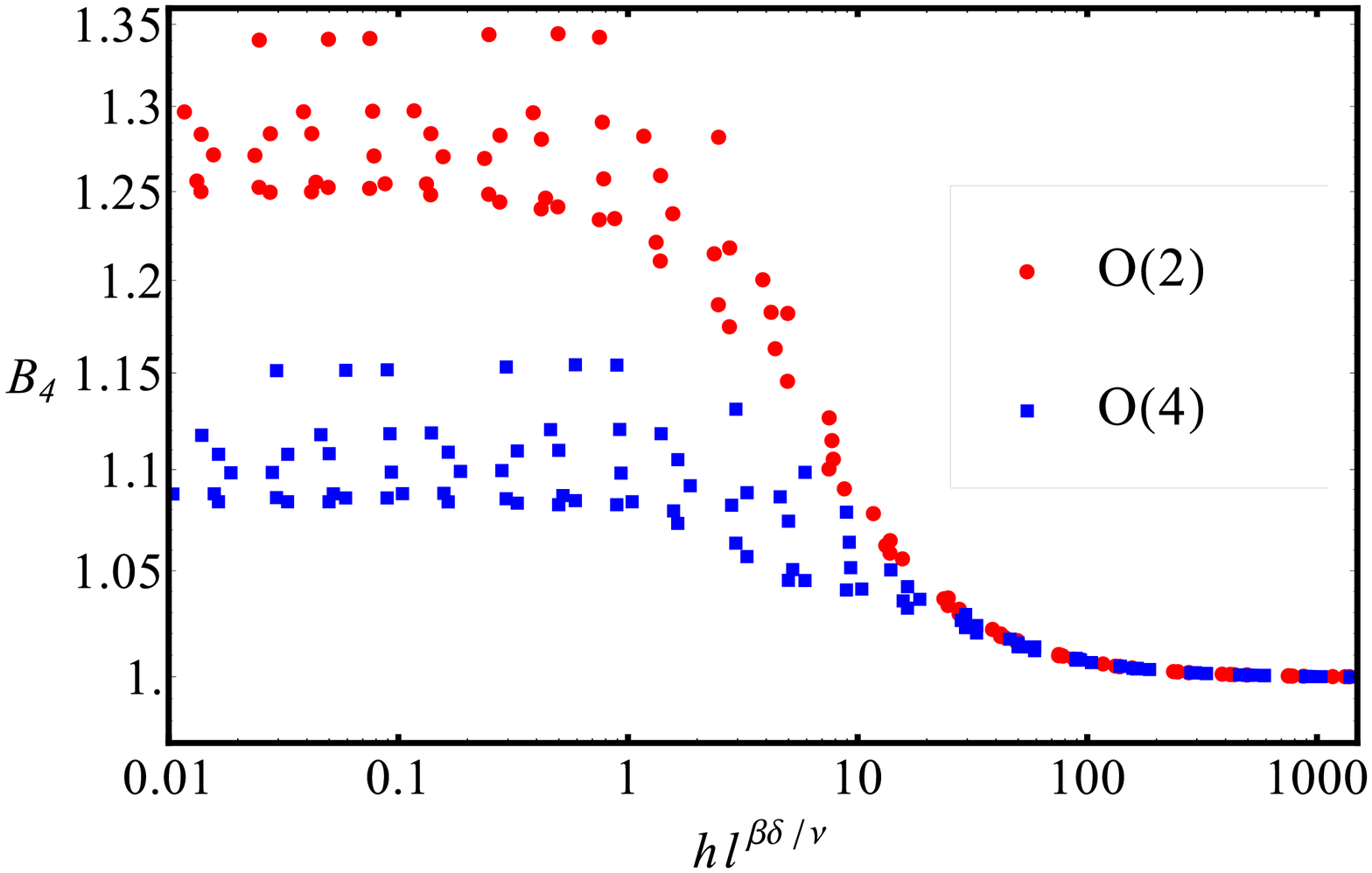}
\end{center}
\caption{$4^{\mathrm{th}}$ order Binder cumulant $B_4(T_c)$ for both the $\mathrm{O}(2)$ model (\emph{red dots}) and $\mathrm{O}(4)$ model (\emph{blue squares}) in $d=3$ at the critical temperature as a function of the symmetry-breaking field $h$ rescaled with the volume from~\cite{Springer:2015kxa}.  The points in this plot correspond to volumes from $L=10~\mathrm{fm}$ (largest values of $B_4$) to $L=5000~\mathrm{fm}$ (smallest values of $B_4$). The central observation is that, \emph{even in the presence of} \emph{significant non-universal finite-size corrections}, it is possible to distinguish between $O(2)$ and $O(4)$ critical behavior by means of the Binder cumulant $B_4$.}
\label{fig:BinderO2O4comparison}
\end{figure}

An extension of  these investigations of the O(4) model to the O(2) model in $d=3$ is provided in~\cite{Springer:2015kxa}. In addition to a comparison of the finite-size scaling functions and scaling regions for these models, for the first time the functional RG method has been used to calculate $4^{\mathrm{th}}$ order Binder cumulants~\cite{Binder:1981sa} in this work.
The Binder cumulant is a correlation measure which exhibits at the critical point in the infinite-volume limit a value specific to a particular universality class, and it is therefore very useful for the determination of a critical point and its universality class~\cite{Karsch:2001nf,deForcrand:2003hx,Cucchieri:2002hu}. It is in and of itself already a finite-size scaling function~\cite{Cucchieri:2002hu}
\be
B_4(T, L) &=& Q_B(tL^{1/\nu}, L^{-\omega}, \ldots) = a_0 + a_1 tL^{1/\nu}+ a_2 L^{-\omega}+ \ldots, 
\label{eq:B4expansion}
\ee
which reduces to the universal value $a_0$ at the critical point ($t=0$, $L\to \infty$). Non-universal finite-size scaling corrections proportional to $L^{-\omega}$ ($\omega>0$) vanish in this limit. Different normalizations for the Binder cumulant are sometimes used, we choose
\be
B_4&=& \frac{\langle(\phi^2)^2\rangle}{\langle\phi^2\rangle^2}, 
\ee 
where $\phi^{\mathrm{T}} = (\sigma, \pi_i)$ and $i$ runs over the total number of Goldstone modes. The Binder cumulant is essentially a measure of non-trivial fourth order correlations, since it compares all contributions of possible fourth-order fluctuations to the contributions from the trivial Gaussian part. 
In the ordered phase ($T\ll T_c$), fluctuations are suppressed by powers of $1/V$, and in leading order
\be
\langle(\phi^2)^2\rangle \to M^4 \quad \mathrm{and} \quad \langle\phi^2\rangle \to M^2,
\ee
 such that 
 \be
 B_4(T\ll T_c) \to 1.
 \ee
 On the other hand, for $T\gg T_c$, $B_4$ is dominated by fluctuation effects, and its value must depend strongly on the number of degrees of freedom in the system. For $\mathrm{O}(N)$ models, analytic calculations~\cite{Springer:2015kxa} indicate that
 \be
 B_4(T\gg T_c) \to \frac{N+2}{N},
 \ee
 which is in good agreement with the numerical results. 
The values found in the numerical calculations for the universal values of the Binder cumulant at the critical temperature $T_c$ and in the infinite-volume limit are
 \be
 B_4(\mathrm{O}(2)) &=& 1.2491(39)\nn\\
 B_4(\mathrm{O}(4)) &=&  1.0836(10), 
 \ee
 which is within $1\%$ 
 of the values obtained in lattice spin model simulations for the $\mathrm{O}(2)$ model~\cite{Cucchieri:2002hu} and the  $\mathrm{O}(4)$ model~\cite{Kanaya:1994qe}. 
These results are again obtained over a very large range of volume sizes and values for the symmetry-breaking field $H$. They show that despite the presence of large non-universal finite-volume corrections in the finite-size scaling region, the Binder cumulant can still be used successfully to distinguish $N=2$ from $N=4$ in $\mathrm{O}(N)$ scaling behavior. 
 However, as a second result, one finds that the size of the finite-size scaling region in both models is very similar. Expressed in terms of the longitudinal correlation length, $\xi_L =1 /m_\sigma$, we find
 \be
 \xi_L(\mathrm{O}(2))/L &=& 0.395(5)\nn\\
  \xi_L(\mathrm{O}(4))/L &=& 0.372(2).
 \ee
 By using $m_\pi < m_\sigma$ as a bound, we obtain the more useful estimates
 \be
 (m_\pi L)(\mathrm{O}(2)) &=& 2.01(1)\nn\\
  (m_\pi L)(\mathrm{O}(4)) &=& 2.12(2)
 \ee
 for the size of the finite-size scaling regions in the respective models. These values are likely too similar to use them in any way as a distinction of the universality classes. However, they provide a useful estimate for the size of the region in which finite-size scaling effects can be reasonably expected to occur.
Figs.~\ref{fig:BinderO2} and~\ref{fig:BinderO2O4comparison} show the results for the Binder cumulant at the critical temperature ($t=0$) as a function of the symmetry-breaking field $H$, plotted against the finite-size scaling variable $hl^{\beta\delta/\nu}$. As usual, $h=H/H_0$ and $l=L/L_0$ are normalized according to the conventions \eqref{eq:Hnormal} and \eqref{eq:lengthnormal}. 
Dimensional scales are set by the parameter choices at the UV scale of the model. In Fig.~\ref{fig:BinderO2}, we observe a nice collapse of the data onto a single scaling curve when plotting $B_4$ against the scaling variable $hl^{\beta\delta/\nu}$ and a convergence of the $B_4$ value for small symmetry-breaking field $h$ with increasing volume size against the expected value at the critical temperature. 
From Fig.~\ref{fig:BinderO2O4comparison}, we learn that even for small volumes sizes, with sizeable finite-size corrections to the scaling behavior, it is still possible to distinguish between the $\mathrm{O}(2)$ and $\mathrm{O}(4)$ universality classes: The groups of points from different volume sizes are clearly distinct for the  $\mathrm{O}(2)$ and $\mathrm{O}(4)$ models. In particular, we see that the large-volume limit is always approached from above (implying a positive coefficient $a_2$ in Eqn. \eqref{eq:B4expansion}). If this holds in general also for other systems, this would yield a useful criterion to exclude scaling behavior: For example, for $t=0$ and in small volumes $m_\pi L \lesssim 2$, a value $B_4(t=0, L) < 1.25$ would clearly exclude the possibility of $\mathrm{O}(2)$ scaling behavior.

\begin{figure}
\begin{center}
\hspace*{-4mm}\includegraphics[scale=0.82,=true]{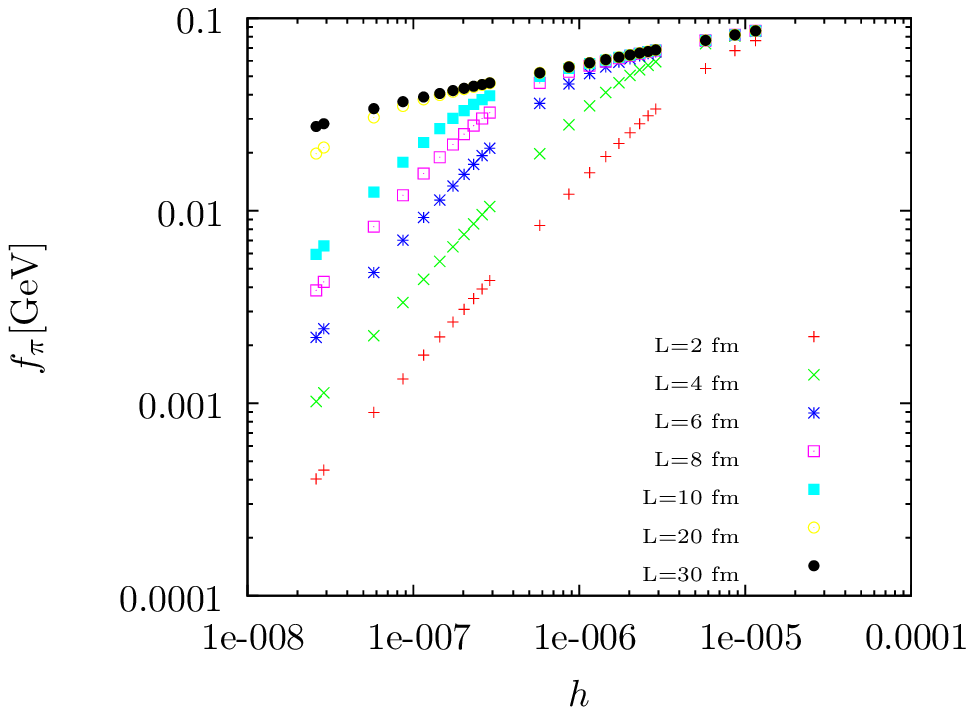}\\
\includegraphics[scale=0.7, clip=true]{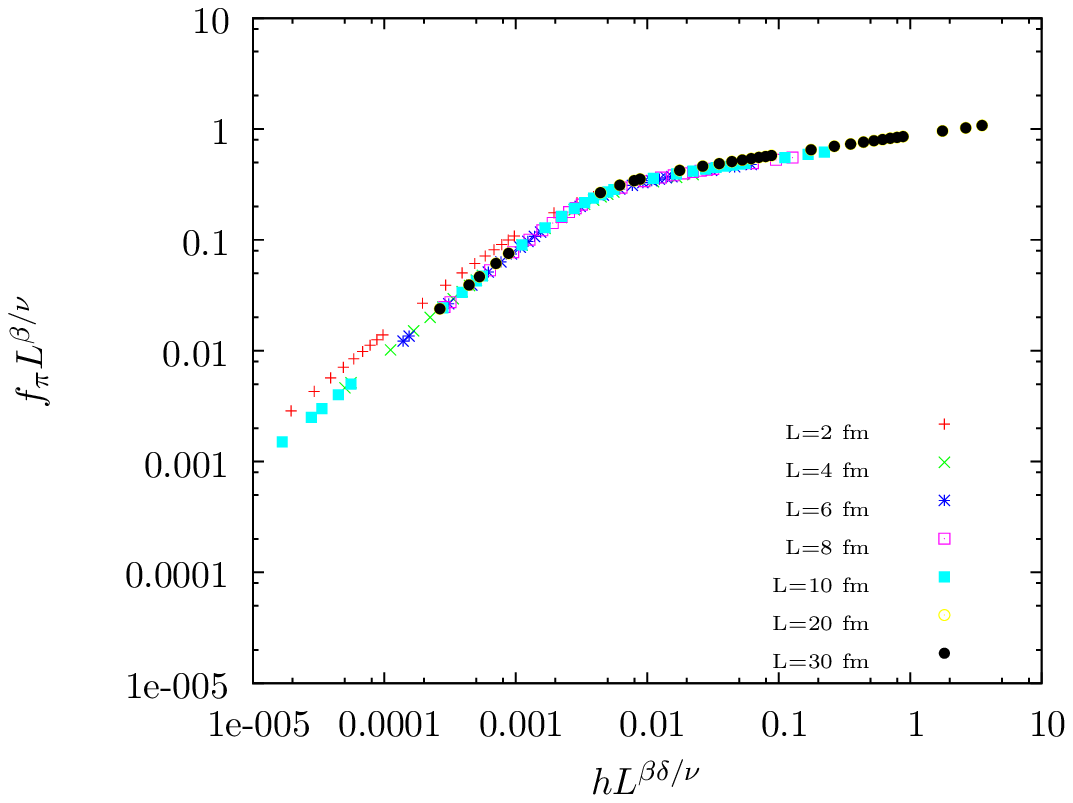}
\end{center}
\caption{Finite-size scaling behavior of the order parameter $M=f_\pi$ at the critical temperature, $z=0$, for a quark-meson model in $d=3+1$ dimensions at finite temperature~\cite{Braun:2010vd}. The scaling behavior conforms to the one in $d=3$ dimensions, with the appropriate values for the critical exponents. The theory is effectively reduced to $d=3$. The first panel shows the order parameter $M=f_\pi$ as a function of $h$ for different volume sizes. The second panel shows the finite-size rescaled order parameter $L^{\beta/\nu} M$ as a function of $h L^{\beta \delta/\nu}$. Figures from J.~Braun \emph{et al.}, Eur. Phys. J. C71, 1576 (2011), Copyright (2011), reprinted with kind permission from Springer Science and Business Media.}
\label{fig:qmmfss}
\end{figure}

In order to make more direct contact with QCD lattice simulation results, we have repeated a finite-size scaling analysis for the quark-meson model~\cite{Braun:2010vd}, where we adjusted parameters and mass scales to match the physical values. 
This makes it possible to address several questions that are of relevance for the interpretation of lattice simulations. The scaling region for a finite-size scaling analysis can be estimated from the results.
The pion mass in the model can be easily adjusted from the physical value to the chiral limit of exactly massless pions. The volume size can be adjusted over a wide range of values as well. 
Fig.~\ref{fig:qmmfss} shows the order parameter $M= f_\pi$ as a function of the symmetry-breaking field $h$, as for the $\mathrm{O}(4)$ model in Fig.~\ref{fig:o4-fss}. The rescaled data $L^{\beta/\nu}M$ as a function of $hL^{\beta\delta/\nu}$ collapse onto the scaling function. The result most relevant for practitioners is the size of the region where finite-size scaling effects dominate. In Fig.~\ref{fig:qmmfss}, this region is limited by the point where the slope of the curve changes and deviates from the asymptotic infinite-volume behavior. The dimensionless variable $m_\pi L$ at the bend point takes values in the range from $m_\pi L = 2 - 3$. For a volume size of $L= 4 $~fm, the value of $h$ at the bend point corresponds to a vacuum pion mass of $m_\pi = 139$ MeV and $m_\pi L = 2.82$. Values for several other volume sizes are given in~\cite{Braun:2010vd}. Overall one can conclude that the finite-size scaling region is currently not explored in QCD lattice simulation runs with $m_\pi L \approx 4 - 5$, and consequently a finite-size scaling analysis cannot be brought to bear on these results. Other finite-volume effects are however present in such calculations, as will be discussed below. 

\subsection{Scaling analysis of lattice data}
\label{sec:scalinglattice}
Lattice QCD simulation results are necessarily obtained in a finite simulation volume. This makes it very difficult to observe critical (second-order) phase transition behavior. Actually divergent quantities can only appear in the thermodynamic limit which requires infinite volume size $V \to \infty$. 
Due to the specific challenges of implementing fermions on a discretized lattice, chiral symmetry is always broken and quark masses in the simulations are finite. 
This makes a scaling analysis absolutely necessary for a classification of the observed phase transition behavior. 
Once the pion masses become smaller in improved simulations and approach the chiral limit more closely, in any given volume the effects of finite-volume effects become again more acute: Smaller pion masses lead to a longer range of fluctuations and stronger influence of the finite volume size. 
On the one hand, this requires to account for finite-volume effects in the (assumed) infinite-volume scaling behavior. On the other hand, this opens up the opportunity to perform a finite-size scaling analysis which can provide additional information about the scaling behavior.

The famous renormalization-group based calculation by Pisarski and Wil\-czek~\cite{Pisarski:1983ms}  from a perturbative treatment of matrix models shapes our expectations of the phase transition behavior: If the phase transition is dominated by the restoration of chiral symmetry, for $N_\mathrm{f}=2$ massless flavors, the transition should be of second order and fall into the $\mathrm{O}(4)$ universality class. For $N_\mathrm{f}=3$ massless flavors, fluctuations should drive the transition to be first order~\cite{Pisarski:1983ms}. A decreasing instanton density at high temperature can effectively restore the $U_{\mathrm{A}}(1)$ symmetry broken by the anomaly, which would lead to an additional change in the transition order, if the axial symmetry is effectively restored before the chiral flavor symmetry. 

Scaling arguments have been applied in the analysis of QCD phase transitions for any number of quark flavors and values of the quark mass. 
A very successful application of finite-size scaling analysis to lattice QCD was achieved in the case of $N_\mathrm{f}=3$ quarks with degenerate masses.
For $N_\mathrm{f}=3$ massless or very light quarks, the transition was indeed found to be of first order. With increasing quark mass, the line of first-order transitions terminates in a critical end point at a quark mass $m_\mathrm{c}$.
This critical end point in the Ising universality class  $\mathrm{Z}(2)$ was unambiguously identified in $N_\mathrm{f}=3$ lattice simulations~\cite{Karsch:2001nf,deForcrand:2003hx,deForcrand:2006pv}. 

For the chiral phase transition with $N_\mathrm{f}=2$ light quarks, following the theoretical arguments in~\cite{Pisarski:1983ms}, a second-order phase transition in the O(4) universality class for $d=3$ is expected. 
An early analysis \cite{Iwasaki:1996ya} with $N_\mathrm{f}=2$ Wilson fermions found behavior that was consistent with the expected $\mathrm{O}(4)$ scaling behavior, although volumes were very small ($8^3\times 4$) and pion masses very large. The results were obtained mainly for temperatures above the transition, and the behavior was compatible both with mean-field and $\mathrm{O}(4)$ scaling behavior, which was slightly favored.

Additional scaling analysis results from lattice QCD for $N_\mathrm{f}=2$ flavors were obtained by the JLQCD~\cite{Aoki:1998wg}, MILC~\cite{Bernard:1999fv} and CP-PACS~\cite{AliKhan:2000iz} collaborations.  
For Wilson fermions, apparent $\mathrm{O}(4)$ scaling behavior was observed by the CP-PACS collaboration~\cite{AliKhan:2000iz}, but only for temperatures \emph{above} the phase transition temperature, where there is very little dependence on the symmetry group, since there are no Goldstone modes.
The JLQCD study \cite{Aoki:1998wg} for two flavors with staggered quarks found significant finite-size effects on small lattices and effectively ruled out a first-order phase transition. Scaling behavior was observed, but the coefficients were inconsistent with either  $\mathrm{O}(4)$ or  $\mathrm{O}(2)$ scaling behavior, which is expected as a possibilities due to the symmetries of staggered fermions on the lattice.
For staggered fermions, the MILC collaboration did not find  $\mathrm{O}(4)$ scaling behavior~\cite{Bernard:1999fv}, also above the transition temperature.

The absence of the expected scaling behavior or any other (first-order) scaling behavior in the transition region was a long-standing problem, and in fact even the CP-PACS results were somewhat unsatisfactory, since they did not extend to a region where Goldstone mode effects would be observable and the critical fluctuation behavior associated with the transition would undoubtably be present. 

Another early finite-size scaling analysis of (then still unpublished) lattice simulation results for $N_\mathrm{f}=2$ from the JLQCD collaboration \cite{Aoki:1998wg} and the Bielefeld group \cite{Laermann:1998gf} was performed in \cite{Engels:2001bq}. The finite-size scaling behavior obtained from spin model simulations was compared to the lattice simulation results and an appropriate finite-size scaling analysis was performed. As far as any conclusion could be drawn in this study, the lattice results for the peak position of the susceptibility conformed to the asymptotic large-volume behavior of the finite-size scaling functions. Interpreting these results in the light of our investigations, the data seem to fall into a range with only small finite-volume effects, where infinite-volume scaling is expected, and not properly into the finite-size scaling region. This suggests that pion masses are too large to lead to finite-size scaling effects at the given lattice volumes for these data sets. 

Kogut and Sinclair have simulated lattice QCD with two flavors of staggered quarks and a chirally modified action to investigate the expected $\mathrm{O}(2)$ scaling behavior by comparing to an $\mathrm{O}(2)$ spin model \cite{Kogut:2006gt}. This modified staggered action allow simulations in the chiral limit. The comparison between QCD results and $\mathrm{O}(2)$ spin model results is done carefully in a way that matches results with the same ratio $\xi/L$. They identify a misleading, apparent tri-critical scaling for very small volume size which might lead to false conclusion in the analysis of lattice QCD data.  At the same time, they find a finite-size scaling behavior consistent with the $\mathrm{O}(2)$ universality class. Although not conclusive, the study clearly establishes that the finite-volume effects for critical scaling are not negligible.

\begin{figure}
\includegraphics[scale=0.5]{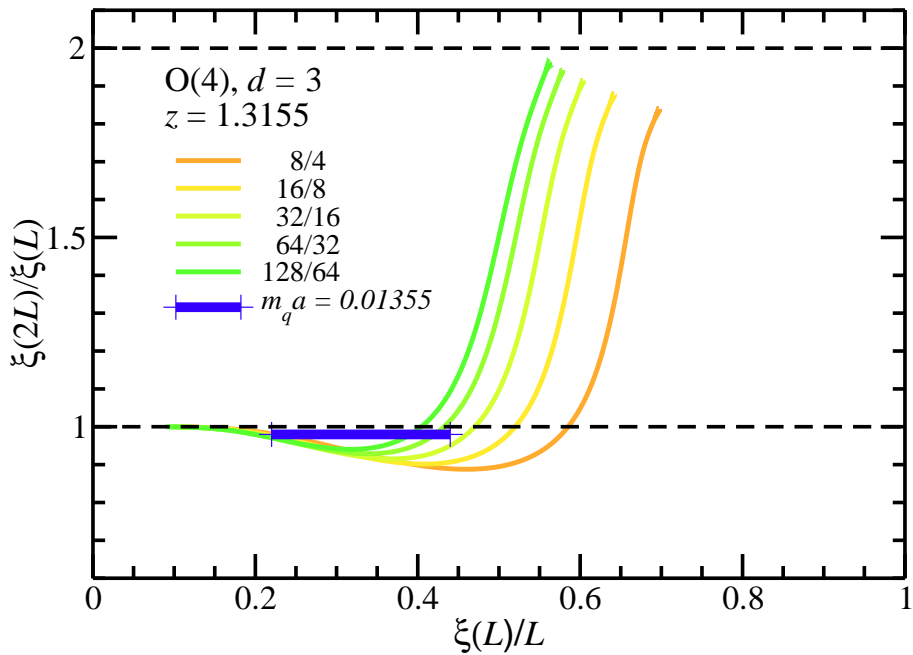}
\caption{Ratios of the correlation lengths $\xi(2L) / \xi(L)$ in volumes of linear extent $L$ and $2L$ as a function of volume size $L$. The volume range marked blue in the plot is the one estimated by us for simulations with the lightest pion mass undertaken in~\cite{D'Elia:2005bv}. $m_q a$ is the quark mass in lattice units. Our results imply that finite-size scaling effects are still very small in this region and likely not observable in the lattice simulation data of~\cite{D'Elia:2005bv,Cossu:2007mn}.}
\label{fig:giacomocomparison}
\end{figure}
An extremely careful analysis of finite-size scaling effects for QCD with two quark flavors ($N_f=2$) was more recently carried out in~\cite{D'Elia:2005bv,Cossu:2007mn}. In these papers, the authors put forward the hypothesis that the phase transition observed in two-flavor QCD might be of first order, and not of second order in the $O(4)$ universality class, as the arguments from fluctuations compatible with the chiral flavor symmetry imply \cite{Pisarski:1983ms}. The authors argue that the phase transition in QCD might be dominated by the confinement--deconfinement phase transition, instead of chiral effects, and therefore be of first order. 
They find some evidence in support of this thesis, but cannot establish any clear scaling behavior, neither in the quark mass dependence nor in the finite-size dependence, which would allow definitive conclusions.  
The question raised by these results were an important inspiration for the RG investigations into the finite-size scaling behavior of the $O(4)$ universality class.  

A significant problem in the comparison of finite-size scaling functions to simulation data is the determination of a length scale normalization. A practical way around this problem is to consider only dimensionless ratios of the same quantity for different volume sizes at a fixed ratio. This idea is widely used in condensed matter applications, where scaling behavior in terms of a ratio of quantities at volumes of size $L$ and $2L$ is considered~\cite{Caracciolo:1994ed,Cucchieri:1995yd,Caracciolo:2003nq}. 
Results for the ratio of correlation lengths $\xi(2L)/\xi(L)$ as a function of the dimensionless ratio $\xi(L)/L$ for an $\mathrm{O}(4)$ model in $d=3$ from~\cite{Braun:2008sg} are shown in Fig.~\ref{fig:giacomocomparison}. For increasing values of $L$, the ratios for pairs of volumes $L$ and $2L$ approach a universal scaling curve. One finds that $\xi(L)/L$ approaches some maximal value, and hence the ratio $\xi(2L)/\xi(L)$ converges towards the value $2$. For large volumes, $\xi(L) \to \xi(\infty)$ and the ratio approaches $1$. 
In the plot, the situation at the peak of the longitudinal susceptibility is considered. The scaling variable $z=z_\mathrm{p}$ is chosen such that for any value of the symmetry breaking parameter, the temperature corresponds to the one at the peak of the susceptibility. 
We can identify the region in $\xi(L)/L$ for which finite-size scaling effects become significant.

In order to compare to the results from the lattice QCD simulation~\cite{D'Elia:2005bv}, it is still necessary to estimate the correlation length, for which no direct measurement is available. For a rough estimate, we can use the correlation length from the lightest fluctuations as an upper bound: Since the pion is the lightest particle, no other fluctuation can lead to correlations with a longer range, and the pion mass can provide an estimate of the lower bound $1/(m_\pi(L) L)$ for the ratio $\xi/L$. 
We have used the results from~\cite{D'Elia:2005bv} to estimate this bound, and the result for $1/(m_\pi(L) L)$for the smallest pion mass and the smallest quark mass $m_\mathrm{q}a =0.01355$ (in lattice units) is indicated by the horizontal bar in Fig.~\ref{fig:giacomocomparison}. One can conclude that the regime in quark mass and volume size considered in the lattice analysis likely does not exhibit significant finite-volume effects or falls into the finite-size scaling region. For the given pion mass, this would actually require a \emph{smaller} volume.   

An important result of this investigation is the observation that it requires very small pion masses, compared to the hadronic UV scale of the problem, to obtain convergent results for the scaling functions. 
We first observed this in \cite{Braun:2007td}, where an initial UV scale on the order of $\sim 1$ GeV was used for integrating out critical fluctuations in an $O(4)$-model. As is customary for universal quantities, the subsequent analysis was performed in terms of dimensionless parameters which assure comparability between different systems in the infrared regime. 

In the investigation in \cite{Braun:2010vd} of a quark-meson model in $d=3+1$ Euclidean space-time dimensions, we also find that the pion masses required to observe scaling behavior are extremely small. Deviations from the scaling behavior are only negligible if one decreases the pion masses by at least one order of magnitude from its physical value.

An extensive scaling study of the Brookhaven-Bielefeld group of the magnetic equation of state~\cite{Ejiri:2009ac} for $2+1$ quark flavors with staggered fermions and with pion masses down to about $75$ MeV appears to validate $\mathrm{O}(N)$ scaling behavior. $\mathrm{O}(4)$ and $\mathrm{O}(2)$ scaling behavior expected for staggered fermions cannot be distinguished in this analysis. For large quark masses (and hence pion masses) in these simulations, one observes significant deviations of the results from the scaling functions. 
These effects due to large quark masses and the observed scaling deviations depend on the choices (which data points are included in the analysis) for setting the scales $T_0$ and $H_0$  and determining the critical temperature $T_c$ \cite{Springer:2011bt}.

\begin{figure}
\begin{center}
\includegraphics[scale=0.8]{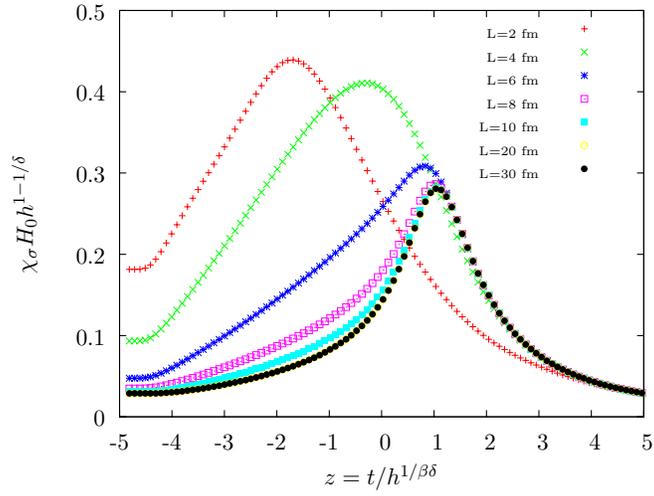}
\end{center}
\caption{Rescaled longitudinal susceptibility as a function of the scaling variable $z$ for $M_\pi = 75$~MeV for different volume sizes from the quark-meson model RG calculation in \cite{Braun:2010vd}. For smaller pion masses, the finite-volume effects become significantly larger. Significant deviations from the infinite-volume behavior appear for $L\le 6$~fm. Figure from J.~Braun \emph{et al.}, Eur. Phys. J. C71, 1576 (2011), Copyright (2011), reprinted with kind permission from Springer Science and Business Media.}
\label{fig:inffvscaling}
\end{figure}

The results for finite-volume effects in~\cite{Braun:2010vd} for the quark-meson model are also relevant for the analysis of these lattice results. It is discussed what happens when we assume infinite-volume scaling behavior in a finite volume and the effects of deviations from the expected scaling behavior. 
Results for these effects for a pion mass of $75$ MeV (as in~\cite{Ejiri:2009ac}) can be seen in Fig.~\ref{fig:inffvscaling}.
For a quark-meson model we find that the scaled susceptibility in the finite volume is actually \emph{larger} than that in infinite volume, and significant effects appear already for volumes as large as $L=6 - 8 $~fm, which is comparable to the volume sizes in~\cite{Ejiri:2009ac}.  
A recent paper of the HotQCD collaboration~\cite{Bazavov:2012jaa} invokes the results from the calculation~\cite{Braun:2010vd} as a possible explanation of differences for the chiral susceptibility between lattice results from different simulation runs with different fermion discretizations and different simulation volumes in terms of finite-volume effects.

\begin{figure}
\includegraphics[scale=0.5]{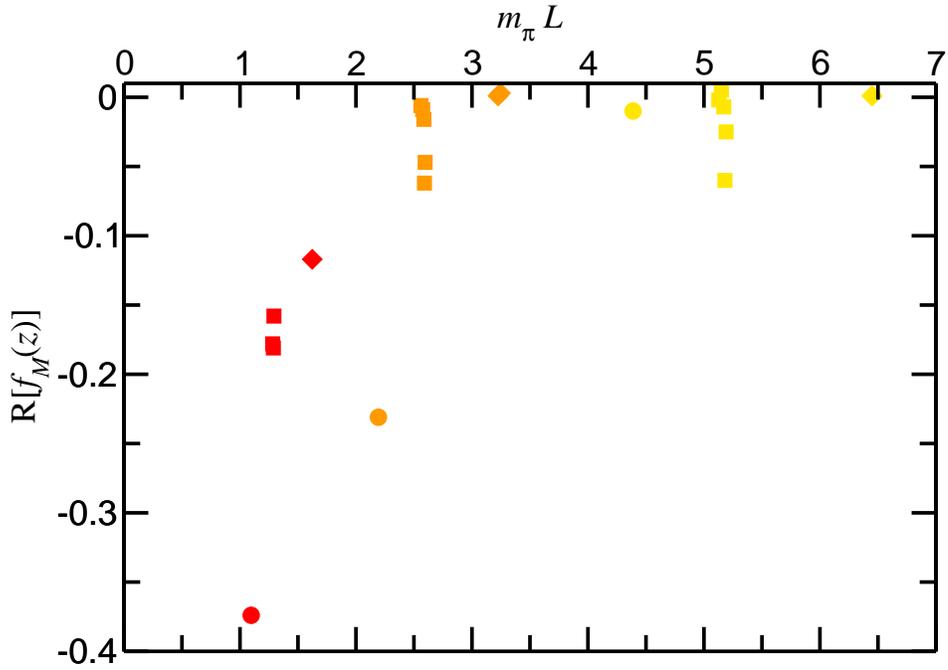}
\caption{Analysis of lattice QCD data from \cite{Ejiri:2009ac} for the deviation of the scaling behavior of $M(t, h, L)$ from the scaling function $f_M(z)$ in a finite volume $R[f_M(z)]  = \Big[M(t, h, L) h^{-1/\delta} - f_M(z)\Big]/f_M(z)$  performed in \cite{Probst:2011bt}. The lattice simulation is done for $N_\mathrm{f}=2+1$ flavors for $H=m_u/m_s \in \{1/80, 1/40, 1/20\}$ (circles, squares, diamonds) and lattice volumes $L/a\in \{8, 16, 32\}$ (red, orange, yellow). While the results do not fall on a neat curve, the scaling deviations clearly decrease as a function of $m_\pi L$, which is expected if they are due to finite-volume effects.
\label{fig:Rfz-mpil}}
\end{figure}

With some reasonable inferences about the pion masses in the simulation~\cite{Ejiri:2009ac}, making use of explicit results from~\cite{Cheng:2007jq}, it is possible to analyze the deviations from the infinite-volume scaling behavior of the order parameter $M$ in a finite volume~\cite{Probst:2011bt}.
The deviations in the rescaled order parameter 
\be
R[f_M(z, L)]  = \frac{M(t, h, L) h^{-1/\delta} - f_M(z)}{f_M(z)}
\ee
from the infinite-volume scaling function $f_M(z)$ decrease as a function of $m_\pi L$~\cite{Probst:2011bt} (see Fig.~\ref{fig:Rfz-mpil}). In Fig.~\ref{fig:Rfz-mpil}, the results for this quantity are shown. While they do not fall onto a neat curve, the monotonous decrease of the deviations as a function of $m_\pi L$ demonstrate clearly that these deviations are indeed due to finite-volume effects. 

In conclusion, these results on finite-size scaling and the finite-volume effects on the scaling behavior at the chiral phase transition contribute valuable context to the analysis of lattice simulation results: They provide guidance on the regime in pion mass and volume size in which finite-size scaling can be observed. In relation to QCD lattice simulations such as \cite{D'Elia:2005bv, Cossu:2007mn}, they indicate that for the realized pion masses \emph{smaller} simulation volumes are required to see finite-size scaling effects. With regard to the scaling analysis of simulation data under the assumption of infinite-volume scaling behavior, they indicate that scaling violations should still be significant at current pion masses, and that finite volume effects might affect in particular susceptibilities significantly.  
With further progress of lattice QCD simulations and ever decreasing pion masses, finite-volume effects will on the one hand become larger again, but also allow for a finite-size scaling analysis. This brings a powerful additional method to bear on the analysis.

\section{Phenomenological effects at finite density}

\subsection{Curvature of the phase transition line and position of critical end point}
\label{sec:curvature}
The phase diagram of QCD in the plane of finite baryon chemical potential and temperature has been studied extensively with many methods. 
Lattice QCD simulations, which are an extremely important benchmark method, have also been used to explore the phase diagram, although this application presents a particular problem: the complex-valued fermion determinant in the presence of a finite quark chemical potential makes a direct application of Monte-Carlo algorithms difficult. This so-called \emph{sign problem} can be overcome by different calculation techniques: The probabilities at finite chemical potential $\mu$ can be evaluated by using reweighting~\cite{Fodor:2001au,Fodor:2001pe}, the partition function can be expanded in a Taylor series around $\mu=0$~\cite{Allton:2002zi,Schmidt:2002uk,Allton:2003vx,Allton:2005gk}~\cite{Gavai:2003mf,Gavai:2004sd,Gavai:2008zr}, or it can be evaluated for imaginary chemical potential, with results continued to real values of $\mu$~\cite{deForcrand:2002ci,deForcrand:2002yi,deForcrand:2003bz}.
Reviews of these lattice methods can be found e.g. in~\cite{Laermann:2003cv,Philipsen:2005mj,Schmidt:2006us,Philipsen:2008gf,deForcrand:2010ys}.

For small chemical potential, the phase transition line extending in the $\mu$-$T$--plane can be characterized in terms of a Taylor expansion around $\mu=0$. Because the QCD partition function is symmetric under $\mu \to -\mu$, the first non-trivial term in the expansion is quadratic $\sim \mu^2$, and the curvature $\kappa$ appears as the coefficient.
The definition of the curvature $\kappa$ from the expansion of the chiral phase transition line $T_{\mathrm{c}}(\mu)$ as a function of the chemical potential is given by
\be
T_\mathrm{c}(\mu) &=& T_\mathrm{c} \left [1+ \kappa \left(\frac{\mu}{\pi T_\mathrm{c}}   \right)^2 + \ldots \right]
\ee
where $T_\mathrm{c} = T_\mathrm{c}(0)$. Note that different conventions are used for including factors of $\pi$ in the  expansion, which in turn changes the value of $\kappa$. We use the convention from \cite{deForcrand:2002ci,deForcrand:2006pv}. We also speak loosely of a transition temperature $T_\mathrm{c}$ at finite $m_\pi$, and later finite $L$, where only a smooth crossover exists and provides only a pseudo-transition temperature.

Lattice simulation results for the curvature have been obtained using an imaginary chemical potential \cite{deForcrand:2002ci,deForcrand:2006pv} from Taylor expansion and reweighting \cite{Karsch:2003va}, and from a scaling analysis of the magnetic equation of state \cite{Kaczmarek:2011zz}.
The Wuppertal group has  obtained results from a Taylor expansion method at the physical quark mass~\cite{Endrodi:2011gv}.
More recently, in new calculations based on the imaginary potential formalism~\cite{Bonati:2015bha, Bellwied:2015rza,Cea:2015cya}, several different groups have also attempted an extrapolation of the results for the curvature to the continuum limit.  
For comparison, results in the continuum in the chiral limit $m=0$ for QCD with $N_\mathrm{f}=1$ massless quark flavor have been obtained in~\cite{Braun:2008pi} in a functional RG calculation.

Results from the different lattice determinations of the curvature differ significantly. This is not necessarily surprising, since the different lattice simulations use different fermion discretizations, a different number of time slices in the Euclidean time direction, different quark masses, and different simulation volume sizes. It is clear that all of these  parameters will affect the phase transition, which has e.g. a well-investigated dependence on the quark mass~\cite{Karsch:2000zv,Bernard:2004je}.

\begin{figure}
\includegraphics[scale=0.95]{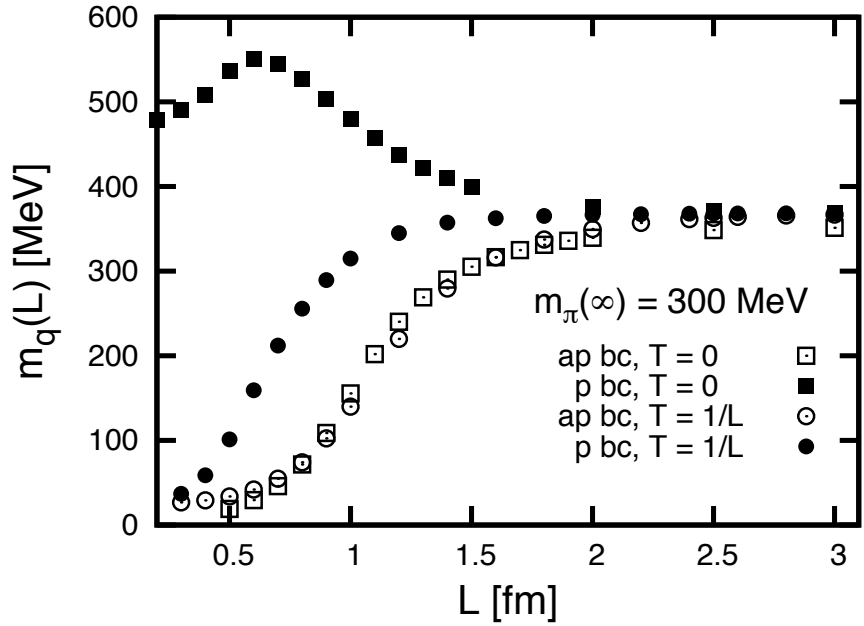}
\caption{Constituent quark mass $m_{\mathrm{q}}(L)$ from the quark-meson model in a finite volume $V=L^3\times 1/T$ as a function of the volume size $L$, for $T=1/L$ (circles) and $T=0$ (squares). (Open) solid symbols indicate results for (anti-)periodic boundary conditions. For periodic quark boundary conditions, fermionic zero-mode effects lead to an \emph{increase} in the constituent quark mass in intermediate volume size \cite{Braun:2005gy}. The system then becomes less sensitive to a change in the quark chemical potential $\mu$.}
\label{fig:constituentm}
\end{figure}

Specifically, it is an interesting question whether the phase transition line for finite $\mu$ and $T$ is sensitive to the simulation volume, and whether discrepancies between different lattice simulation runs can also be attributed to different simulation volumes. 
This is not as far-fetched as it might at first seem: Model calculations show that a finite volume affects the chiral condensate~\cite{Braun:2005gy} and hence the constituent quark masses, and a change in the quark mass leads to a change in the sensitivity of the system to a change in the chemical potential. As the chemical potential corresponds to the energy that it takes to add additional quarks to the system, it stands to reason that the phase transition temperature will be less affected by an increase of the chemical potential if the quark mass is increased.
In Fig.~\ref{fig:constituentm}, the volume dependence of the constituent quark mass in the quark-meson model in finite volume from the calculation in~\cite{Braun:2005gy}is shown, for both anti-periodic and periodic spatial boundary conditions for the quark fields. The latter are more interesting and also more relevant for lattice simulations, since periodic boundary conditions are more commonly used.
The results in the figure demonstrate that there is a small, intermediate-size volume region in which the constituent quark mass is \emph{larger} than in infinite volume, and hence a \emph{decreased} sensitivity of the model to a change in the chemical potential should be expected. This would translate into a \emph{decreased} curvature. 

\begin{figure}
\includegraphics[scale=0.50]{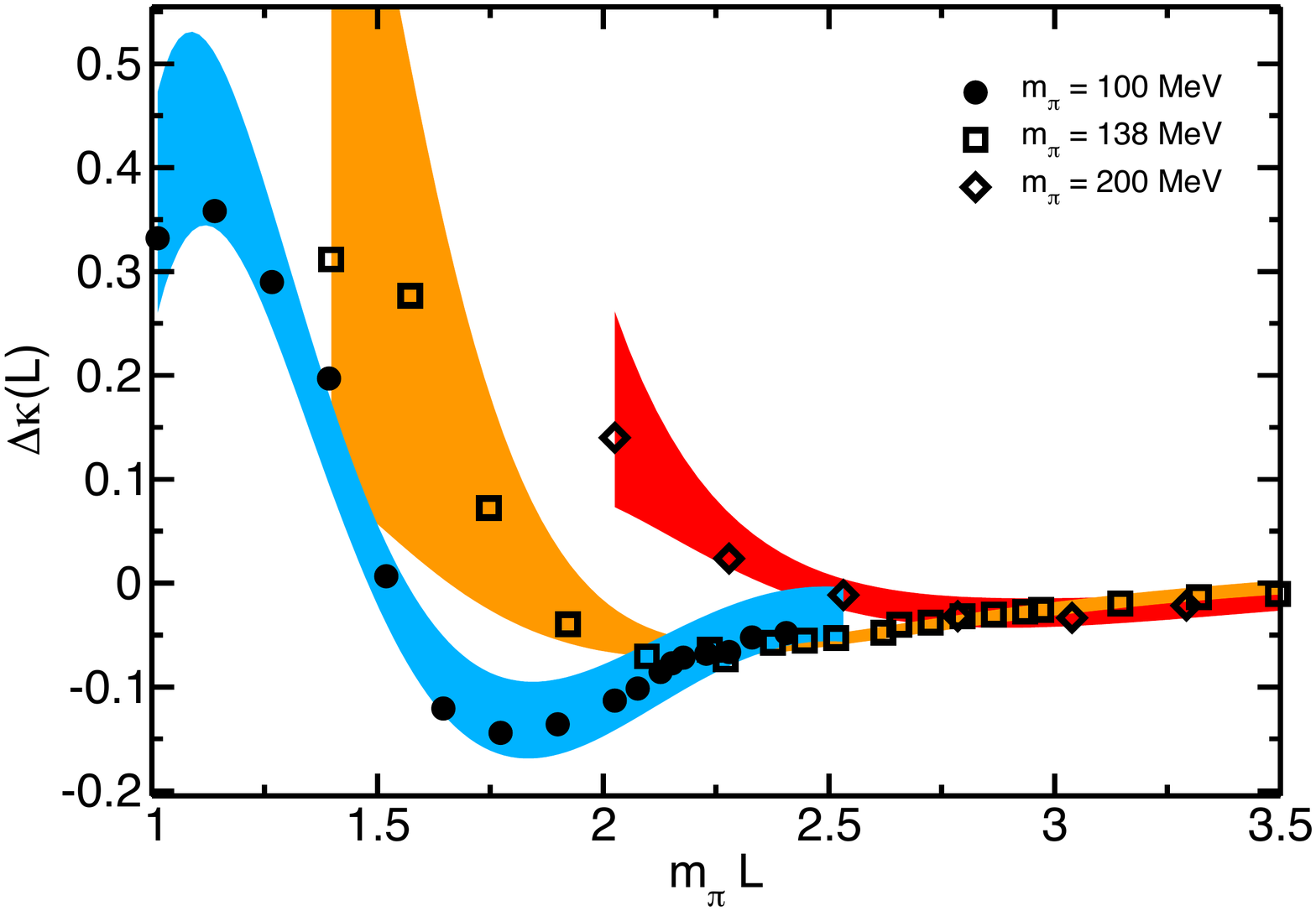}
\caption{Change of the curvature in a finite volume $\Delta\kappa(L)=({\kappa(L)-\kappa(\infty)})/
  {\kappa(\infty)}$ as a function of the dimensionless variable $m_\pi L$ from the quark-meson model \cite{Braun:2011iz}. Results are shown for pion masses of $100$ MeV, $138$ MeV and $200$ MeV. Error bands are estimated from different polynomial fits to the transition line $T_\mathrm{c}(\mu)$. Figure reprinted from J. Braun \emph{et al.}, Phys. Lett. B713, 216 (2012), Copyright (2012), with permission from Elsevier.}
  \label{fig:curvature}
\end{figure}

The effect of the volume on the curvature can be quantified by looking at the relative deviation
\be
\Delta\kappa(L) &=& \frac{\kappa(L) - \kappa(\infty)}{\kappa(\infty)},
\ee 
which will in turn depend on the pion mass. Results for this quantity from the study~\cite{Braun:2011iz} are shown in Fig.~\ref{fig:curvature} as a function of the dimensionless quantity $m_\pi L$. For the pion masses $m_\pi=100, 138$, and $200$ MeV, the largest effect can be seen for the smallest pion mass. In accordance with the arguments given above, the curvature \emph{decreases} for intermediate volume size, with a minimum at $m_\pi L = 1.84(1)$ approximately 15\% below the infinite-volume value for $m_\pi = 100 $ MeV. 
For very small volume size, chiral symmetry is approximately restored, constituent quark masses drop, and the curvature becomes very large. 
For pion masses $m_\pi \approx 100 - 200$ MeV, the effect on the curvature $\Delta\kappa(L)$ is strongest for $m_\pi L \approx 2 - 3$.
For anti-periodic boundary conditions, the curvature is also affected. In this case, the constituent quark mass always decreases with decreasing volume, and in general the effects on the transition temperature are even stronger~\cite{Braun:2005fj}. This case is of less importance for a comparison to lattice simulations.
As argued in~\cite{Braun:2011iz}, we expect that the effect changes little in the presence of gauge fields. This is supported by RG arguments for QCD in the investigations~\cite{Braun:2011fw,Braun:2012zq} of the interplay of chiral symmetry breaking and gauge fields. 

By implication, a change of the phase transition line $T_\mathrm{c}(\mu)$ for small chemical potential must also have consequences for the overall structure of the phase diagram in finite volume. In particular, a smaller curvature and a flatter transition line imply that a possible critical end point of the first-order line at large chemical potential could move to larger $\mu_c$ in a finite volume. Since the first-order transition is accessible in the model and this calculational framework (see e.g.~\cite{Schaefer:2004en,Schaefer:2006ds}), these consequences can also be investigated directly.

Finite-size effects for the overall phase structure of the Nambu--Jona-Lasinio model at finite temperature and chemical potential have been investigated in~\cite{Abreu:2006pt,Kiriyama:2006uh} in a mean-field approach.

Similarly, finite-size effects for the phase diagram from a quark-meson model have been considered in~\cite{Palhares:2009tf,Palhares:2011jf,Fraga:2011hi,Fraga:2011rh}, including a discussion of effects on the critical endpoint of the first-order line at finite chemical potential and its possible signatures 
and finite-size scaling effects for the critical endpoint. Here the effects of different spatial boundary conditions for the fermions are also compared. 

Aspects of confinement have been included through a study of the PNJL model in a finite volume within the mean field approximation~\cite{Bhattacharyya:2012rp}. In this study, boundary conditions have not been considered explicitly, and the effect of the finite volume has been included solely through an IR cutoff $k_{\mathrm{IR}} =\pi/L$ in the gap equations. This corresponds to an implicit assumption of anti-periodic boundary conditions for the fermion fields in the spatial directions, because periodic boundary conditions would require an additional explicit treatment of the fermionic zero mode.

In general, mean-field studies of chiral symmetry breaking in a finite volume have to be taken with a \emph{caveat}: Since Goldstone mode fluctuations tend to restore chiral symmetry in small volumes, and since in general finite-volume effects are dependent on the presence of Goldstone mode fluctuations or critical fluctuations,   the finite-volume behavior in a mean field calculation will differ significantly from the full result. For small pion masses in particular, these effects will be large, with the size of the effect depending on the relative sizes of $m_\pi$ and $L$.  
However, as far as the phase structure is mainly determined by quark condensation effects, mean-field studies allow for an exploration of finite-volume effects. 

In~\cite{Palhares:2009tf}, the effects of different spatial boundary conditions for the quark fields on the first-order phase transition at large chemical potential and on the position of the critical end point are investigated. It is observed that the critical end point moves to larger values of $\mu_\mathrm{c}$ and smaller $T_\mathrm{c}$ with decreasing volume size $L$. This happens more rapidly for periodic boundary conditions, compared to the case with anti-periodic boundary conditions.  

In~\cite{Bhattacharyya:2012rp}, likewise a reduction of the temperature $T_\mathrm{c}$ at the critical endpoint with decreasing volume size $L$ has been found. 

The results for the shift of the crossover temperature in finite volume at $\mu=0$ are in general agreement with the finite-volume PNJL-model results in~\cite{Cristoforetti:2010sn}, which include zero-mode fluctuations. In both investigations, a decrease of the chiral crossover temperature by about the same magnitude has been observed, with a slightly larger effect when fluctuations are included~\cite{Cristoforetti:2010sn}. In contrast, in both studies the deconfinement transition temperature is essentially unaffected by a change in volume size. Since the finite volume affects only quarks and mesons and there is no back-coupling to the Polyakov loop potential, this is as expected. Lattice simulations of pure $\mathrm{SU}(3)$ gauge theory indicate that, depending on the boundary conditions, the deconfinement transition does in fact depend on the lattice size and receives sizable corrections for the volume sizes considered here~\cite{Bazavov:2007ny}. Dyson-Schwinger Equation studies also find significant finite volume effects which affect the confinement-deconfinement transition~\cite{Fischer:2007pf}.

The results for the finite-volume shift of the critical endpoint (CEP) from~\cite{Palhares:2009tf,Bhattacharyya:2012rp} are in general agreement with observations in the RG studies~\cite{Klein:2010tk,Braun:2011iz}, which include critical fluctuations and Goldstone mode effects. 

In~\cite{Tripolt:2013zfa}, a dedicated investigation of the first-order line and its critical end point in finite volume is undertaken with functional RG methods for a quark-meson model with $N_\mathrm{f}=2$ flavors. In order to investigate the first-order transition, the effective order parameter potential is discretized and solved on a grid~\cite{Adams:1995cv,Schaefer:2004en,Schaefer:2006sr}.
\begin{table}[t!]
\begin{center}
\begin{tabular}{|c|c||c|c|}
\hline 
$L\,[\text{fm}]$&
$m_{\pi}L$&
$\Delta\mu_{\rm CEP}$&
$\Delta T_{\rm CEP}$
\tabularnewline
\hline
\hline 
$10$&
$3.75$&
{$0.00(1)$}&
{$-0.06(6)$}
\tabularnewline
\hline 
${8}$&
${3.00}$&
{$0.00(1)$}&
{$\phantom{-}{0.00(6)}$}
\tabularnewline
\hline 
$6$&
$2.25$&
{$0.02(1)$}&
{$-0.26(5)$}
\tabularnewline
\hline 
$5$&
$1.88$&
{$0.04(1)$}&
{$-0.49(4)$}
\tabularnewline
\hline 
$4$&
$1.50$&
{$0.12(1)$}&
{$-0.74(4)$}
\tabularnewline
\hline
\end{tabular}
\end{center}
\caption{Relative shifts~$\Delta\mu_{\rm CEP}=(\mu_{\rm CEP}^{\rm max.}/\mu_{\rm CEP}^{(\infty)}-1)$ 
and~$\Delta T_{\rm CEP}=(T_{\rm CEP}^{\rm max.}/T_{\rm
  CEP}^{(\infty)}-1)$ of the critical endpoint (CEP) coordinates~$(\mu_{\rm CEP}^{\rm max.},T_{\rm CEP}^{\rm max.})$
in a finite volume~$V=L^3$ compared to the CEP position ~$\big(\mu_{\rm CEP}^{(\infty)},T_{\rm CEP}^{(\infty)}\big)$ in the infinite-volume limit. Results are taken from~\cite{Tripolt:2013zfa}.
The infinite-volume pion mass in this calculation is
$m_{\pi}=75\,\text{MeV}$. All fields have periodic boundary conditions in the spatial
directions. Errors are due to the uncertainty of the determination of the maximum of the susceptibility.}
\label{tab:CEP} 
\end{table}

In principle, critical behavior and a critical point are ill-defined in a finite volume, since no actually divergent susceptibility can be found, and all quantities are bounded by the finite volume. For the susceptibility relevant for the CEP in the model, we expect from dimensional arguments
\be
\chi_\sigma \sim L^2 \sim V^{2/3}.
\ee
(In the exact relationship $\chi_\sigma \sim L^{\gamma/\nu}$, the exponent $\gamma/\nu = 2 - \eta$ may receive corrections from an anomalous dimension $\eta$, which however is zero in the present approximation). Therefore it is a plausible choice to define the CEP in a finite volume as the point in the $(\mu, T)$-plane at which the susceptibility attains its maximum. In the infinite-volume limit, the result then coincides with the usual critical end point.

\begin{figure}
\includegraphics[clip=true,width=\columnwidth]{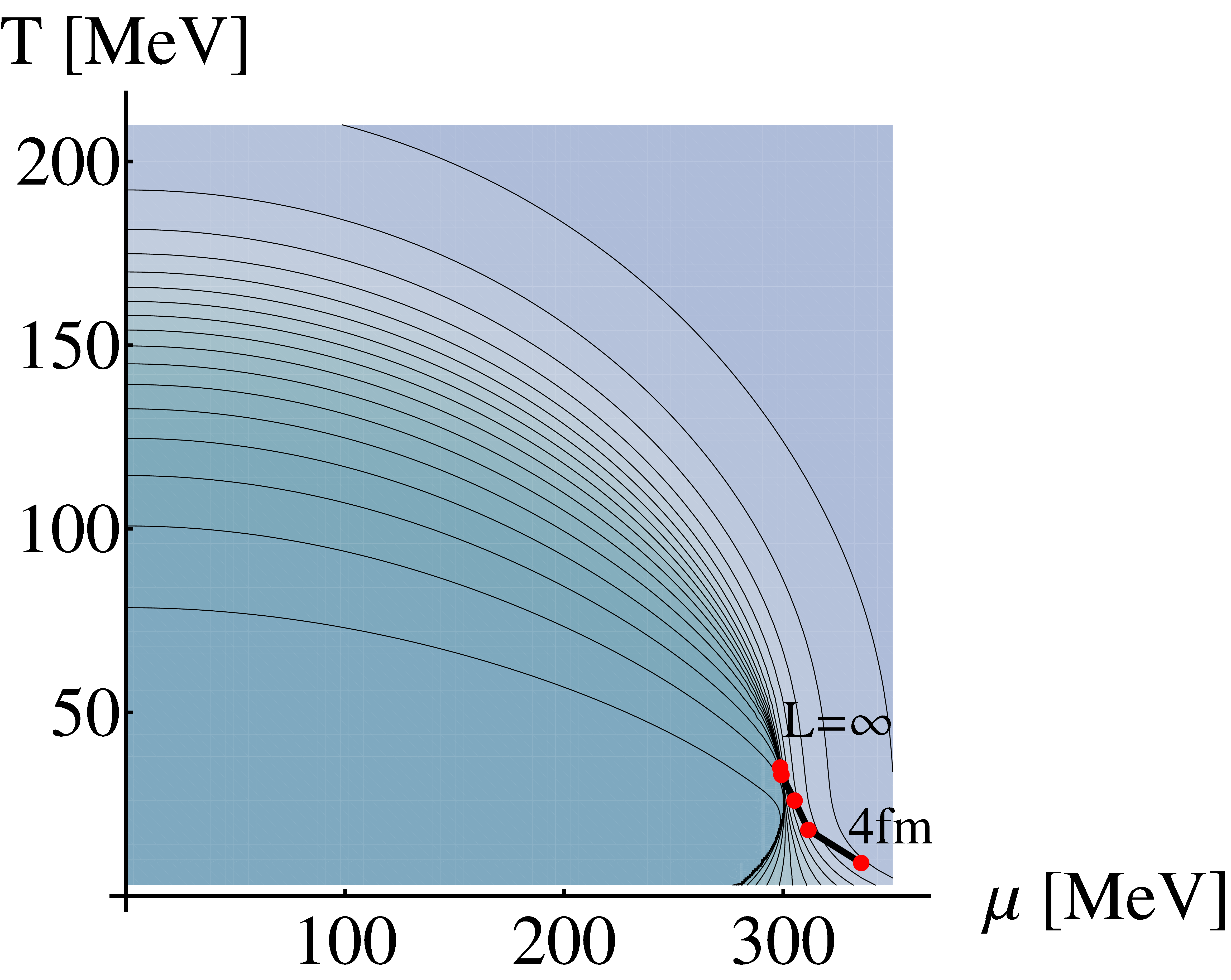}\llap{\makebox[3.5cm][l]{\raisebox{3.5cm}{\includegraphics[height=3cm]{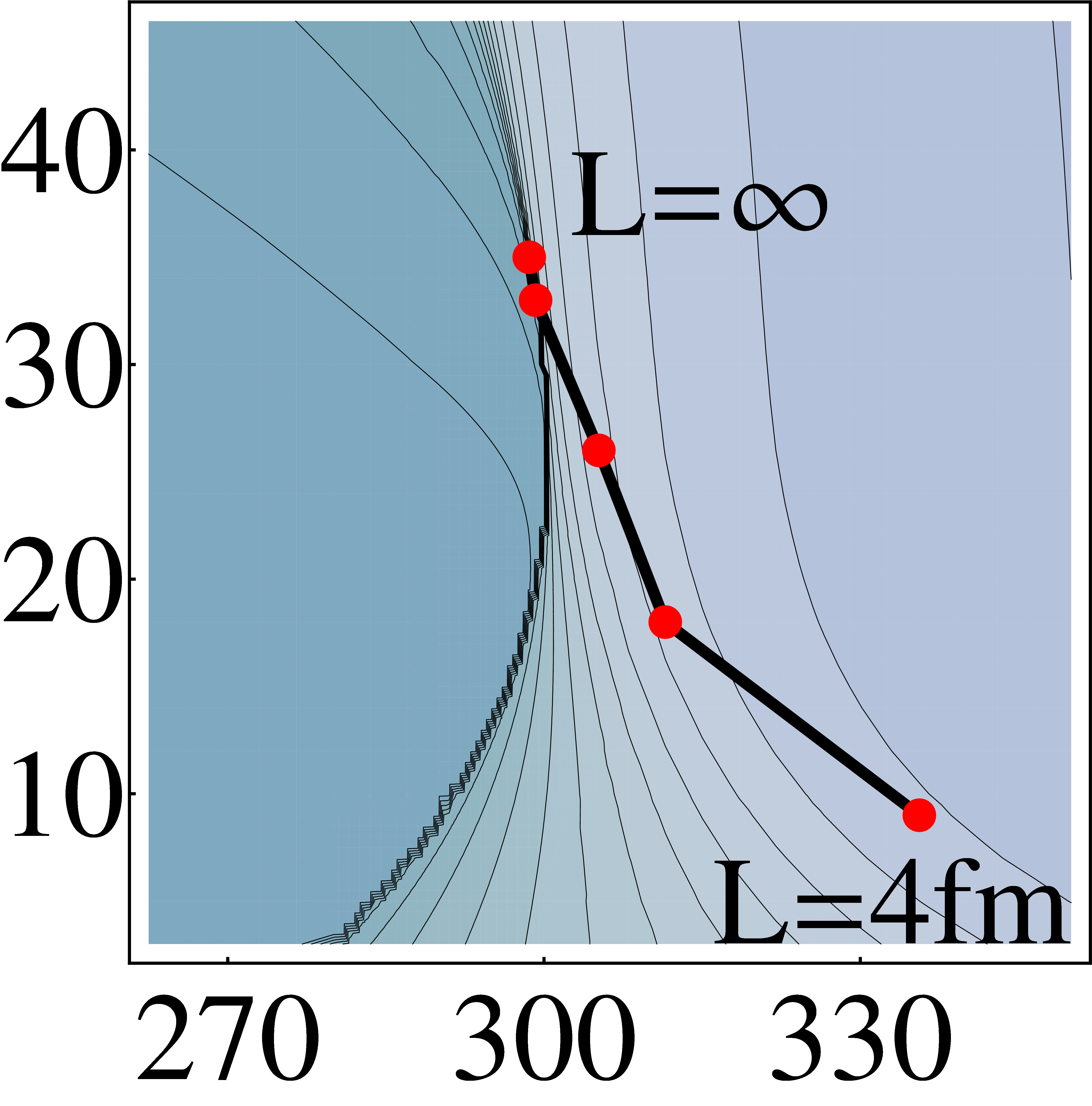}}}}
\caption{Contour plot of the magnitude of the chiral
  order parameter~$\sigma_0=f_{\pi}$ in the plane spanned by
  temperature and quark chemical potential {for
  $L$ towards infinity}, taken from \cite{Tripolt:2013zfa}: The order-parameter
  decreases from dark to light shading. The
  inlayed (red) dots show the behavior of the critical end point  (CEP) (associated with a
  Z($2$) symmetry) as a function of the volume size for periodic
  boundary conditions for the quark fields in spatial directions. One observes a shift of the  CEP towards larger values of the
  chemical potential and smaller values of the temperature
  with decreasing volume size.}
\label{fig:pd}
\end{figure}

The results for the shift of the critical end point of the first-order line are given in Tab.~\ref{tab:CEP}. Here these results are given as shifts relative to the infinite-volume position of the CEP, since the model results cannot be considered as a reliable guide for the actual position of the critical end point in QCD, should it exist.
For completeness, the infinite-volume results for the position of the CEP ($L \to \infty$) in this model calculation are  $(\mu_{\mathrm{CEP}}^{(\infty)}, T_{\mathrm{CEP}}^{(\infty)}) = (299\pm 1\,\mathrm{MeV}, 35 \pm 1\,\mathrm{MeV})$ for an (unphysical) pion mass of $m_\pi=75 \,\mathrm{MeV}$.
One finds that the effect on the temperature $T_{\mathrm{CEP}}$ in relative as well as absolute terms is much stronger than on the chemical potential $\mu_{\mathrm{CEP}}$.
These results are shown graphically in Fig.~\ref{fig:pd}, which is also taken from~\cite{Tripolt:2013zfa}. The shift of the critical end point of the first order line with a change of the volume size can be seen clearly in this plot.
These results confirm the observations from~\cite{Palhares:2009tf,Bhattacharyya:2012rp} for the overall shift of a first-order transition line and its critical end point qualitatively. While the effect of the fluctuations does not change the qualitative behavior, it leads to a larger shift already at larger volume sizes and a significantly larger broadening of the susceptibilities, compared to the mean-field results. While all these results have a certain model dependence, they demonstrate the importance of the fluctuations in a finite volume and their significant impact on the system. Overall, these results would indicate that a relatively large volume ($L\gtrsim 10\,\mathrm{fm}$) is required to effectively resolve the position of the critical end point. 

However, these values are clearly model dependent, since the absolute position of the CEP ultimately depends on the choice of model parameters. In a sense, the scales $T$, $\mu$, and $L$, which all appear in the quark propagator, are in competition with each other. Only if $1/L \sim \mu^{(\infty)}_{\mathrm{CEP}}$ or $1/L \sim T^{(\infty)}_{\mathrm{CEP}}$ do we expect significant finite-volume effects on the position of the CEP. If however both  $1/L \ll \mu^{(\infty)}_{\mathrm{CEP}}$ and $1/L \ll T^{(\infty)}_{\mathrm{CEP}}$, there should be little effect of the volume size on the observed CEP position. For QCD lattice simulations, this implies that the effects of the volume are small and the infinite-volume CEP position can be determined already for small box size, if the CEP is located at sufficiently large $\mu$ and $T$. If the position should be similar to the one in this model, the finite-volume shift would set in already for relatively large volume sizes, which would further complicate the search for the CEP in QCD lattice simulations.

\subsection{Quark number susceptibilities}
\label{sec:qsusc}
For the description of phase transition behavior at finite chemical potential, susceptibilities with regard to the quark chemical potentials $\mu_\mathrm{q}$ are of great importance: They are a measure of fluctuation effects in a critical region~\cite{Stephanov:1998dy,Stephanov:1999zu}.
These susceptibilities appear as the coefficients of a Taylor expansion of the thermodynamic potential~\cite{Gavai:2003mf} and are therefore very important for QCD lattice calculations at finite $\mu_\mathrm{q}$~\cite{Gavai:2004sd,Gavai:2005yk,Gavai:2005sd,Gavai:2008zr,Allton:2005gk}.
Susceptibilities of higher order and the convergence of the Taylor series have been explored in~\cite{Wagner:2009pm,Karsch:2010hm}.
Effects of the finite simulation volume on the radius of convergence of the expansion have been investigated in~\cite{Gavai:2004sd}.

QCD lattice simulations find a complex behavior for the flavor non-diagonal quark susceptibility
\be
\chi_{ud} &=& \frac{1}{VT} \frac{\partial^2}{\partial \mu_u \partial \mu_d} \log Z
\ee
as a function of the temperature $T$ \cite{Gavai:2005yk,Gavai:2005sd,Gavai:2008zr,Allton:2005gk}: there is a pronounced \emph{decrease} in this quantity, which peaks around the (pseudo-) critical temperature at the phase transition.
Such density fluctuations couple to vector mesons, so that the vector couplings become important for a description of related phenomena. 
In addition to the vector mesons, there are also pion fluctuations that carry an isospin charge and which can contribute to the flavor non-diagonal quark susceptibility $\chi_{ud}$.

Both in a quasi-particle model calculation \cite{Bluhm:2008sc} and a PNJL-model calculation in the saddle point approximation \cite{Roessner:2009zz}, the leading contributions to this susceptibility vanish. In the absence of vector mesons and pion fluctuations of finite isospin charge, only the Polyakov loop couples different quark flavors in a PNJL model, and the resulting effects are relatively small \cite{Roessner:2009zz,Cristoforetti:2010sn}. However, the effects of Polyakov loop fluctuations are independent of the volume size and always couple different quark flavors. Depending on the size of a simulation volume, they can be the determining contribution to the susceptibility.

A similar calculation was performed by Sasaki \emph{et al.}~\cite{Sasaki:2006ws}. The authors include in addition to the pion and sigma fields also vector mesons and a vector interaction, which supports fluctuations in the flavor non-diagonal susceptibilities $\chi_{\mathrm{ud}}$. The results depend on the choice for the value of the vector coupling, compared to the scalar coupling. A phenomenological model  with vector couplings conversely estimates the coupling strength from lattice QCD from non-zero flavor off-diagonal susceptibilities \cite{Ferroni:2010xf}. These calculations are volume independent.

The finite-volume calculation in~\cite{Cristoforetti:2010sn} in effect models fluctuations of a static, mean pion field in a PNJL model. Such fluctuations will be absent in the large-volume limit. This calculation captures mean-field fluctuations beyond the saddle point approximation in a finite volume. In the infinite-volume limit, where such fluctuations are absent, the saddle-point approximation becomes exact again. However, for small (realistic) pion masses and small volumes, these fluctuations can be quite significant. 

The pion contribution to the mean field fluctuations can be calculated from a static partition function, involving the pion fields which carry isospin charge and couple to the isospin chemical potential.
The static part of the chiral Lagrangian, coupled to the isospin chemical potential according to the symmetries of QCD, is given by~\cite{Kogut:1999iv,Kogut:2000ek,Son:2000xc,Son:2000by,Splittorff:2002xn}
\be
{\mathcal L }^{\mathrm{static}} &=&  \frac{1}{2}m_\pi^2 \pi^a \pi^a - 2 \mu_I^2 (\pi^1\pi^1 + \pi^2 \pi^2)
\ee
when expanded in powers of the pion field. 
We have defined the isospin chemical potential as $\mu_{I} = (\mu_u -\mu_d)/2$ in terms of the chemical potentials of the two lightest quark flavors. We use for the pion field the conventions $\vec{\pi}=(\pi^1,\pi^2,\pi^3)$ and  $\pi^\pm=\frac{1}{\sqrt{2}}(\pi^1\pm i\pi^2)$, $\pi^0=\pi^3$ with $\tau^\pm=\frac{1}{2}(\tau^1\pm i\tau^2)$, so that
\be\nonumber
	&\vec{\tau}\cdot\vec{\pi}=\sqrt{2}(\tau^+\pi^-+\tau^-\pi^+)+\tau^3\pi^0.&
\ee

In lattice QCD calculations, the aspect ratio of the Euclidean volume between the spatial directions and the Euclidean time direction is kept fixed, while the temperature is adjusted by changing the lattice spacing and thus the overall scale. Hence both temperature $T$ and spatial volume size $L$ change effectively at the same time. For this reason, we also choose to keep the product $LT$ constant, and the spatial volume is given by
\be
V = k/T^3 \quad \mathrm{with} \quad k =  (N_\mathrm{s}/N_\mathrm{t})^3, 
\ee
where $N_\mathrm{s}$ is the number of lattice sites in the spatial direction and $N_\mathrm{t}$ in the Euclidean time direction. For  $N_s/N_t=4$  we have $k=64$, which is the aspect ratio used in the simulations~\cite{Gavai:2008zr,Allton:2005gk}.

\begin{figure}
       \begin{center}
      \includegraphics[width=0.9\textwidth]{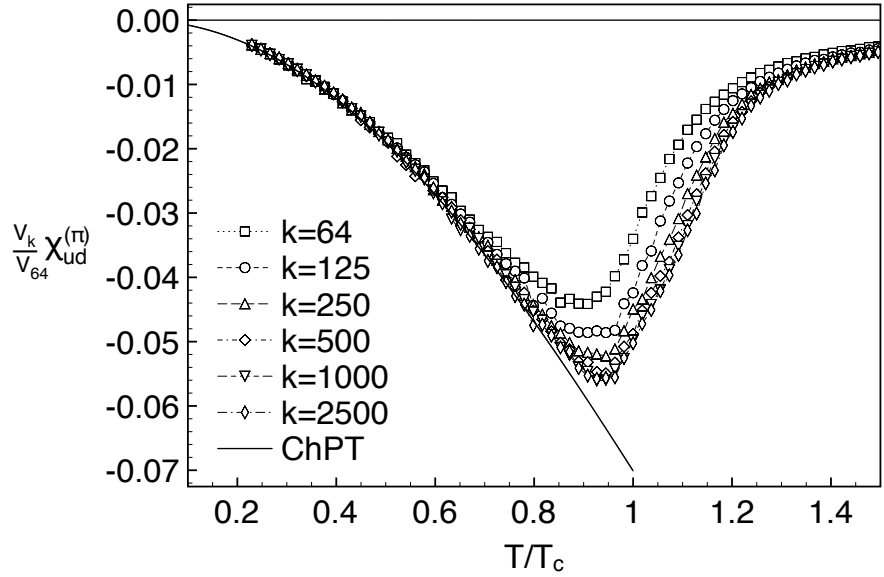}
      \end{center}
      \caption{Pion contribution to the off-diagonal quark susceptibility $\chi_{ud}$ from the PNJL model (symbols), compared to the finite-volume ChPT prediction (solid line) using a temperature-dependent pion mass~\cite{Cristoforetti:2010sn}. The finite-volume PNJL results are rescaled by the volume ratio $V_k/V_{64}$ and thus scale with the curve corresponding to the volume with $k=64$. Figure reprinted with permission from M. Cristoforetti \emph{et al.}, Phys. Rev. D81, 114017 (2010). Copyright (2010) by the American Physical Society.}\label{fig:c2chpt}
\end{figure}

Making use of these results, the static partition function 
\be
Z^{\mathrm{static}}&=& \int \prod_{a=1}^3 \mathrm{d} \pi^a\; \exp \left\{ -\frac{V}{T} \left[\frac{1}{2} m_\pi^2 \pi^a \pi^a - 2 \mu_I^2 (\pi^1\pi^1+\pi^2\pi^2)  \right] \right\}
\ee
can be trivially evaluated. The static pion contribution to the second-order flavor non-diagonal expansion coefficient in terms of the quark chemical potentials is given by
\be
\chi^{(\pi, \mathrm{static})}_{ud} &=& \frac{1}{VT} \frac{\partial^2}{\partial \mu_u \mu_d} \log Z = - \frac{2}{VT} \frac{1}{m_\pi^2} = -\frac{2 T^2}{k} \frac{1}{m_\pi^2}
\ee
where $V=k/T^3$. Using the one-loop result for the temperature dependence of the pion mass~\cite{Gasser:1986vb}, below the chiral transition temperature $T_c$ this result nicely agrees with that of the Monte-Carlo evaluation of the mean-field fluctuations in the PNJL model. A comparison can be found in Fig.~\ref{fig:c2chpt} taken from~\cite{Cristoforetti:2010sn}, where the prediction from the static chiral Lagrangian and the numerical results are seen to agree nicely. 

A comparison of the results to those from lattice QCD simulations~\cite{Gavai:2008zr,Allton:2005gk} show that the pion effects have the right size to explain the observed flavor off-diagonal susceptibilities.
However, static fluctuations are damped to zero in the infinite-volume limit, so these static effects will play a diminishing role for increased simulation volumes. Since contributions due to Polyakov loop fluctuations are unaffected, these results predict a decreasing magnitude for the susceptibility $\chi_{ud}$ with increasing volume size, bounded by the prediction from the Polyakov loop contribution found from PNJL- or PQM-type models. 

In the recent investigation~\cite{Bhattacharyya:2014uxa}, some of the same authors of~\cite{Bhattacharyya:2012rp} extend their framework for the finite-volume analysis of the $N_\mathrm{f}=2$ PNJL model to quark number susceptibilities. The same \emph{caveats} as before apply: The analysis is restricted to a mean-field calculation, and the finite volume is introduced solely through an IR cutoff $k_{\mathrm{IR}} =\pi/L$, which implies anti-periodic spatial boundary conditions for the quark fields. Even static finite-volume fluctuations as treated in~\cite{Cristoforetti:2010sn} are absent in this framework.
Susceptibilities for baryon (quark) number ($X=B$) and for isospin ($X=I$) 
\be
c_n^{(X)}(T) &=& \left .\frac{1}{n!} \frac{\partial^n}{\partial \left(\frac{\mu_X}{T} \right)^n} \frac{1}{T^4}\Omega(T, \mu_B, \mu_I) \right |_{\mu_B = 0, \mu_I = 0}
\ee
are considered up to 6$^\mathrm{th}$ order. 
The main effect of the finite volume is a dampening and broadening of the susceptibilities close to the transition. As the main observation in this work, it is found that the quark number susceptibilities scale as naively expected with the volume only away from the crossover region, but violate the expected volume scaling behavior in the transition region. As a consequence, certain ratios of susceptibilities, e.g. the kurtosis $c_4/c_2$,  which are independent of volume in the bulk of a phase, become volume-dependent in the crossover region. Since this is the region in which also the strongest critical fluctuations are expected, and since already static pion fluctuations in this region affect the isospin susceptibility~\cite{Cristoforetti:2010sn}, it would be useful to test the robustness of these results by including such fluctuation effects beyond the mean field approximation.

In the strong-coupling limit lattice study~\cite{Ichihara:2015kba} of baryon number fluctuations at finite density, indeed significant finite-volume effects for these susceptibilities are observed. They are qualitatively similar to those found in~\cite{Bhattacharyya:2014uxa}, insofar as they appear e.g. in the kurtosis primarily in the transition region.
For a discussion of the fluctuation effects on higher-order cumulants, see e.g.~\cite{Friman:2011pf,Skokov:2011yb}, which also makes use of functional RG methods and applies them to a Polyakov-loop extended quark-meson model, but does not consider finite volumes.

In the study~\cite{Skokov:2012ds}, effects of a fluctuation in volume size (as they might perhaps occur in a heavy-ion collision) on baryon number fluctuations are calculated in a PQM-model, using functional RG methods. The goal of the study is to assess sources of non-critical fluctuations, in order to distinguish their effects from those of critical fluctuations in the vicinity of a phase transition. The authors find that for moments of the fluctuation of net baryon number, volume fluctuations tend to suppress the signal from critical fluctuations at the phase transition.

\section{Conclusion}

Because chiral symmetry is spontaneously broken and pions appear as light degrees of freedom -- light compared to other hadrons -- , significant finite-size effects can be observed in the QCD vacuum. These effects can be studied by a variety of different methods, such as Random Matrix Theory or Chiral Perturbation Theory. They provide important information for our understanding of chiral symmetry breaking and its mechanisms in QCD and can be used as analytical tools. 

For very small volume size or large temperature, chiral symmetry is effectively restored, and methods are needed which are capable of dealing with a transition from the phase with spontaneously broken symmetry to one with restored symmetry. 
One possible approach is the use of \emph{models} for chiral symmetry breaking in order to study the transition regime. The quark-meson model is such a model of dynamical chiral symmetry breaking which is very useful for certain purposes and which can be used to include mesonic fluctuations.

The analysis of finite-volume effects in QCD is of considerable interest, since finite-volume lattice simulations are an important source of our knowledge about QCD. In recent years significant progress has been achieved in lowering the pion masses in these lattice simulations. Smaller pion masses imply larger finite-volume effects at the same lattice size, and thus a thorough understanding of such effects becomes even more important. 
At the chiral phase transition, where the behavior of the system is dominated by the light critical fluctuations, the effects of a finite volume on the critical scaling behavior also become very important. 

In recent years, functional Renormalization Group methods have been brought to bear on these problems: They account for the effects of fluctuations, correctly describe critical behavior, and can be applied over a wide range of momentum scales. They are not limited by the crossing of a phase transition line and the accompanying critical fluctuations or a possible change in the degrees of freedom in a theory. In the course of these investigations, the necessary Renormalization Group techniques for finite-volume calculations have been developed in two different  cutoff schemes. 

On the one hand, it is possible to make contact with finite-volume results from chiral perturbation theory, which is an important benchmark. In this context, it is found that the results are dependent on the choice of spatial boundary conditions for the fields, and unexpected results for periodic spatial boundary conditions for quarks have been found,  a very common choice in lattice simulation results.
This has additional consequences for the phenomenology of models for chiral symmetry breaking and by implication also for QCD: We find effects on the chiral susceptibility, and even on the behavior of the phase transition line at finite temperature and chemical potential.

On the other hand, the finite-size scaling behavior in the O(4) and O(2) symmetry classes has been explored, which are expected to be the relevant ones for QCD with $N_\mathrm{f}=2$ light quark flavors. We have identified the regime in pion mass and volume size in which finite-size scaling analysis ought to be applicable, and we have investigated possible deviations from the scaling behavior due to small volume size and large quark masses. In particular with regard to current lattice simulations with ever decreasing pion masses, these results can provide useful guidance for a finite-size scaling analysis.

The investigations and the results reviewed here demonstrate that finite-volume effects in QCD are an important consideration in the interpretation of lattice QCD simulation results, and that they can be a very useful tool for the analysis of the chiral phase transition.

\section*{Acknowledgements}
I acknowledge support of the DFG cluster of excellence "Origin and Structure of the Universe".\\

I would like to thank all of my collaborators for the enjoyable and fruitful collaborations over the past years, namely Jens Braun, Marco Cristoforetti, Matthias Drews, Kenji Fukushima, Thomas Hell, Kazuhiko Kamikado, Piotr Piasecki, Hans-J\"urgen Pirner, Jonas Probst, Amir Rezaeian, Bernd-Jochen Schaefer, Paul Springer, Dominique Toublan, Arno Tripolt, Jac Verbaarschot, and Wolfram Weise.\\

In particular, I wish to thank Wolfram Weise for his support and the opportunity to pursue independent research interests while  working in his group.\\

I am grateful to Hans-J\"urgen Pirner for first introducing me to functional Renormalization Group methods and their application to finite-volume problems, which has proved an inspiration. \\

Particular thanks also go to Jens Braun for exploring many of the interesting phenomena considered here together with me, for innumerable useful discussions, and for a thoroughly enjoyable collaboration over the past years. I would also like to thank him for a reading of the manuscript. Any remaining mistakes and oversights are, however, mine alone.\\

Last but not least, I would like to thank my wife, Anna Br\"uggemann, for her love and support and her patience, and our children Clara and Clemens for the joy they bring to our lives. 

\newpage





\appendix

\section{Threshold functions for the PTRG}
The RG flow equations can be written in a compact way by using so-called threshold functions. These incorporate the contributions of individual degrees of freedom to the RG flow. They depend on the cutoff function and its parameters, temperature, volume size and boundary conditions, and correspondingly distinguish between fermionic and bosonic degrees of freedom. For completeness, we list the individual threshold functions for the PTRG cutoff scheme.
With the proper-time cutoff function
\be
f_{a}(\tau k^2) &=& \frac{\Gamma(a+1, \tau k^2)}{\Gamma(a+1)}, 
\ee
the PTRG threshold function in infinite volume in $d$ space-time dimensions is
\be
\ell_{a+1}^{(d)}(\omega)&=& \frac{1}{(4\pi)^{d/2}} \frac{\Gamma(a+1-d/2)}{\Gamma(a+1)} \frac{1}{\omega^{(a+1-d/2)}}.
\label{eq:PTRGthreshold_iv}
\ee
In  a $d$-dimensional finite volume $V=L^d$, the corresponding threshold function becomes
\be
\ell_{a+1}^{(d)}(L, \omega)&=& \frac{1}{(4\pi)^{d/2}}\frac{1}{\Gamma(a+1)} \left(\frac{L^2}{4 \pi}\right)^{a+1-d/2} \times \el
&&\quad \quad \times  \int_0^\infty \mathrm{d}s \, s^a \exp(-s \left(\frac{L^2}{4 \pi}\right) \omega) (\Theta_\mathrm{l}(s))^d .\el
\ee
For periodic boundary conditions in the spatial directions, the function $\Theta_\mathrm{p}(s)$ that appears in the cutoff is given by
\be
\Theta_\mathrm{p}(s) &=& \vartheta_3(0, \mathrm{e}^{-\pi s}) =\sum_{n=-\infty}^\infty \mathrm{e}^{-\pi n^2 s}
\ee
and expressed in terms of a Jacobi $\vartheta$-function. 
For anti-periodic boundary conditions, the corresponding function is
\be
\Theta_{\mathrm{ap}}(s) &=& \vartheta_2(0, \mathrm{e}^{-\pi s}) = \sum_{n=-\infty}^\infty \mathrm{e}^{-\pi s (n+\frac{1}{2})^2}.
\ee
Using $\Theta_{\mathrm{l}}(s) \simeq s^{-1/2}$ for $s \to 0$ for both $\mathrm{l}= \mathrm{p}$ and $\mathrm{l}= \mathrm{ap}$, one recovers the infinite-volume threshold function in the limit $L \to \infty$.

The Jacobi $\vartheta$-functions used here have the following series expansions:
\be
\vartheta_2(z, q) &=& \sum_{n=-\infty}^\infty q^{(n+\frac{1}{2})^2} \mathrm{e}^{(2n + 1) \mathrm{i} z}\nn\el
\vartheta_2(0, \mathrm{e}^{-\pi s}) &=& \sum_{n=-\infty}^\infty \mathrm{e}^{-\pi s (n+\frac{1}{2})^2} \nn\el
\vartheta_3(z, q) &=& \sum_{n=-\infty}^\infty q^{n^2} \mathrm{e}^{2n \mathrm{i} z}\nn\el
\vartheta_3(0, \mathrm{e}^{-\pi s}) &=& \sum_{n=-\infty}^\infty \mathrm{e}^{-\pi s n^2} \nn
\ee
For a $d$-dimensional Euclidean  volume $V=\frac{1}{T}\times L^{d-1}$ at finite temperature $T$, the threshold functions depend on the boundary conditions for bosons (B) and fermions (F) in the Euclidean time direction, and for the freely-chosen boundary conditions in the spatial directions. 
In total, we therefore need to distinguish between three cases of practical relevance and obtain the following three threshold functions ($t=TL$):
\be
\ell_{a+1}^{(\mathrm{B}, \mathrm{p})(d)}(t, L, \omega)&=& \frac{1}{(4\pi)^{d/2}}\frac{1}{\Gamma(a+1)}t  \left(\frac{L^2}{4 \pi}\right)^{a+1-d/2} \times \el
&& \times \int_0^\infty \mathrm{d}s \, s^a \exp(-s \left(\frac{L^2}{4 \pi}\right) \omega) \Theta_{\mathrm{p}}(s t^2) (\Theta_{\mathrm{p}}(s))^{d-1}\el
\ell_{a+1}^{(\mathrm{F}, \mathrm{p})(d)}(t, L, \omega)&=& \frac{1}{(4\pi)^{d/2}}\frac{1}{\Gamma(a+1)}t  \left(\frac{L^2}{4 \pi}\right)^{a+1-d/2} \times \el
&& \times \int_0^\infty \mathrm{d}s \, s^a \exp(-s \left(\frac{L^2}{4 \pi}\right) \omega) \Theta_{\mathrm{ap}}(s t^2) (\Theta_{\mathrm{p}}(s))^{d-1}\el
\ell_{a+1}^{(\mathrm{F}, \mathrm{ap})(d)}(t, L, \omega)&=& \frac{1}{(4\pi)^{d/2}}\frac{1}{\Gamma(a+1)}t  \left(\frac{L^2}{4 \pi}\right)^{a+1-d/2} \times \el
&& \times \int_0^\infty \mathrm{d}s \, s^a \exp(-s \left(\frac{L^2}{4 \pi}\right) \omega) \Theta_{\mathrm{ap}}(s t^2) (\Theta_{\mathrm{ap}}(s))^{d-1}.\el
\ee
All these functions again reduce to the function~\eqref{eq:PTRGthreshold_iv} in the limit of $T\to 0$ and $L\to \infty$.
Making once again use of $\Theta_{\mathrm{l}}(s) \simeq s^{-1/2}$ for $s \to 0$, one finds that 
\be
\lim_{L \to \infty} \ell_{a+1}^{(\mathrm{B}, \mathrm{p})(d)}(t, L, \omega)
&=& \frac{1}{(4\pi)^{d/2}}\frac{\Gamma(a+1-d/2)}{\Gamma(a+1)}\frac{1}{\omega^{(a+1-d/2)}}  \el
&=& \ell_{a+1}^{(d)}(\omega).
\ee
(note that $t = TL$ is kept fixed, so that the limit $L \to 0$ implies $T \to 0$ simultaneously).
To obtain this result, one makes use of the equality
\be
\lefteqn{\lim_{L \to \infty} \int_0^\infty \mathrm{d}s\, s^a \exp(-s \left(\frac{L^2}{4 \pi}\right) \omega) \Theta_{\mathrm{l}}(s t^2) (\Theta_{\mathrm{l^\prime}}(s))^{d-1} }\el
&=& \lim_{L \to \infty} \frac{1}{t}  \int_0^\infty \mathrm{d}s\, s^{a - d/2} \exp(-s \left(\frac{L^2}{4 \pi}\right) \omega) \el
&=&  \lim_{L \to \infty} \frac{1}{t} \left(\frac{4 \pi}{L^2 \omega}\right)^{(a+1-d/2)} \Gamma(a+1-d/2).
\ee

\newpage

\bibliographystyle{h-physrev}
\bibliography{FV_review_report.bib}








\end{document}